\documentclass[journal]{IEEEtran}
\usepackage{graphicx}
\usepackage{amsmath,amssymb,dsfont,bbm,epstopdf,pgfplots,mathtools,enumitem,mathrsfs, bbm, algpseudocode , graphicx, dblfloatfix, booktabs, physics, algorithm,algcompatible, derivative, physics, rotating, subcaption} 
\usepackage[normalem]{ulem}
\usepackage{color}
\usepackage{cite}%
\usepackage{url}
\usepackage[hidelinks]{hyperref}
\usepackage{graphics}
\usepackage{tikz}
\usepackage{pgfplots}
\usepackage{subcaption}
\usepackage[utf8]{inputenc}
\usepackage[T1]{fontenc}
\usepackage[left=0.6in,right=0.6in,top=0.71in, bottom = 0.94in]{geometry}
\usepackage[absolute,overlay]{textpos}
\usetikzlibrary{shapes, arrows, decorations.markings, arrows.meta}
\usetikzlibrary{patterns}
\usetikzlibrary{positioning, fit, backgrounds, calc}
\allowdisplaybreaks
\graphicspath{{.}{./Figures/}}
\allowdisplaybreaks[4]

\usepackage{xcolor}
\usepackage[normalem]{ulem}

\usepackage{multirow}

\begin{document}

\begin{textblock*}{7.3in}(0.6in,0.2in)
\noindent\fbox{\parbox{\dimexpr7.3in-2\fboxsep-2\fboxrule}{\footnotesize
H.~Nikbakht, M.~Ying, D.~Shakya, D.~Abraham, and T.~S.~Rappaport, ``ARIA: An AI-powered radio-channel inference and analysis pipeline,'' submitted to \emph{IEEE Trans. Wireless Commun.}, 2026, under review.}}
\end{textblock*}

\title{ARIA: An AI-Powered Radio-Channel Inference and Analysis Pipeline }

\author{\IEEEauthorblockN{\hspace{-0.5cm} Homa Nikbakht, Mingjun Ying, Dipankar Shakya,  Daniel Abraham, and Theodore S. Rappaport, \IEEEmembership{Life Fellow,~IEEE}}
\thanks{This work was supported by the NYU WIRELESS Industrial Affiliates Program and NYU Technology Acceleration and Commercialization Award.}
\thanks{H.~Nikbakht,  M.~Ying, D.~Shakya, D.~Abraham, and T.~S.~Rappaport are with NYU WIRELESS, New York University, Tandon School of Engineering, Brooklyn, NY, USA (emails: \{homa.n, yingmingjun, dshakya, dan.abraham, tsr\}@nyu.edu). }}

\maketitle
\begin{abstract}
Statistical channel models play a key role in the design and standardization of next-generation wireless systems. The models substitute for costly and time-consuming field measurements when fitted distributions accurately reproduce the measured channel impulse responses. Future wireless systems will operate in global spectrum allocations from $1$ to $150$\,GHz. Over the full span, analytically chosen distribution families fitted to scarce measurement data fail to deliver the required accuracy. Generative artificial intelligence (AI) models instead learn each distribution from the radio-propagation data itself, adapt to the measured structure at every frequency, and scale across the entire band. We present \emph{ARIA}, an AI-powered Radio-channel Inference and Analysis pipeline that learns each channel-parameter distribution directly from the measurements. The pipeline first extracts power-delay profiles from the raw data and recovers the primary channel parameters through time-cluster spatial-lobe (TCSL) analysis. Three neural density estimators, namely mixture density networks (MDN), neural spline flows (NSF), and denoising diffusion implicit models (DDIM), then learn each parameter's distribution, with kernel density estimation (KDE) as a classical baseline. To identify the best estimator, we use a composite rank across goodness-of-fit metrics with Friedman--Nemenyi significance testing. A maximum-likelihood step, anchored to the measured samples, then converts the density learned by the best estimator into closed-form parametric distribution families. We instantiate ARIA on the NYU~WIRELESS $16.95$\,GHz indoor hotspot (InH) campaign and find that NSF outperforms the other estimators on every metric. We also show that a conditional NSF based on feature-wise linear modulation (FiLM), trained jointly on the $6.75$\,GHz and $16.95$\,GHz measurements, improves over training on the single frequency alone. We feed the distributions fitted by ARIA into the NYUSIM channel simulator and generate $10{,}000$ channel realizations at $16.95$\,GHz, reproducing the measured omnidirectional root-mean-square (RMS) delay spread medians within about $2$\,ns, the per-lobe azimuth and zenith spread medians within $1^\circ$, and the omnidirectional zenith spread medians within $0.5^\circ$.
\end{abstract}


\section{Introduction}
\label{sec:introduction}

Next-generation wireless systems will operate over an unprecedented span of spectrum. Sixth-generation (6G) cellular targets the sub-$7$\,GHz Frequency~Range~1 (FR1)~\cite{3GPP_TS38101_1}, the upper mid-band FR3~\cite{Bazzi2025}, the millimeter-wave FR2~\cite{3GPP_TS38101_2}, and the sub-terahertz (sub-THz) regime above $100$\,GHz~\cite{Rappaport2019, Wang2025_6G}, while Wi-Fi operates in the $2.4$, $5$, and $6$\,GHz unlicensed bands and the $60$\,GHz band of IEEE~802.11ad/ay~\cite{IEEE80211ad, IEEE80211ay}. Channel models for system design and standardization must cover the entire range, across which propagation behavior varies considerably with carrier frequency.
At higher carrier frequencies, weaker diffraction, together with higher penetration and scattering losses, widens the gap between line-of-sight (LOS) and non-line-of-sight (NLOS) received power~\cite{Pi2011, Rangan2014, Rappaport2019}. The multipath component also becomes sparser as frequency increases, with received energy concentrated in fewer clusters~\cite{Shakya2024, Shakya2025InH, Wang2025_6G}.
 Capturing the frequency-dependent behavior in a single, tractable simulation framework is essential for link-level and system-level evaluation of multiple-input multiple-output (MIMO) and beamforming designs, multiple-access methods, and emerging use cases such as integrated sensing and communication (ISAC) studied at 3GPP, ITU, IEEE~802.11, and related industry forums~\cite{3GPP38901, Poddar2023}.

Statistical channel models provide such a simulation framework, describing the channel impulse response through probability distributions over primary channel parameters such as cluster counts, multipath delays and powers, and arrival and departure angles, and form the basis of the 3GPP TR~38.901 channel model~\cite{3GPP38901} and of widely used simulators such as NYUSIM~\cite{Samimi2016, Sun2017, Poddar2023, Jariwala2026_NYUSIM}, QuaDRiGa~\cite{Jaeckel2014}, and COST~2100~\cite{Liu2012_COST2100}. The simulators share a common workflow that first extracts the primary channel parameters from measurements, and then fits each parameter to a closed-form parametric distribution, and generates sample functions of the channel impulse response by Monte~Carlo sampling from the fitted distributions. Simulator outputs are usually validated against second-order statistics of the same measurements, namely the root-mean-square (RMS) delay and angular spreads.

The described workflow suffers from two limitations. First, the simulators are limited to a small library of closed-form distributions, such as exponential, lognormal, or Gaussian, and the chosen family is often inherited unchanged across frequency bands~\cite{Samimi2016, Sun2017}. The limited library cannot represent the heavy tails, multimodality, or discrete structure that channel statistics exhibit at the higher bands. Second, the sample size available for each fit is small. Some parameters, especially the number of time clusters and spatial lobes in LOS, yield only a few observations per scenario in typical FR3, mmWave, and sub-THz campaigns~\cite{Shakya2025InH, Shakya2024}. Maximum-likelihood estimation (MLE) applied directly to the few measured samples is biased and unstable, often returning degenerate parameter values that distort the simulator output. When the limited library is combined with a small data set, it becomes difficult to identify which of the two factors, if not both, corrupts the simulation output.

Machine learning research has meanwhile developed neural density estimators that learn arbitrary distributions from data. Mixture density networks~\cite{Bishop1994}, normalizing flows~\cite{Rezende2015, Papamakarios2021, Durkan2019}, denoising diffusion models~\cite{SohlDickstein2015, Ho2020, Song2021}, and flow-matching models~\cite{Lipman2023} are now standard tools in probabilistic generative modeling for vision and speech. A growing line of work brings generative AI (GenAI) tools to wireless channels for channel estimation~\cite{Fesl2024, Zhou2025_HighDim}, MIMO channel synthesis~\cite{Orekondy2022}, and sub-THz channel generation~\cite{Xia2024}. Recent work~\cite{Huang2022_AIradioI, Huang2022_AIradioII, Zemen2025} argues that artificial intelligence (AI) can replace analytically chosen channel representations with learned ones, and industry whitepapers have begun describing data-driven methods~\cite{Samsung2025}.
To the best of our knowledge, no published work has employed neural density estimators to learn the primary channel parameter distributions from raw measurements and export validated closed-form distributions for statistical channel simulators.

In this paper, we introduce \emph{ARIA}, an end-to-end \textbf{A}I-powered \textbf{R}adio-channel \textbf{I}nference and \textbf{A}nalysis pipeline that addresses both limitations. ARIA replaces the limited distribution library with neural density estimators that adapt to the measured data, and overcomes the small-sample limitation through synthetic data augmentation, where additional samples are drawn from the learned density for each parameter. The pipeline applies to any carrier frequency for which measurement data is available, covering all cellular and Wi-Fi bands relevant for next-generation wireless networks. ARIA proceeds through the following stages.
\begin{itemize}

  \item \emph{Measurement processing and parameter extraction:} Starting from raw measurement data, ARIA applies the channel-sounder calibration to extract the system gain
  and the receiver's linear dynamic range; from each calibrated measurement, ARIA then extracts the power-delay profile (PDP), identifies multipath components (MPCs), aligns the PDPs in time, and fits standard large-scale path-loss models. A time-cluster spatial-lobe (TCSL) analysis~\cite{MacCartney2017, Ju2021, Shakya2024, Shakya2025InH} then yields primary channel parameters per LOS and NLOS scenario.

  \item \emph{Distribution fitting with density
  estimators:} Four density estimators fit each channel parameter's empirical distribution, namely kernel density estimation (KDE)~\cite{Parzen1962, Silverman1986} as a classical baseline, mixture density networks (MDN)~\cite{Bishop1994}, neural spline flows (NSF) built on monotonic rational-quadratic spline transforms~\cite{Durkan2019, Papamakarios2021}, and denoising diffusion implicit models (DDIM)~\cite{Song2021}.

 \item \emph{Model selection:} For each distribution fitting task, the density estimator models are ranked on five goodness-of-fit metrics, namely the
  Kolmogorov--Smirnov (KS)~\cite{Massey1951},
  Cram\'er--von Mises (CvM)~\cite{Anderson1962},
  Anderson--Darling (AD)~\cite{Anderson1954}, $W_1$
  Wasserstein~\cite{Villani2009}, and
  energy~\cite{Szekely2013} distances, and the normalized
  per-metric ranks are averaged into a composite
  rank~\cite{Demsar2006, Kansal2023}. The model with the lowest
  composite rank wins the task. Across all tasks, a Friedman
  omnibus test followed by Nemenyi post-hoc
  comparisons~\cite{Demsar2006} then determines whether the
  observed differences between models are statistically
  significant.

 \item \emph{Closed-form parametric fitting:} Once the best
  model is selected for each task, ARIA converts the learned
  density into a closed-form parametric distribution through an
  augmented MLE step that fits a parametric family to the
  combined set of measured and synthetic samples.
  Physical constraints on each parameter (e.g., positive support
  for delays) restrict the candidate parametric families, and
  ARIA selects the family with the smallest KS distance to the
  measurements.

  \item \emph{Multi-frequency conditioning:} When neighboring
  frequency bands offer additional measurements, ARIA trains a
  single conditional density model on all bands jointly,
  improving the fit at the target frequency while keeping the
  distribution at each band distinct. The
  conditioning is implemented through Feature-wise Linear
  Modulation (FiLM)~\cite{Perez2018_FiLM}, which modulates the
  internal layers of the density model on a learned embedding
  of the carrier frequency.

  \item \emph{End-to-end validation:} The exported closed-form
  distributions feed a NYUSIM Monte Carlo channel generator that
  draws $10{,}000$ synthetic channel realizations per scenario.
  ARIA then compares the omnidirectional, directional, and
  per-lobe RMS delay and angular spreads from the generated realizations
  against the corresponding measured second-order statistics.
  Because the RMS spreads are not fit directly but emerge from
  the joint sampling of the primary first-order distributions,
  a close match confirms both the individual per-task fits and
  the joint behavior produced by the simulator.
\end{itemize}

\begin{figure*}[!t]
\centering
\resizebox{\textwidth}{!}{%
\begin{tikzpicture}[
  font=\small,
  >={Latex[length=2.4mm,width=1.8mm]},
  block/.style={
    rectangle, draw, line width=0.8pt, rounded corners=2pt,
    minimum height=1.0cm, minimum width=1.8cm,
    align=center, fill=white, inner sep=3pt
  },
  stagebox/.style={
    rectangle, draw, dashed, line width=1.0pt, rounded corners=3pt,
    inner xsep=6pt, inner ysep=24pt,
    fill=gray!14
  },
  substagebox/.style={
    rectangle, draw, dashed, line width=1.0pt, rounded corners=2pt,
    inner xsep=8pt, inner ysep=6pt,
    fill=white
  },
  stagelabel/.style={
    font=\small, anchor=south, yshift=2pt
  },
  substagelabel/.style={
    font=\footnotesize, anchor=south, yshift=2pt
  },
  arr/.style={->, line width=0.8pt, >={Latex[length=2.4mm,width=1.8mm]}},
  line/.style={line width=0.8pt},
  flowlabel/.style={
    font=\small\itshape
  }
]

\node[block] (data) {Measurement\\Data};

\node[block, right=1.0cm of data]  (cal)   {Calibration};
\node[block, right=0.3cm of cal]   (pdp)   {PDP / MPC\\Alignment\\PL Model};
\node[block, right=0.3cm of pdp]   (tcsl)  {TCSL};

\node[block, right=3.6cm of tcsl, yshift=0.85cm] (sfit) {Single-frequency\\fitting};
\node[block, below=0.40cm of sfit]               (cfit) {Multi-frequency\\fitting};
\node[block, right=1.6cm of sfit, yshift=-0.85cm] (pf)  {Parametric\\Fit };

\node[block, right=2.6cm of pf]      (nyusim) {NYUSIM};
\node[block, right=0.3cm of nyusim]  (val)    {Validation};

\draw[arr] (data)  -- (cal);
\draw[arr] (cal)   -- (pdp);
\draw[arr] (pdp)   -- (tcsl);

\coordinate (split) at ($(sfit.west)+(-0.5cm,-0.85cm)$);
\draw[line] (tcsl.east) -- (split);
\draw[arr]  (split) |- (sfit.west);
\draw[arr]  (split) |- (cfit.west);
\node[flowlabel, anchor=south]
  at ($(tcsl.east)!0.5!(split) + (0.10cm,0.08cm)$)
  {primary};
\node[flowlabel, anchor=north]
  at ($(tcsl.east)!0.5!(split) + (0.10cm,-0.08cm)$)
  {parameters};

\coordinate (mergeNDE) at ($(pf.west)+(-0.5cm,0)$);
\draw[line] (sfit.east) -| (mergeNDE);
\draw[line] (cfit.east) -| (mergeNDE);
\draw[arr]  (mergeNDE) -- (pf.west);

\draw[arr] (pf) -- (nyusim);
\node[flowlabel, anchor=south]
  at ($(pf.east)!0.5!(nyusim.west) + (0,0.08cm)$)
  {fitted};
\node[flowlabel, anchor=north]
  at ($(pf.east)!0.5!(nyusim.west) + (0,-0.08cm)$)
  {distributions};

\draw[arr] (nyusim) -- (val);

\coordinate (stage2leftpad) at ($(sfit.west)+(-0.6cm,0)$);
\begin{scope}[on background layer]
  \node[stagebox, fit=(cal)(pdp)(tcsl)]                     (stage1) {};
  \node[stagebox, fit=(stage2leftpad)(sfit)(cfit)(pf)]      (stage2) {};
  \node[substagebox, fit=(sfit)(cfit)]                      (nde)    {};
  \node[stagebox, fit=(nyusim)(val)]                        (stage3) {};
\end{scope}
\node[stagelabel]    at (stage1.north) {Initial Data Processing};
\node[stagelabel]    at (stage2.north) {GenAI Distribution Fitting};
\node[substagelabel] at (nde.north)    {Neural Density Estimation (NDE)};
\node[stagelabel]    at (stage3.north) {Channel Modeling and Validation};

\end{tikzpicture}%
}%
\caption{End-to-end overview of the ARIA pipeline. Measurement data flows left-to-right through three stages, namely initial data processing (calibration, PDP/MPC extraction, alignment, TCSL clustering), GenAI distribution fitting, and channel modeling with a NYUSIM-style Monte Carlo generator validated against the measured second-order statistics.}
\label{fig:pipeline}
\vspace{-0.5cm}
\end{figure*}
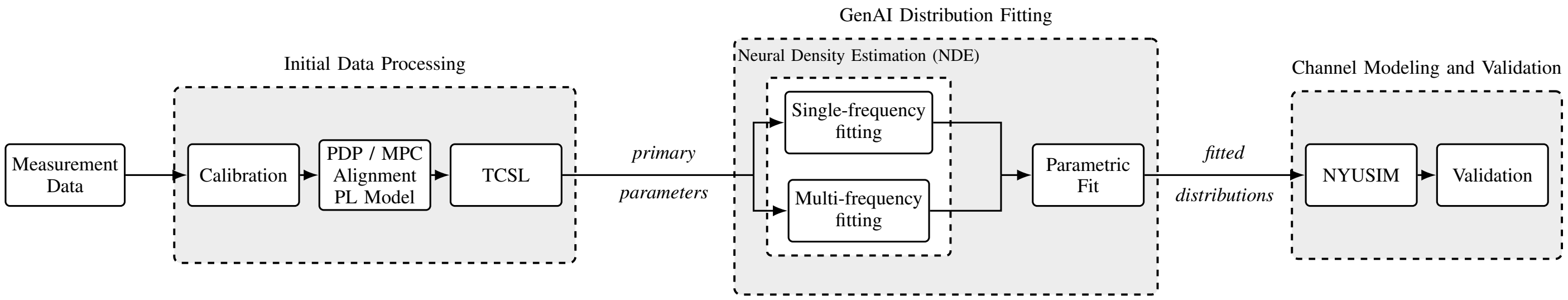

Fig.~\ref{fig:pipeline} summarizes the end-to-end ARIA pipeline. We instantiate ARIA on the NYU~WIRELESS indoor hotspot (InH)
measurement campaigns at $16.95$\,GHz (FR3) and $6.75$\,GHz
(FR1(C), the C-band portion of FR1)~\cite{Shakya2025InH}. The $16.95$\,GHz dataset provides
$28$ distribution-fitting tasks across LOS and NLOS scenarios,
and the $6.75$\,GHz dataset is used jointly with $16.95$\,GHz to
train the multi-frequency conditional models. The main findings
are as follows:
\begin{itemize}
  \item We evaluate every model under two protocols. Protocol-A
  is an in-sample protocol that trains and evaluates each model on the same $N$
  samples of a task, establishing a best-case fit. Protocol-B is an
  out-of-sample protocol where the $N$ samples are split into five
  equally sized parts, and the model is trained on four of them
  and evaluated on the fifth, repeating over the five parts and
  averaging the per-metric values.

  \item Under Protocol-A, NSF and MDN form the top tier of unconditional density estimators, with KDE and DDIM in a clearly separated lower tier on the KS, CvM, and Energy distances and DDIM alone separated on $W_1$. Under Protocol-B, KDE closes the gap to MDN, since the held-out evaluation removes the in-sample fit advantage that MDN's mixture flexibility provides, while KDE is less prone to overfitting on the limited training data per task. NSF's advantage over the other three estimators holds across both protocols.

  \item Across the four models, the per-category winner varies by parameter shape. NSF wins on powers and most delays, where its piecewise-spline transformations capture heavy-tailed and skewed shapes well. MDN wins on most of the discrete count parameters, where its Gaussian-mixture inductive bias places one component per integer value. DDIM wins on most of the symmetric unimodal angular offsets, where diffusion sampling is accurate. KDE wins on a few LOS tasks with the smallest sample counts, where its smoothing protects against overfitting.

  \item A FiLM-conditional NSF trained jointly on the
  $6.75$\,GHz and $16.95$\,GHz campaigns outperforms the
  single-frequency NSF baseline trained on $16.95$\,GHz alone.
  Pooling samples across frequencies without conditioning, by
  contrast, consistently degrades the single-frequency fit.

  \item The augmented parametric step achieves KS $\leq 0.1$ against the measured empirical cumulative distribution functions (ECDFs) on the subpath powers, intra-cluster delay, and lobe counts, with the remaining parameters within KS $\leq 0.19$, confirming that the standard closed-form families remain a faithful representation of the primary parameters when the fitting set is augmented with neural density estimation (NDE) samples.

  \item End-to-end validation confirms that the full ARIA pipeline reproduces the measured channel statistics, with the simulator's omnidirectional RMS delay spread medians within about $2$\,ns, the per-lobe azimuth and zenith spread medians within about $1^\circ$, and the omnidirectional zenith spread medians within $0.5^\circ$. Because the RMS spreads emerge from the joint sampling of the first-order distributions rather than being fitted directly, the agreement confirms both the individual per-task fits and the joint behavior of the simulated channel.
\end{itemize}

The rest of the paper is organized as follows.
Section~\ref{sec:initial} covers the initial
data-processing steps. Section~\ref{sec:models} introduces the
density estimators and the FiLM-conditional variant.
Section~\ref{sec:methodology} formalizes the model-selection
methodology and the augmented parametric step.
Section~\ref{sec:results} illustrates the distribution-fitting
results on the $16.95$\,GHz indoor campaign.
Section~\ref{sec:validation} validates the full pipeline
against the measurements.
Section~\ref{sec:conclusion} concludes the paper and outlines
future work.

\section{Initial Data Processing}
\label{sec:initial}

In this section, we briefly summarize the stages that turn raw
measurement data into the primary channel parameters. Each stage follows standard practice in the NYU~WIRELESS
measurement and modeling framework~\cite{MacCartney2017,
Samimi2016, Sun2017, Ju2021, Poddar2023, Shakya2024,
Shakya2025InH, Shakya2025UMi, Shakya2025InF}; the cited references provide the full details.

\subsection{Channel Sounder and Calibration}

The measurement system employs a wideband sliding-correlator
channel sounder with $1$\,GHz null-to-null RF bandwidth and
$39$\,dB processing gain, using directional horn antennas at
both transmitter (TX) and receiver (RX) to characterize
angle-of-departure (AOD) and angle-of-arrival (AOA) through
mechanical rotation~\cite{Shakya2024, MacCartney2017}. Each
measurement day starts and ends with a three-phase calibration
covering power linearity, timing, and spatial alignment, which
establishes absolute references for received power, multipath
delay, and angular orientation~\cite{Shakya2024}.

\subsection{PDP Extraction, Alignment, and Path-Loss Modeling}

The pipeline extracts PDPs from the raw measurements, applies a dual-threshold denoising step (the stricter of $25$\,dB below the PDP maximum and $5$\,dB above the noise floor), identifies MPCs as local maxima separated by at least the width of the pseudorandom noise (PN) autocorrelation peak, aligns the directional PDPs in time~\cite{Shakya2024, Shakya2025InH}, and fits standard large-scale path-loss models~\cite{MacCartney2017, Samimi2016, Sun2017}.

\subsection{Time-Cluster Spatial-Lobe (TCSL) Analysis}

The omnidirectional PDP and the power angular spectrum (PAS) are
partitioned through the TCSL framework~\cite{Samimi2016, Ju2021,
3GPP38901}. A \emph{time cluster} (TC) is a group of MPCs
arriving within a minimum time interval (MTI) and corresponds to a
distinct scattering mechanism such as a building reflection. A
\emph{spatial lobe} (SL) is a contiguous angular region of the
PAS within $10$\,dB of its peak. Each \emph{subpath} (SP) is an
individual MPC assigned to a unique TC and SL, assuming independent TC and SL.

\subsection{Primary Channel Parameters}
The double-directional channel impulse response is~\cite{Steinbauer2001}:
\begin{IEEEeqnarray}{rCl}
    \lefteqn{h(t, \vec{\Theta}, \vec{\Phi})} \notag \\
&& = \sum_{n=1}^{N} \sum_{m=1}^{M_n} a_{m,n} e^{j\varphi_{m,n}} \delta(t - \tau_{m,n}) \, \delta(\vec{\Theta} - \vec{\Theta}_{m,n}) \, \delta(\vec{\Phi} - \vec{\Phi}_{m,n}) \notag \\
    \label{eq:cir}
\end{IEEEeqnarray}
where $N$ is the number of TCs, $M_n$ is the number of SPs in the $n$-th TC, and $a_{m,n}$, $\varphi_{m,n}$, $\tau_{m,n}$ are the amplitude, phase, and delay of the $m$-th SP. The vectors $\vec{\Theta}_{m,n} = (\phi_{\text{AOD}}, \theta_{\text{ZOD}})$ and $\vec{\Phi}_{m,n} = (\phi_{\text{AOA}}, \theta_{\text{ZOA}})$ give the departure and arrival directions, with the zenith angles $\theta_{\text{ZOD}}$ and $\theta_{\text{ZOA}}$ measured relative to the horizontal plane, following~\cite{Ju2021}.
The following primary parameters describe the spatio-temporal behavior of the channel~\cite{Samimi2016}:

\begin{enumerate}
    \item \textit{Number of time clusters} ($N$): The total count of TCs in the omnidirectional PDP.

    \item \textit{Number of subpaths per TC} ($M_n$): The count of individual MPCs within the $n$-th TC.

      \item \textit{Intra-cluster subpath excess delay} ($\rho_{m,n}$): The time delay of the $m$-th SP within the $n$-th TC, measured from the first SP in the same TC.

    \item \textit{Cluster excess delay} ($\tau_n$): The time from the first arriving SP in the PDP to the first SP in the $n$-th TC. 
    It is computed as $\tau_n = \tau_{n-1} + \rho_{M_{n-1},n-1} + \Delta\tau_n + \text{MTI}$, where $\Delta\tau_n$ is the inter-cluster excess delay with the MTI removed. The SP delay in Eq.~\eqref{eq:cir} decomposes as $\tau_{m,n} = \tau_n + \rho_{m,n}$.

    \item \textit{Cluster power} ($P_n$): The total power of all SPs in the $n$-th TC. 

    \item \textit{Subpath power} ($\Pi_{m,n}$): The power of the $m$-th SP in the $n$-th TC.

    \item \textit{Subpath phase} ($\varphi_{m,n}$): The phase of the $m$-th SP in the $n$-th TC.

    \item \textit{Number of spatial lobes} ($L$): The count of SLs in the PAS, for both AOD and AOA.

    \item \textit{Spatial lobe mean angles} ($\phi_i$, $\theta_i$): The mean azimuth and zenith pointing angles of the $i$-th SL, for both AOD and AOA.

    \item \textit{Subpath angular offsets} ($\Delta\phi_{m,n}$, $\Delta\theta_{m,n}$): The azimuth and zenith offsets of each SP from its assigned SL mean angle, for both AOD and AOA directions.
\end{enumerate} 
The empirical distributions of the primary parameters across all measured links form the fitting tasks of Section~\ref{sec:models}. Counting the AOA and AOD versions separately yields $14$ fitted parameters per scenario; the SP phase and the SL mean azimuths are not fitted, with the phase drawn from $\mathcal{U}(0,\,2\pi)$ and the mean azimuth of the $i$-th of $L$ lobes from $\mathcal{U}(360(i{-}1)/L,\,360i/L)$.

\section{Generative AI (GenAI) Models}
\label{sec:models}

We consider three generative models, namely mixture density networks (MDN), neural spline flows (NSF), and denoising diffusion implicit models (DDIM), along with kernel density estimation (KDE) as a classical non-parametric baseline. The following subsections describe each in turn.
\subsection{Kernel Density Estimation (KDE)}\label{subsec:kde}
KDE is a classical non-parametric method for estimating probability densities from data~\cite{Silverman1986, Parzen1962}. Given $N$ observations $\{x_1, \dots, x_N\}$, the KDE estimate of the density at a point $x$ is
\begin{equation}
    \hat{p}(x) = \frac{1}{Nh} \sum_{i=1}^{N} K\!\left(\frac{x - x_i}{h}\right),
    \label{eq:kde}
\end{equation}
where $K(\cdot)$ is a kernel function and $h > 0$ is the bandwidth. We use a Gaussian kernel $K(u) = (2\pi)^{-1/2}\exp(-u^2/2)$ with Scott's rule for bandwidth selection~\cite{Scott1992}, $h = \hat{\sigma}\,N^{-1/5}$, where $\hat{\sigma}$ is the sample standard deviation. KDE has no learnable parameters and is fitted in closed form, making it a natural baseline whose main limitation is sensitivity to bandwidth. Too small a bandwidth produces a spiky density that overfits the data, while too large a bandwidth over-smooths the distribution and erases multimodal structure.

\subsection{Mixture Density Networks (MDN)}\label{subsec:mdn}

An MDN models the data distribution as a mixture of Gaussian components whose parameters are learned by a neural network~\cite{Bishop1994}. The density is
\begin{equation}
    p(x) = \sum_{k=1}^{K} \pi_k \; \mathcal{N}(x;\, \mu_k,\, \sigma_k^2),
    \label{eq:mdn}
\end{equation}
where $\pi_k$, $\mu_k$, and $\sigma_k$ are the mixing weight, mean, and standard deviation of the $k$-th component.
 The model is trained by maximizing the log-likelihood $\sum_n \log p(x_n)$, which provides exact likelihoods and a strong inductive bias for smooth distributions. The Gaussian mixture form is also well suited to integer-valued count parameters, where each integer is represented by a dedicated component and the discrete histogram is recovered by rounding the generated samples~\cite{Papamakarios2021}.

\subsection{Neural Spline Flows (NSF)}\label{subsec:nsf}

NSF belongs to the family of normalizing flows that model a complex probability density by transforming a simple base distribution through an invertible, differentiable function~\cite{Rezende2015, Papamakarios2021}. Given a noise variable $\mathbf{u} \sim \pi(\mathbf{u})$ (e.g., a standard normal) and an invertible transformation $\mathbf{f}$, the density of $\mathbf{x} = \mathbf{f}(\mathbf{u})$ is given by the change-of-variables formula
\begin{equation}
p(\mathbf{x}) = \pi\!\left(\mathbf{f}^{-1}(\mathbf{x})\right) \left|\det \frac{\partial \mathbf{f}^{-1}}{\partial \mathbf{x}}\right|,
\label{eq:flow_density}
\end{equation}
and the flow is trained by maximizing $\sum_n \log p(\mathbf{x}^{(n)})$.

An NSF~\cite{Durkan2019} builds $\mathbf{f}$ from monotonic rational-quadratic (RQ) splines. Each spline maps $[-B, B]$ to itself through $K$ rational-quadratic segments joined at $K+1$ knot points $\{(x^{(k)}, y^{(k)})\}_{k=0}^{K}$, and acts as the identity outside $[-B, B]$. With bin slope $s^{(k)} = (y^{(k+1)} - y^{(k)})/(x^{(k+1)} - x^{(k)})$ and local coordinate $\xi = (x - x^{(k)})/(x^{(k+1)} - x^{(k)})$, the $k$-th segment is
\begin{equation}
g(\xi) = y^{(k)} + \frac{(y^{(k+1)} - y^{(k)})\left[s^{(k)}\xi^2 + \delta^{(k)}\xi(1-\xi)\right]}{s^{(k)} + \left[\delta^{(k+1)} + \delta^{(k)} - 2s^{(k)}\right]\xi(1-\xi)},
\label{eq:rq_spline}
\end{equation}
where the bin widths, heights, and derivatives $\delta^{(k)}$ are learnable parameters of the flow. The transform is monotonic, with closed-form inverse and Jacobian determinant. NSF is well suited to one-dimensional data because each spline layer independently learns a flexible monotonic transform from knot positions alone, without a coupling or autoregressive split.

\subsection{Denoising Diffusion Implicit Models (DDIM)}\label{subsec:ddim}

A diffusion model generates data by reversing a gradual noise-adding process~\cite{SohlDickstein2015, Ho2020, Song2021}. DDIM~\cite{Song2021} is a diffusion framework that uses a non-Markovian formulation to enable fast deterministic sampling and is composed of the following \emph{forward}, \emph{training}, and \emph{sampling} processes.

\subsubsection{Forward Process}
The forward process takes a data sample $\mathbf{x}_0$ and adds Gaussian noise over $T$ steps according to a variance schedule $\beta_1, \dots, \beta_T$. Defining $\alpha_t = 1 - \beta_t$ and $\bar{\alpha}_t = \prod_{s=1}^{t} \alpha_s$, the noisy sample at any timestep $t$ can be obtained directly from $\mathbf{x}_0$ in closed form as
\begin{equation}
    q(\mathbf{x}_t | \mathbf{x}_0) = \mathcal{N}\!\left(\mathbf{x}_t;\, \sqrt{\bar{\alpha}_t}\,\mathbf{x}_0,\, (1 - \bar{\alpha}_t)\mathbf{I}\right),
    \label{eq:ddim_forward}
\end{equation}
which means $\mathbf{x}_t = \sqrt{\bar{\alpha}_t}\,\mathbf{x}_0 + \sqrt{1 - \bar{\alpha}_t}\,\boldsymbol{\epsilon}$, where $\boldsymbol{\epsilon} \sim \mathcal{N}(\mathbf{0}, \mathbf{I})$.

\subsubsection{Training Process}
A neural network $\boldsymbol{\epsilon}_\theta(\mathbf{x}_t, t)$ is trained to predict the noise component $\boldsymbol{\epsilon}$ from the noisy input $\mathbf{x}_t$ using the mean squared error loss
\begin{equation}
    L_{\boldsymbol{\epsilon}} = \mathbb{E}_{t,\,\mathbf{x}_0,\,\boldsymbol{\epsilon}}\!\left[\left\|\boldsymbol{\epsilon} - \boldsymbol{\epsilon}_\theta\!\left(\sqrt{\bar{\alpha}_t}\,\mathbf{x}_0 + \sqrt{1 - \bar{\alpha}_t}\,\boldsymbol{\epsilon},\, t\right)\right\|^2\right],
    \label{eq:ddim_loss}
\end{equation}
where $t$ is sampled uniformly from $\{1, \dots, T\}$ and $\boldsymbol{\epsilon} \sim \mathcal{N}(\mathbf{0}, \mathbf{I})$.

\subsubsection{Sampling Process}
DDIM generates samples by starting from pure noise $\mathbf{x}_T \sim \mathcal{N}(\mathbf{0}, \mathbf{I})$ and iteratively denoising it using a non-Markovian reverse process. Rather than following all $T$ training timesteps, DDIM uses a shorter subsequence $\tau_1, \tau_2, \dots, \tau_S$ with $S \ll T$. Each reverse step is computed as
\begin{equation}
    \mathbf{x}_{\tau_{i-1}} = \sqrt{\bar{\alpha}_{\tau_{i-1}}}\,\hat{\mathbf{x}}_0 + \sqrt{1 - \bar{\alpha}_{\tau_{i-1}} - \sigma_{\tau_i}^2}\;\boldsymbol{\epsilon}_\theta(\mathbf{x}_{\tau_i}, \tau_i) + \sigma_{\tau_i}\,\mathbf{z},
    \label{eq:ddim_step}
\end{equation}
where $\mathbf{z} \sim \mathcal{N}(\mathbf{0}, \mathbf{I})$ and the predicted clean sample is
\begin{equation}
    \hat{\mathbf{x}}_0 = \frac{\mathbf{x}_{\tau_i} - \sqrt{1 - \bar{\alpha}_{\tau_i}}\,\boldsymbol{\epsilon}_\theta(\mathbf{x}_{\tau_i}, \tau_i)}{\sqrt{\bar{\alpha}_{\tau_i}}}.
    \label{eq:ddim_x0}
\end{equation}
The parameter $\sigma_{\tau_i}$ controls the amount of stochasticity injected at each step. Setting $\sigma_{\tau_i} = 0$ yields a fully deterministic reverse process, which enables fast sampling while preserving sample quality comparable to the full $T$-step Markovian sampler~\cite{Song2021}.

\subsection{Conditional Models for Multi-Frequency Distribution Fitting}
\label{sec:multi_freq}

Channel measurement campaigns are often conducted at several
carrier frequencies in the same or similar
environments~\cite{Shakya2024, Shakya2025UMi, Shakya2025InH}. A
natural question is whether measurements from non-target
frequencies can improve distribution fitting at a given target
frequency. We extend the generative models defined in
Sections~\ref{subsec:kde}--\ref{subsec:ddim} to operate in the following two
multi-frequency modes and compare them against the
single-frequency baselines.

\subsubsection{Frequency-Agnostic Pooling}

In the pooled mode, all samples across the measured frequencies
are combined into a single dataset, and each model is trained on
the combined dataset using the same objective as its
single-frequency counterpart. The
resulting estimator fits one shared distribution across all frequencies, which is accurate at a given frequency only when the per-frequency distributions coincide.

\subsubsection{Feature-wise Linear Modulation (FiLM) Conditioning}
The FiLM-conditional models~\cite{Perez2018_FiLM, Dumoulin2018} share the same backbones as the
unconditional MDN, NSF, and DDIM but receive the carrier
frequency as an additional input. KDE has no trainable
parameters and therefore does not have a FiLM-conditional
variant. Given a hidden representation $\mathbf{h} \in
\mathbb{R}^d$ and a conditioning input $c$ derived from the
carrier frequency, FiLM applies the affine transform
\begin{equation}
    \mathrm{FiLM}(\mathbf{h}; c) = \boldsymbol{\gamma}(c) \odot
    \mathbf{h} + \boldsymbol{\beta}(c),
    \label{eq:film}
\end{equation}
where $\boldsymbol{\gamma}(c),
\boldsymbol{\beta}(c) \in \mathbb{R}^d$ are the scale and shift
vectors predicted from $c$ by a small auxiliary network (the
conditioner), and $\odot$ denotes element-wise multiplication.
The conditioner predicts $\boldsymbol{\beta}$ and an offset
$\Delta\boldsymbol{\gamma}$ with $\boldsymbol{\gamma} =
\mathbf{1} + \Delta\boldsymbol{\gamma}$, and its final layer is
initialized to zero, letting each
conditional model begin as its unconditional counterpart and
learn to use the conditioning signal as training progresses.

For each training sample with carrier frequency $f$ in GHz, the conditioning input is $c = (\log_{10} f - \mu_c)/\sigma_c$, where $\mu_c$ and $\sigma_c$ are the sample mean and standard deviation of $\log_{10} f$ across all training samples.
 Training uses the same objective as the
single-frequency model but with $p(x_i)$ replaced by
$p(x_i \mid c_i)$ for the likelihood-based models, and the
noise-prediction loss replaced by its conditional analogue for
DDIM.

\section{GenAI Distribution Fitting Methodology}
\label{sec:methodology}
We use a two-stage approach to fit statistical distributions to the channel parameters extracted in Section~\ref{sec:initial}. In the first stage, neural density estimation (NDE), a generative model learns each parameter's distribution directly from the measured data without any predefined parametric form. In the second stage, parametric fit with augmentation (PF), the trained NDE generates synthetic samples that are combined with the original measurements, and a classical parametric distribution is fitted to the augmented dataset via maximum likelihood estimation (MLE). The NDE captures the distribution shape without parametric assumptions, and the PF produces a simulator-compatible parametric output with more stable estimates than fitting on the raw measurements alone.
\subsection{Stage 1: Neural Density Estimation}
We consider the generative models defined in Section~\ref{sec:models}. All neural models (MDN, NSF, and DDIM) are trained with the Adam optimizer using a learning rate of $10^{-3}$, weight decay of $10^{-5}$, and gradient clipping with a maximum norm of $1.0$. The batch size is set to $\min(64, \max(4, N/2))$. 
The training loss is the negative log-likelihood for NSF and MDN, and the noise-prediction mean squared error for DDIM (Eq.~\eqref{eq:ddim_loss}). KDE is fitted in closed form as described in Section~\ref{subsec:kde}.

Each model generates 5000 samples and is evaluated against the measured data using five complementary two-sample goodness-of-fit metrics. The \emph{Kolmogorov-Smirnov (KS)} statistic~\cite{Massey1951} measures the maximum absolute difference between the empirical CDFs and is sensitive to shifts in the center of the distribution. The \emph{Cram\'{e}r-von Mises (CvM)} statistic~\cite{Anderson1962} integrates the squared CDF difference across the support and captures overall shape mismatch. The \emph{Anderson-Darling (AD)} statistic~\cite{Anderson1954} is a tail-weighted variant of CvM that places greater emphasis on the distribution tails. The \emph{Wasserstein-1 distance (W1)}~\cite{Villani2009} measures the area between the empirical CDFs and is sensitive to the average displacement between the two distributions. The \emph{energy distance}~\cite{Szekely2013} compares the average between-sample pairwise distance with the average within-sample pairwise distance and is sensitive to any difference between the two distributions across the support. For all five metrics, lower values indicate a better fit.

Because the metrics have different scales (KS is bounded in $[0, 1]$, AD is unbounded, W1 has physical units), absolute values are not directly comparable across metrics. To aggregate them into a single score, we use a normalized-rank approach. For a given task with $k$ candidate models, the raw scores of all models on metric $m$ are ranked from best (lowest, rank 1) to worst (rank $k$), with tied scores receiving their average rank. The ranks are then normalized by dividing by $k$, producing normalized ranks in $[1/k, 1]$. The composite rank $\bar{R}$ for each model is the average of its five normalized ranks, where $\bar{R} = 1/k$ means the model is ranked first on every metric and $\bar{R} = 1$ means it is ranked last on every metric. 

\subsubsection{Statistical Comparison Across Tasks}

The composite rank $\bar{R}$ selects the best model on a per-task basis. To determine whether one model is consistently superior across all tasks in the study, we use the non-parametric statistical testing framework of Dem\v{s}ar~\cite{Demsar2006}, a standard methodology for comparing multiple models across a common set of tasks. Two complementary tests are applied. First, the Friedman test asks whether the $k$ models are equivalent across the $n$ tasks. If the Friedman test rejects the null hypothesis, we proceed to the Nemenyi test, which identifies the specific pairs of models that differ.
 Both tests are applied independently for each of the five goodness-of-fit metrics and additionally on the composite rank $\bar{R}$.

\emph{Friedman test:}
The Friedman test~\cite{Friedman1937} is a rank-based test that checks whether several methods give different results when applied to the same set of tasks. In our setting, the methods are the $k$ candidate generative models and the tasks are the distribution fittings to the primary channel parameters extracted from the TCSL process, resulting in $n = 28$ tasks (LOS and NLOS combined). Within each per-metric score matrix, the $k$ models are ranked per task from $1$ (best) to $k$ (worst). The Friedman test statistic is then computed as
\begin{equation}
    \chi_F^2 = \frac{12n}{k(k+1)} \left[\sum_{j=1}^{k} R_j^{\,2} - \frac{k(k+1)^2}{4}\right],
    \label{eq:friedman}
\end{equation}
where $R_j$ is the mean per-task rank of model~$j$, without the normalization by $k$ used in the composite rank $\bar{R}$. Under the null hypothesis $H_0$ that all models share the same mean rank $R_j = (k+1)/2$, the statistic follows a chi-squared distribution with $k - 1$ degrees of freedom for large $n$. The null hypothesis is rejected at significance level $\alpha = 0.05$.

\emph{Nemenyi post-hoc test:}
When the Friedman test rejects $H_0$, the Nemenyi test~\cite{Nemenyi1963} is used to find which specific pairs of models differ. The test runs all $m = k(k-1)/2$ pairwise comparisons at once, while keeping the overall false-positive rate at $\alpha$~\cite{Demsar2006}. Two models are declared significantly different at level $\alpha$ if their mean ranks differ by more than the critical difference (CD)
\begin{equation}
    \text{CD} = q_\alpha \, \sqrt{\frac{k(k+1)}{6n}},
    \label{eq:nemenyi_cd}
\end{equation}
where $q_\alpha$ is the $(1 - \alpha)$-quantile of the Studentized range distribution divided by $\sqrt{2}$.

\subsubsection{Evaluation Protocols}

We evaluate each model under the following two protocols to capture both in-sample fit and out-of-sample generalization:
\begin{itemize}
\item \emph{Protocol-A (in-sample):} Under Protocol-A, we train each model on all $N$ available samples for a given task and evaluate on the same $N$ samples.
\item \emph{Protocol-B (out-of-sample):} Under Protocol-B, we use five-fold cross-validation. The $N$ samples are shuffled and split into five folds; for each fold, the model is trained on the remaining four folds and evaluated on the held-out fold. Per-metric values are averaged across folds before the rank $\bar{R}$ is computed. Tasks with $N < 50$ automatically fall back to Protocol-A to avoid degenerate fold sizes.
\end{itemize}

\subsection{Stage 2: Parametric Fit with NDE Augmentation}
Once the best NDE model is selected for a given task under each protocol (by composite rank~$\bar{R}$ on the training data for Protocol-A, and by $\bar{R}$ averaged across folds for Protocol-B), it generates $5N$ synthetic samples, where $N$ is the number of original measurements. The synthetic samples are clipped to physically valid ranges (e.g., non-negative for delay parameters, integer-valued for count parameters) and combined with the $N$ measured samples to form an augmented dataset of size $6N$. 
A classical parametric distribution is then fitted to the augmented dataset via MLE.

\begin{figure*}[!t]
  \centering
  \captionsetup[subfloat]{font=scriptsize,labelfont=scriptsize,labelsep=space,skip=0pt}
  \subfloat[$L^{\mathrm{AOA}}$]{\includegraphics[width=0.43\textwidth]{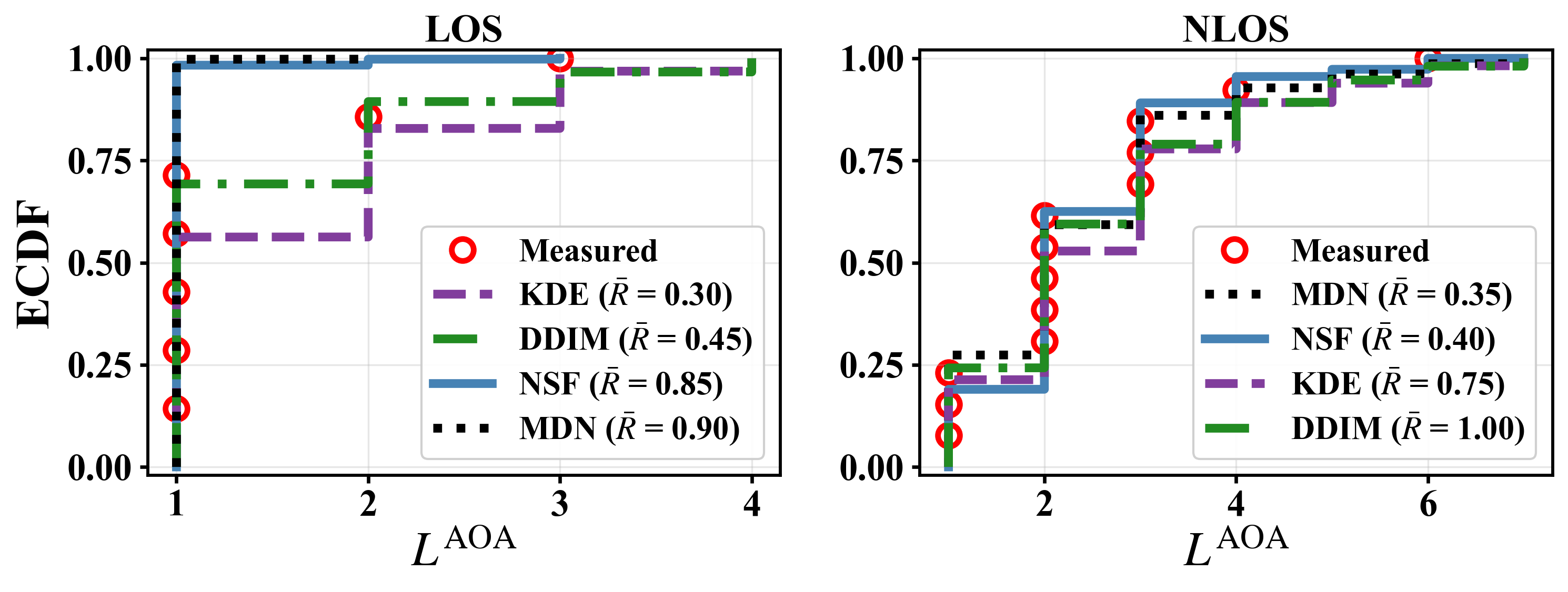}\label{fig:cmp:aoa_lobe}}\hfill
  \subfloat[$L^{\mathrm{AOD}}$]{\includegraphics[width=0.43\textwidth]{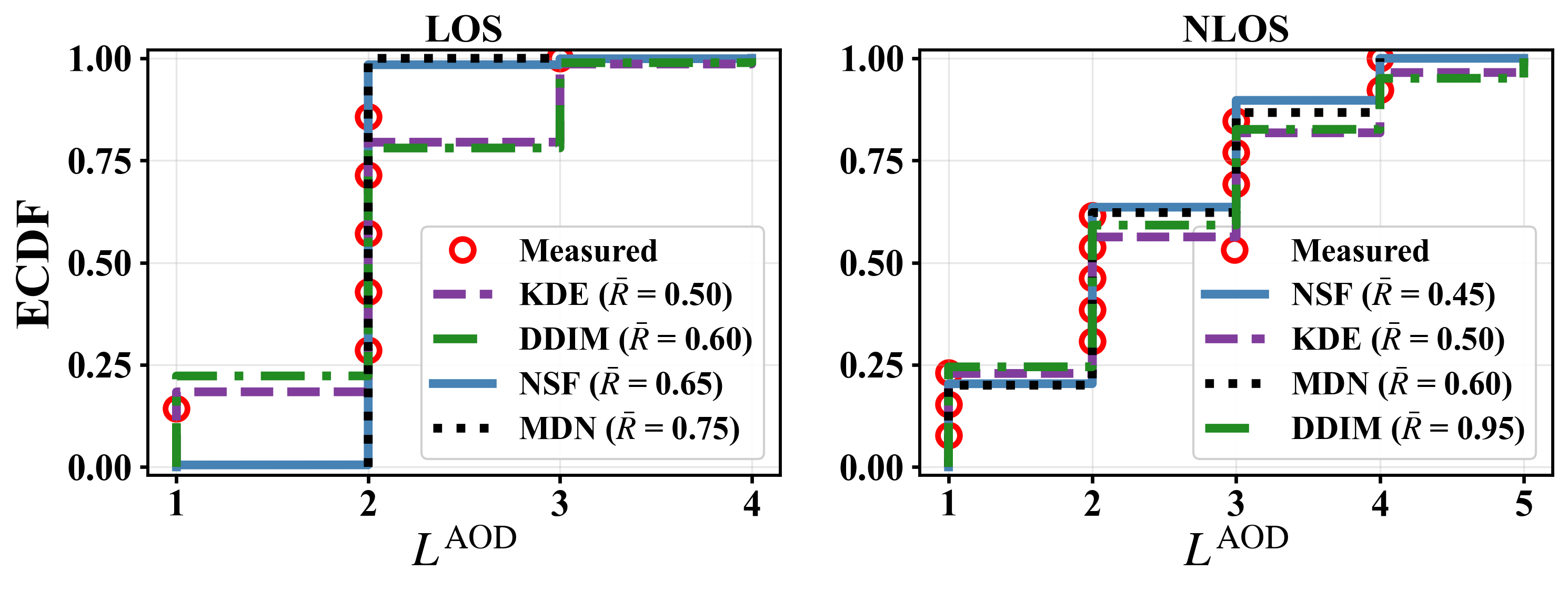}\label{fig:cmp:aod_lobe}}\\
  \subfloat[$N$]{\includegraphics[width=0.43\textwidth]{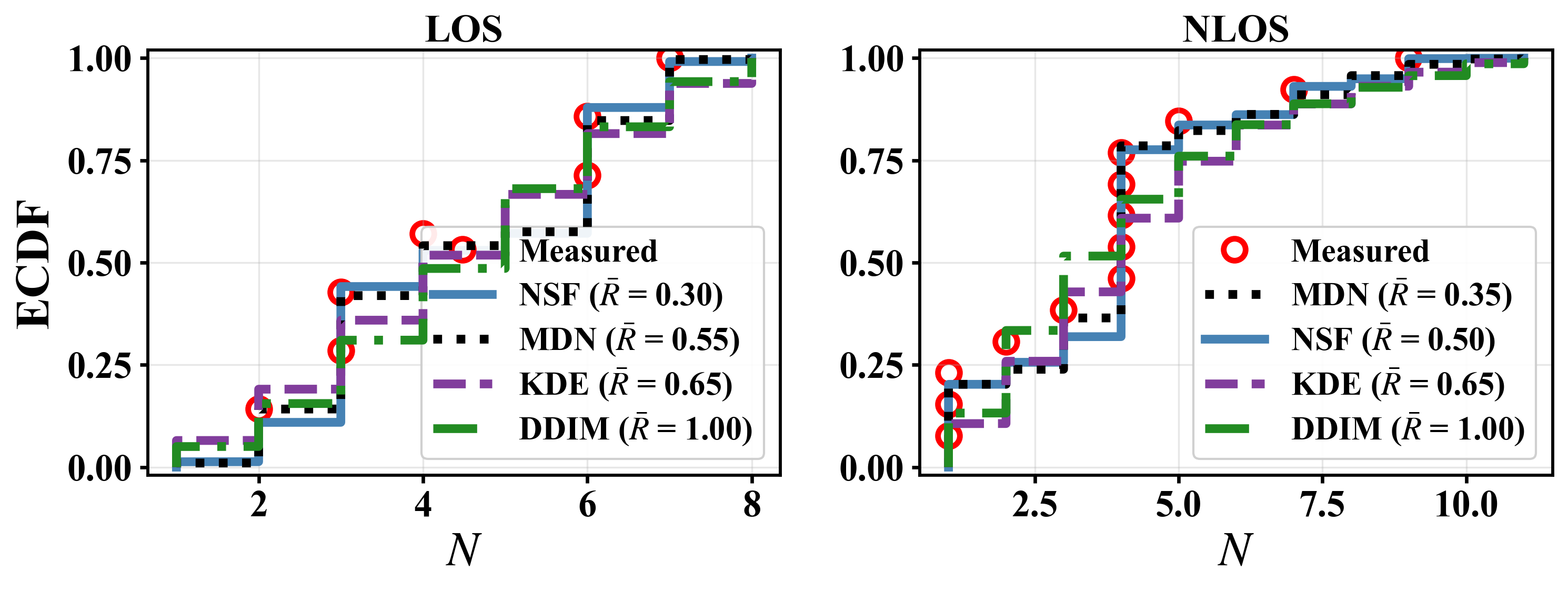}\label{fig:cmp:tc}}\hfill
  \subfloat[$M$]{\includegraphics[width=0.43\textwidth]{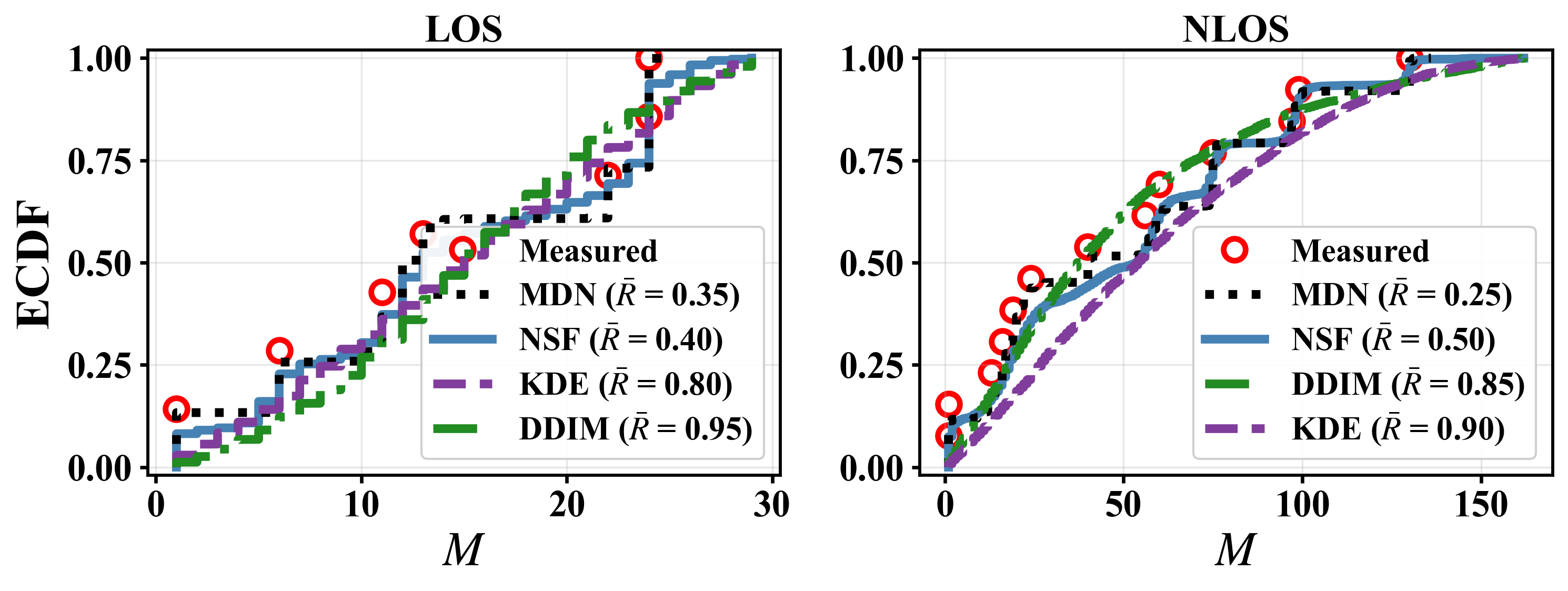}\label{fig:cmp:sp}}\\
  \subfloat[$\Delta\tau$]{\includegraphics[width=0.43\textwidth]{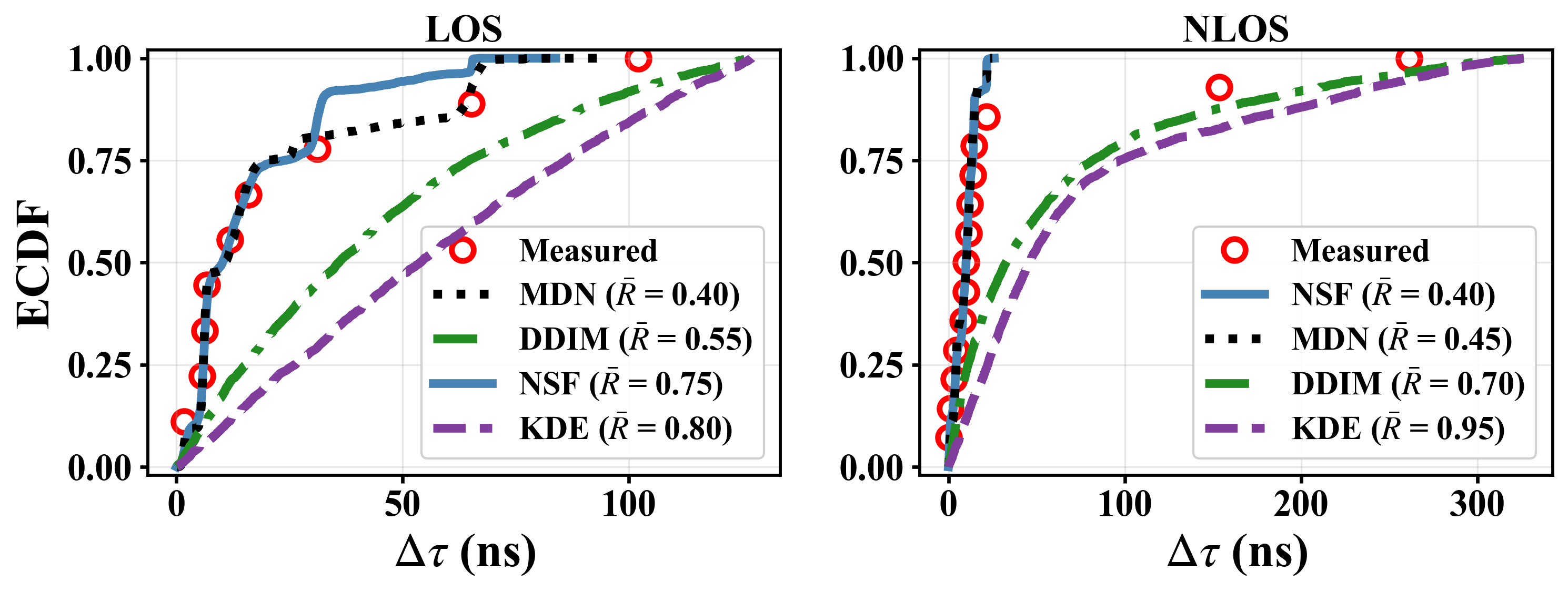}\label{fig:cmp:icd}}\hfill
  \subfloat[$\rho$]{\includegraphics[width=0.43\textwidth]{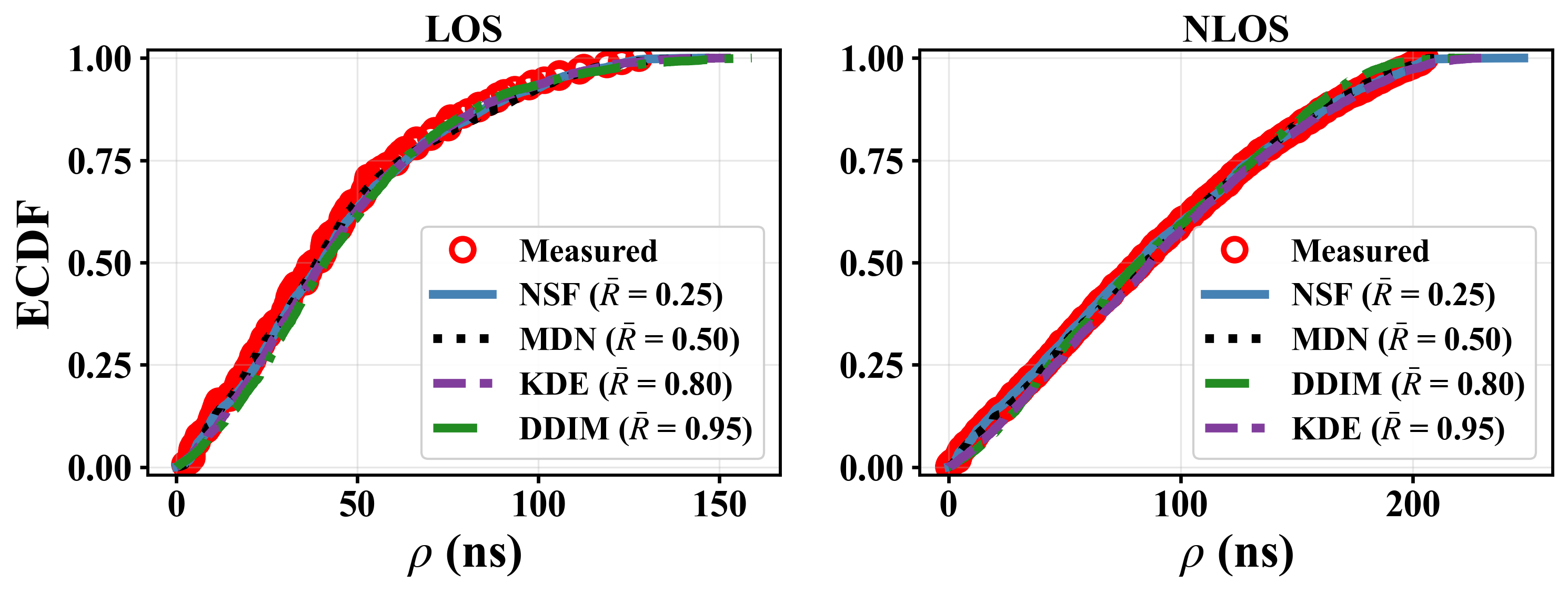}\label{fig:cmp:ind}}\\
  \subfloat[$P$]{\includegraphics[width=0.43\textwidth]{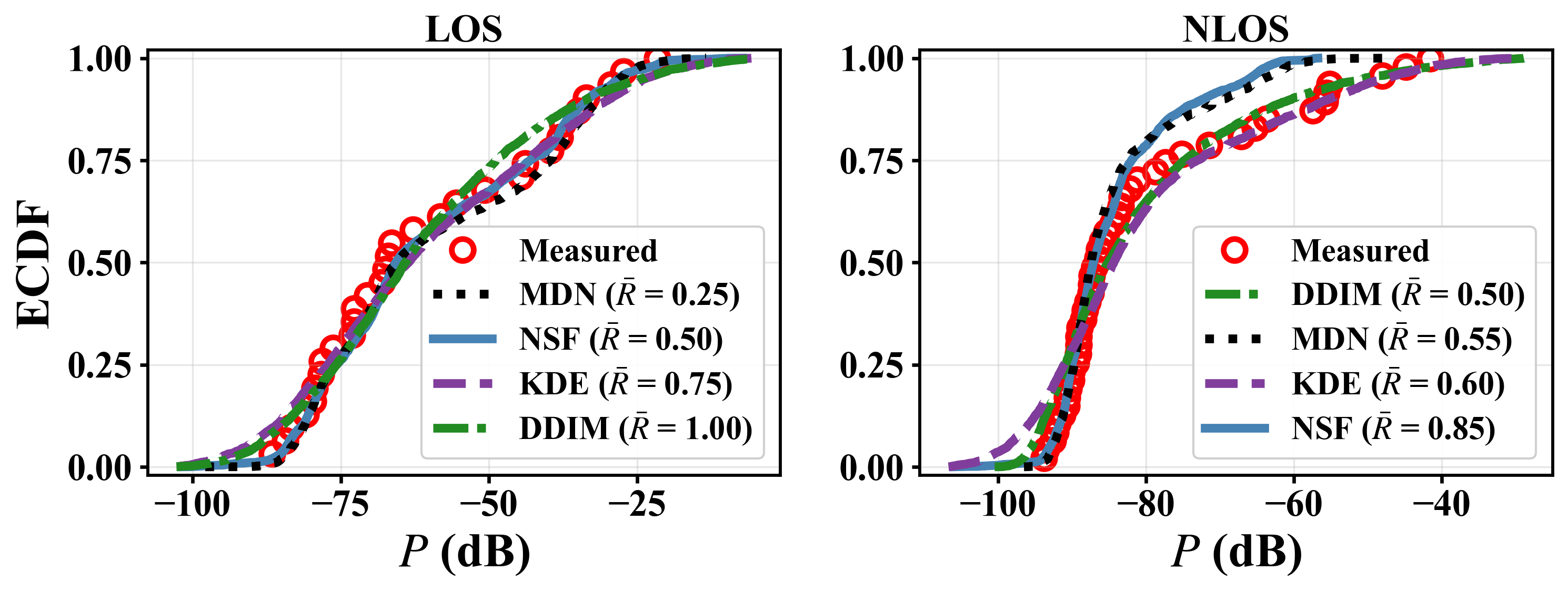}\label{fig:cmp:cpwr}}\hfill
  \subfloat[$\Pi$]{\includegraphics[width=0.43\textwidth]{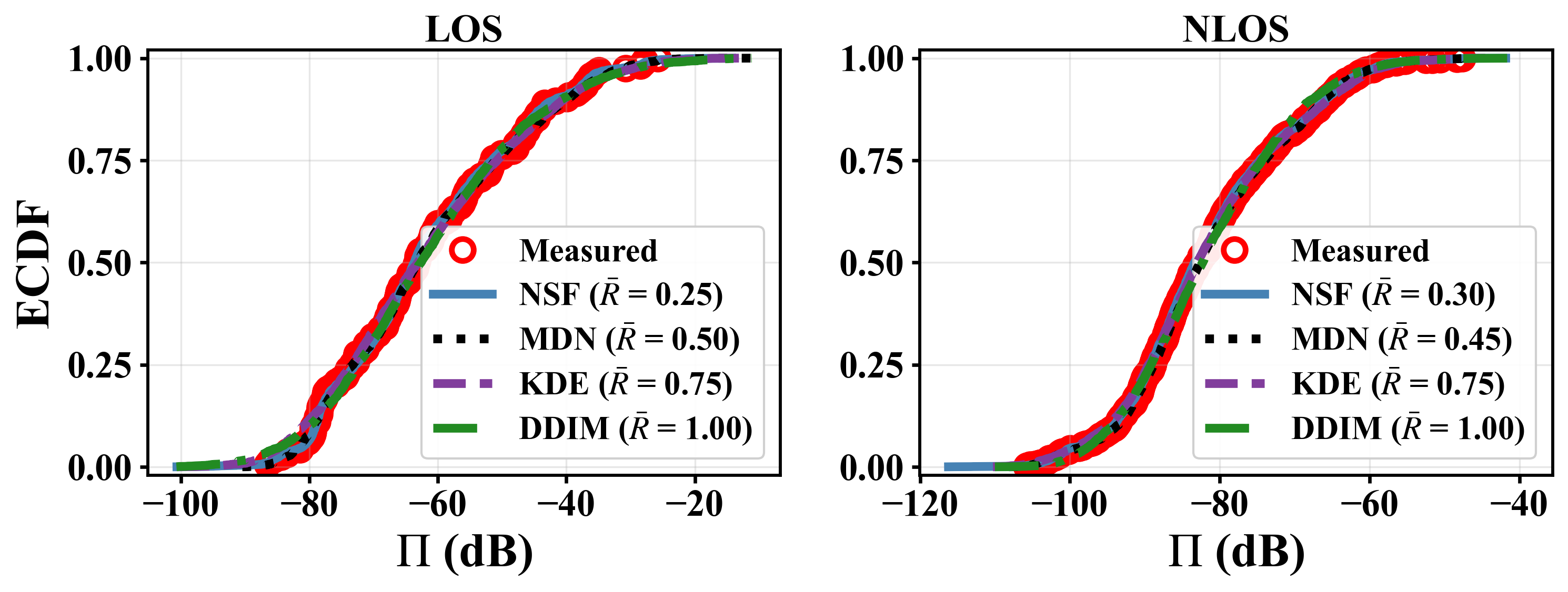}\label{fig:cmp:spwr}}
  \caption{Per-parameter ECDFs from the NDE step against the measured ECDFs for the count, temporal, and power parameters, namely the AOA lobe count $L^{\mathrm{AOA}}$ in (a), the AOD lobe count $L^{\mathrm{AOD}}$ in (b), the time-cluster count $N$ in (c), the subpath count $M$ in (d), the inter-cluster delay $\Delta\tau$ in (e), the intra-cluster subpath delay $\rho$ in (f), the cluster power $P$ in (g), and the subpath power $\Pi$ in (h).}
  \label{fig:cmp-counts}
  \vspace{-0.5cm}
\end{figure*}
\begin{figure}[!t]
  \centering
  \captionsetup[subfloat]{font=scriptsize,labelfont=scriptsize,labelsep=space,skip=0pt}
  \subfloat[$\Delta\phi^{\mathrm{AOA}}$]{\includegraphics[width=0.43\textwidth]{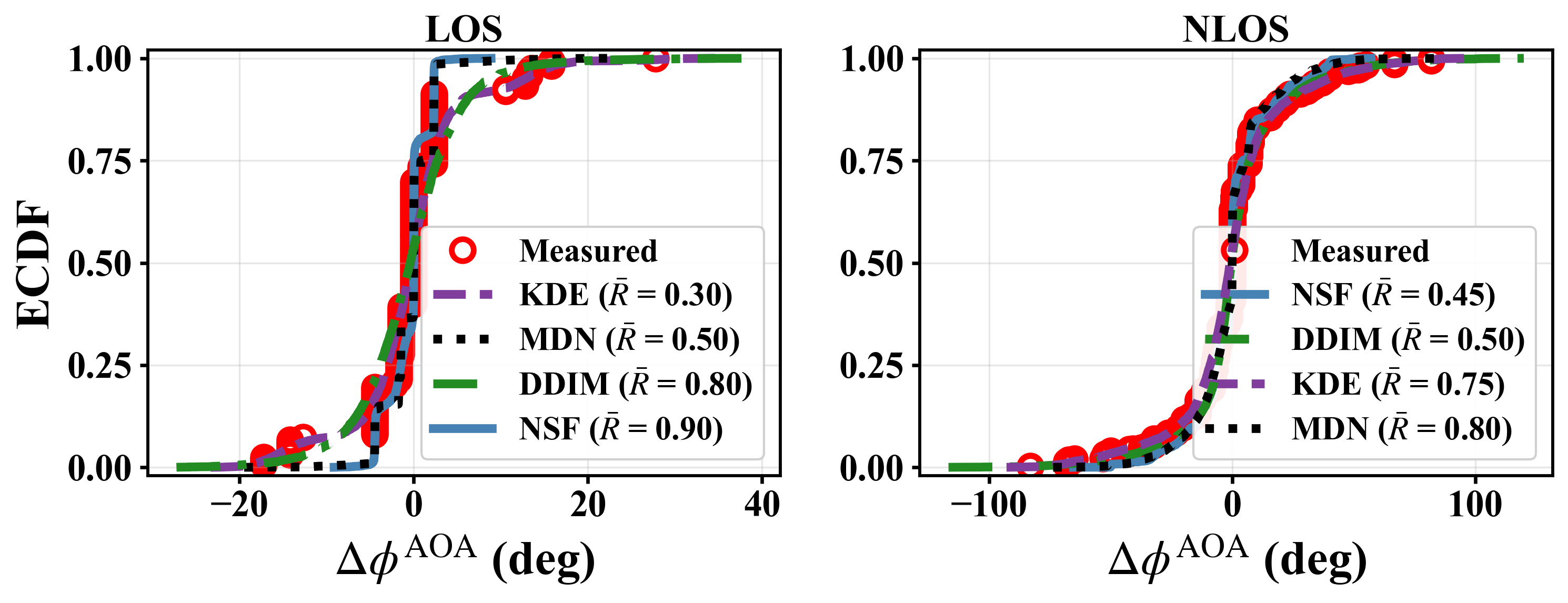}\label{fig:cmp:daoa}}\hfill
  \subfloat[$\Delta\phi^{\mathrm{AOD}}$]{\includegraphics[width=0.43\textwidth]{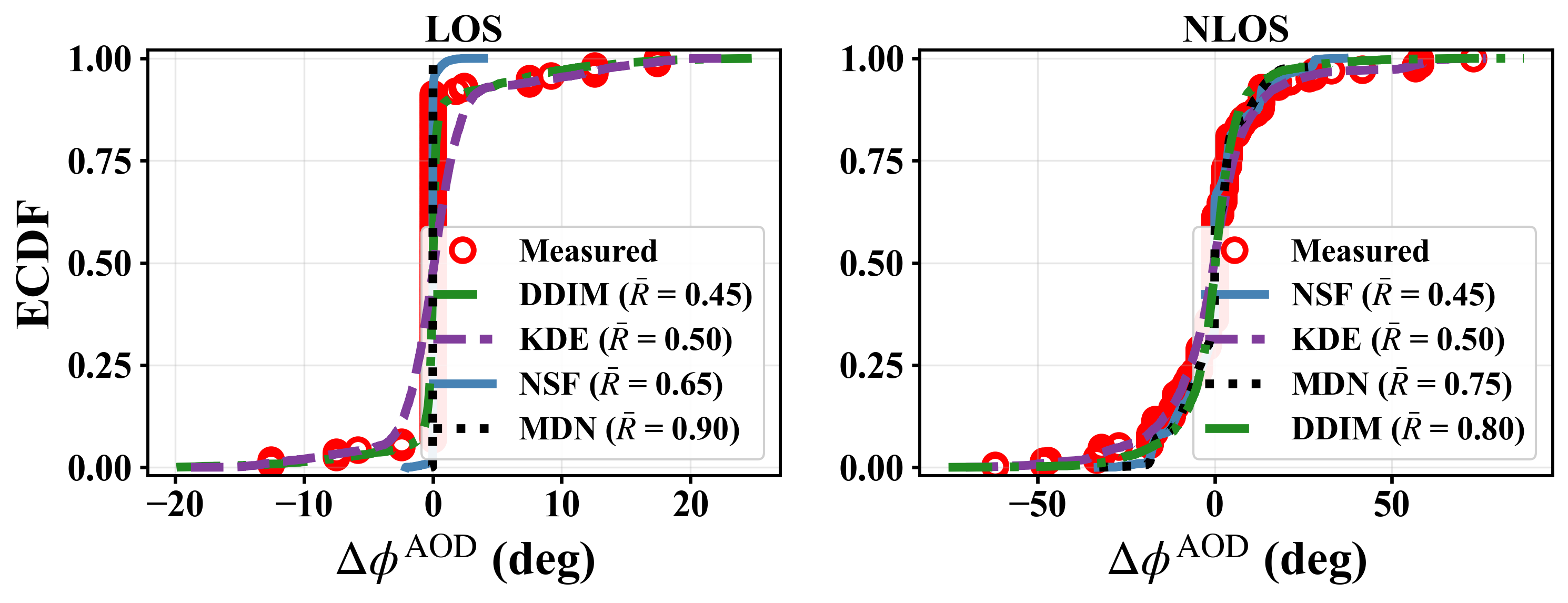}\label{fig:cmp:daod}}\\
  \subfloat[$\Delta\theta^{\mathrm{ZOA}}$]{\includegraphics[width=0.43\textwidth]{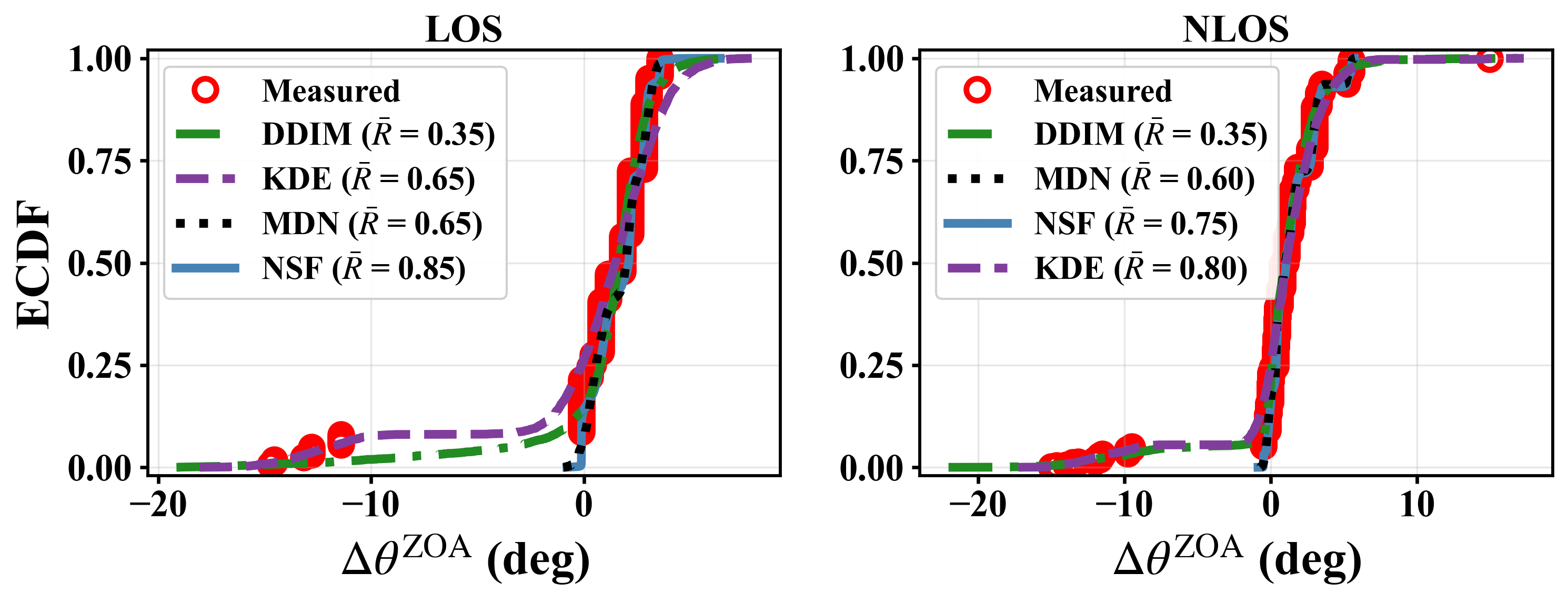}\label{fig:cmp:dzoa}}\hfill
  \subfloat[$\Delta\theta^{\mathrm{ZOD}}$]{\includegraphics[width=0.43\textwidth]{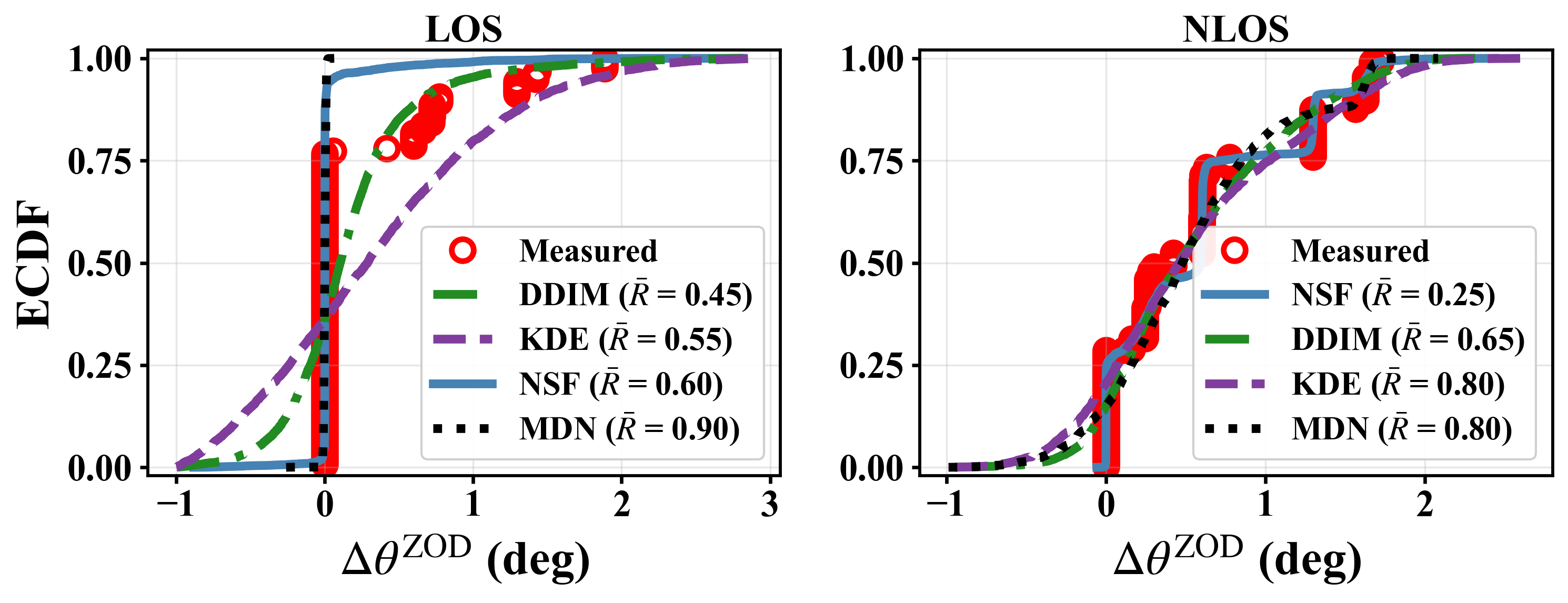}\label{fig:cmp:dzod}}\\
  \subfloat[$\theta^{\mathrm{ZOA}}$]{\includegraphics[width=0.43\textwidth]{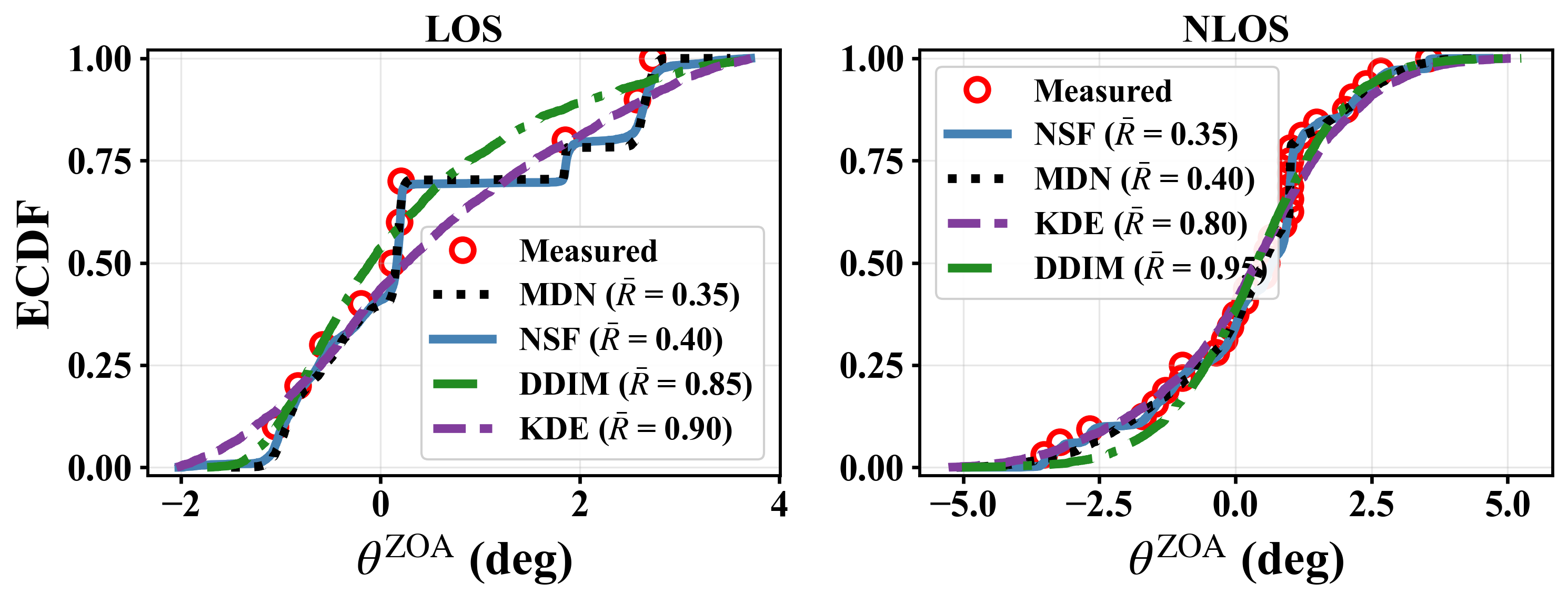}\label{fig:cmp:mzoa}}\hfill
  \subfloat[$\theta^{\mathrm{ZOD}}$]{\includegraphics[width=0.43\textwidth]{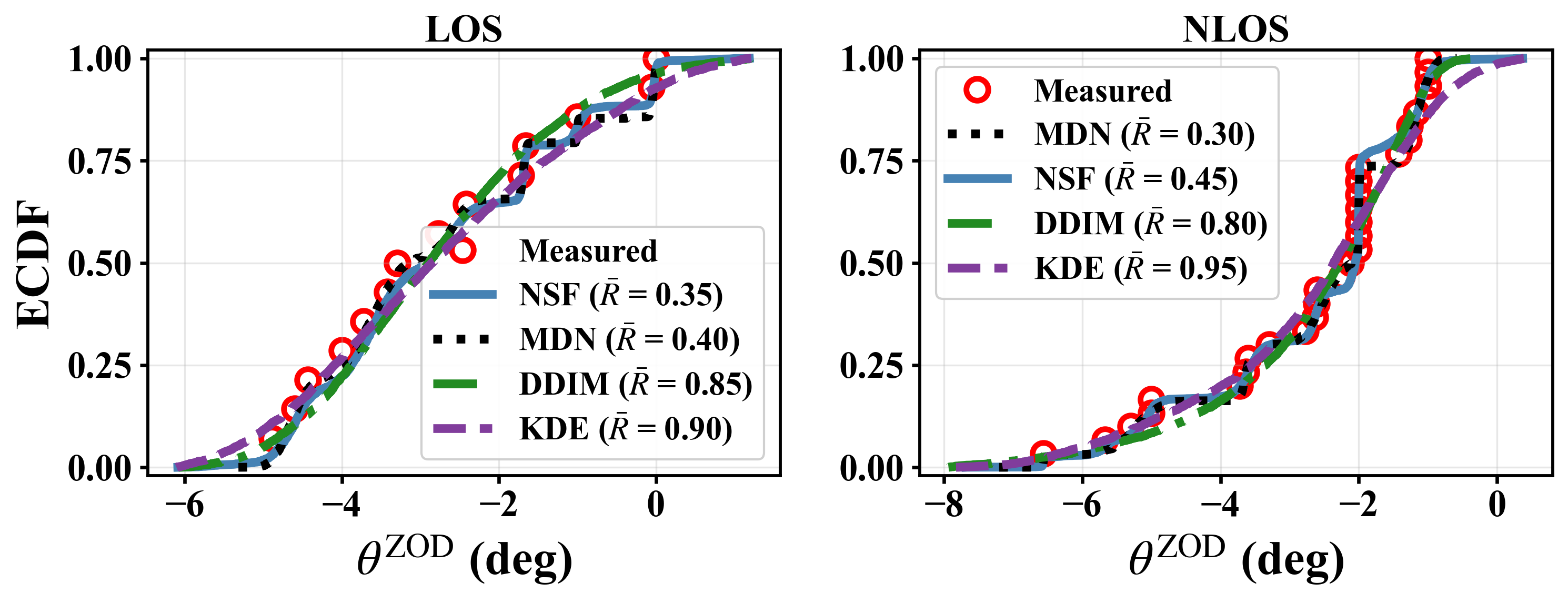}\label{fig:cmp:mzod}}
  \caption{Per-parameter ECDFs from the NDE step against the measured ECDFs for the angular parameters, namely the subpath azimuth offsets $\Delta\phi^{\mathrm{AOA}}$ in (a) and $\Delta\phi^{\mathrm{AOD}}$ in (b), the subpath zenith offsets $\Delta\theta^{\mathrm{ZOA}}$ in (c) and $\Delta\theta^{\mathrm{ZOD}}$ in (d), and the lobe-mean zeniths $\theta^{\mathrm{ZOA}}$ in (e) and $\theta^{\mathrm{ZOD}}$ in (f).}
  \label{fig:cmp-angles}
  \vspace{-0.5cm}
\end{figure}
\begin{figure}[!t]
  \centering
  \includegraphics[width=1\columnwidth]{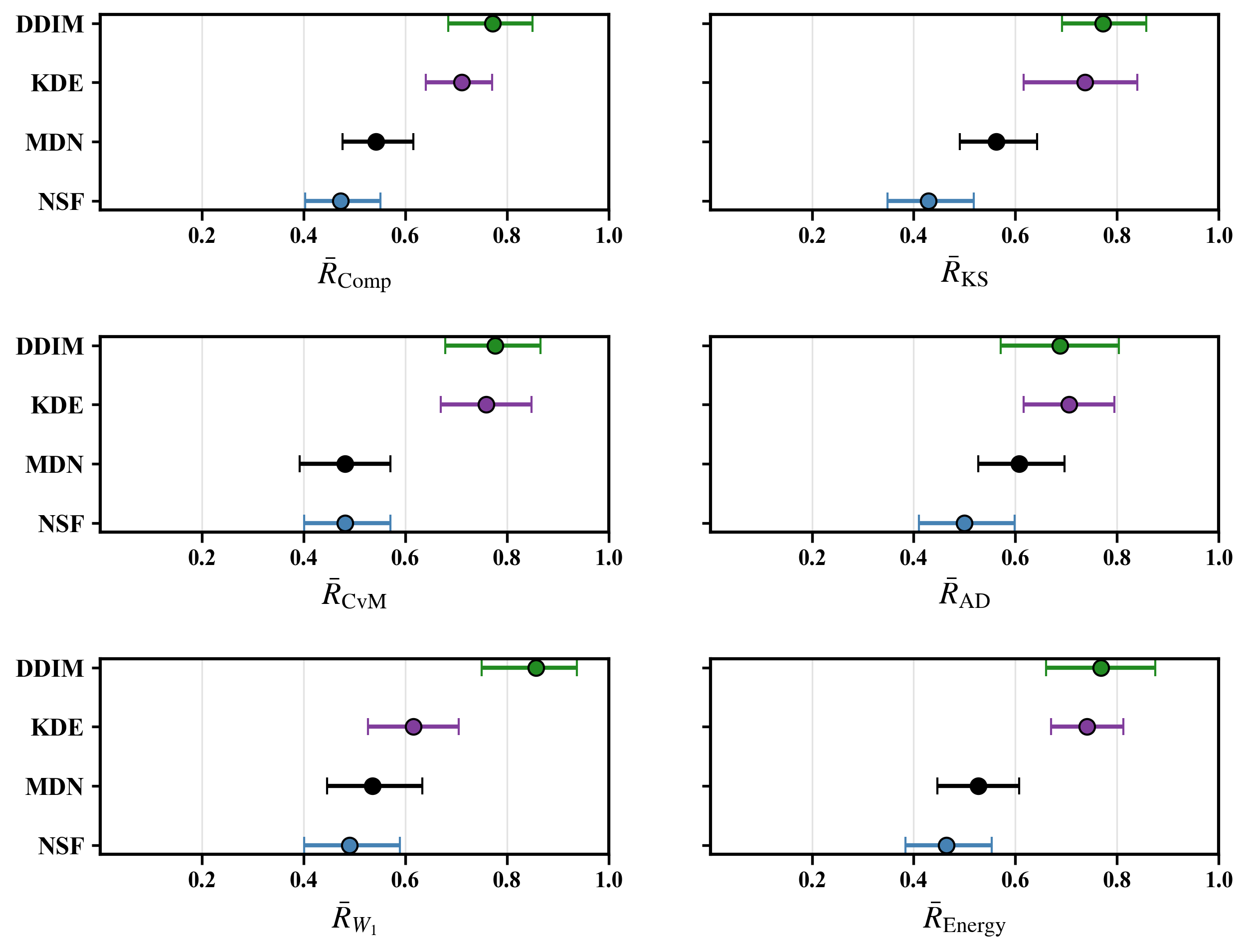}
  \caption{Per-model mean normalized rank across $n=28$ tasks for each goodness-of-fit metric, with $95\,\%$ bootstrap confidence intervals from $1000$ resamples. }%
  \label{fig:forest_all}
  \vspace{-0.5cm}
\end{figure}

\section{Application to Measured Indoor Channels}\label{sec:results}
We apply the proposed GenAI distribution fitting methodology to the InH propagation measurements conducted at $16.95$\,GHz in the FR3 band at the NYU WIRELESS Research Center, 370 Jay Street, Brooklyn, NY~\cite{Shakya2025InH}. A wideband sliding-correlator channel sounder with $1$\,GHz bandwidth and $31$\,dBm effective isotropic radiated power (EIRP) was used with $20$\,dBi gain horn antennas with $15^\circ$ half-power beamwidth (HPBW). A total of 20 TX-RX location pairs were measured using four TX positions, covering TX-RX separation distances from $11$\,m to $97$\,m. Seven of the 20 pairs are in LOS and 13 are in NLOS.
The initial data processing steps of Section~\ref{sec:initial} are applied to the measurements to extract the 14 primary channel parameters for the LOS and NLOS scenarios, yielding the $n=28$ per-parameter datasets that serve as input to the GenAI fitting methodology. The remainder of the section describes the per-task NDE fits, the cross-model statistical comparison, and the augmented parametric step for the InH campaign.

\subsection{NDE: Single-Frequency Fitting}
Figs.~\ref{fig:cmp-counts} and~\ref{fig:cmp-angles} show the ECDF of $5000$ samples from each NDE alongside the measured ECDF for the $14$ primary channel parameters, with LOS and NLOS drawn as internal panels in each subfigure. Each legend lists the per-task composite rank $\bar{R}\in[1/k,1]$ over $k=4$ models.

Across both figures, NSF outperforms the other three models by winning 12 of the 28 tasks, namely both subpath powers, most delays, and most NLOS angular tasks, all distributions whose measured ECDF is heavy-tailed or strongly skewed in dB. Spline flows place knots adaptively along the support, capturing both the steep CDF rise near the mode and the slow decay in the tail without the over-smoothing that affects KDE and DDIM on heavy-tailed shapes. On subpath powers, NSF reaches $\bar{R}=0.25$ in LOS and $\bar{R}=0.30$ in NLOS, with DDIM last at $\bar{R}=1$ in both panels. The cluster-power panel behaves differently, with MDN leading in LOS ($\bar{R}=0.25$) and DDIM in NLOS ($\bar{R}=0.50$); the four model ECDFs lie close together, and the ranking is decided by small differences in tail fit.

MDN wins eight tasks, four in the discrete count family (LOS subpath count, NLOS lobe count AOA, NLOS time clusters, NLOS subpath count). A Gaussian mixture matches the step-function ECDFs of integer-valued data by placing one low-variance Gaussian at each integer, which the smooth CDF of DDIM cannot reproduce. The remaining wins (LOS inter-cluster delay, two lobe-mean zeniths, and LOS cluster power) have ECDFs with separated regions of mass that the mixture matches with one component per region.

DDIM wins five tasks, namely four angular offsets (the LOS azimuth offset AOD, the zenith offset AOA in LOS and NLOS, and the LOS zenith offset AOD) and NLOS cluster power. The four angular distributions concentrate most mass near zero on continuous bounded support, where diffusion sampling fits the bulk shape accurately. The NLOS cluster-power win is decided by a small margin over MDN and KDE, with the three models' composite ranks within $0.10$ of one another. DDIM remains the worst-ranked model on the two subpath powers, on LOS cluster power, and on every count task, where heavy tails and discreteness are the dominant challenges. As a diffusion model, DDIM carries far more parameters than the small per-task sample sizes support, which limits its performance outside the symmetric unimodal regime.

KDE wins the remaining three tasks, all LOS (LOS lobe count AOA, LOS lobe count AOD, LOS azimuth offset AOA). On the two LOS lobe counts (7 samples each over a narrow integer support of 1--4), KDE's Gaussian smoothing appears to fit the few discrete values more closely than the neural models. KDE is, however, the worst-ranked model on the lobe-mean zenith AOD panel (Fig.~\ref{fig:cmp-angles}(f)), at $\bar{R}=0.90$ (LOS) and $\bar{R}=0.95$ (NLOS).

Fig.~\ref{fig:forest_all} shows the per-model mean rank for each goodness-of-fit metric and for the composite, with $95\,\%$ percentile bootstrap confidence intervals from 1000 resamples of the $n=28$ tasks with replacement. The x-axis is the mean normalized rank across the 28 tasks for one metric, while the per-task $\bar{R}$ in Figs.~\ref{fig:cmp-counts} and~\ref{fig:cmp-angles} is a single-task value. NSF has the lowest or tied-lowest mean rank on every metric, tying MDN on CvM (both at $0.48$) and leading MDN on the others. The NSF/MDN gap is largest on KS ($0.43$ for NSF versus $0.56$ for MDN, a gap of $0.13$), where the two bootstrap intervals only marginally overlap. On $W_1$ and Energy the gap is approximately $0.06$ and the intervals overlap. On KS, CvM, and Energy, KDE and DDIM form a separated lower tier. On $W_1$, DDIM is the only model separated from the others, at $0.86$ versus $0.62$ for KDE, indicating that DDIM's bulk-shape error dominates the Wasserstein gap. On AD all four intervals overlap, indicating that AD does not separate the four models on the 28 tasks.

The Friedman omnibus test rejects model equality at $\alpha=0.05$ for KS, CvM, $W_1$, and Energy ($p < 10^{-3}$) and fails to reject for AD ($p = 0.07$). The Nemenyi post-hoc is therefore applied only to KS, CvM, $W_1$, and Energy. Following Dem\v{s}ar~\cite{Demsar2006}, the critical difference on the normalized scale is $\mathrm{CD} = 0.222$ (Eq.~\eqref{eq:nemenyi_cd} divided by $k$). NSF significantly outperforms both KDE and DDIM on KS, CvM, and Energy; on $W_1$, only DDIM separates from the NSF/MDN/KDE cluster. The NSF/MDN gap stays below CD on every metric, and the Nemenyi post-hoc therefore does not separate the two. NSF still has the lower mean rank on every metric and wins 12 of the 28 tasks against MDN's eight, even though the across-task gap falls within CD.
\begin{figure}[!t]
  \centering
  \captionsetup[subfloat]{font=footnotesize,labelfont=footnotesize,labelsep=space,skip=2pt}
  \subfloat[Protocol-A: density]{\includegraphics[width=0.49\columnwidth]{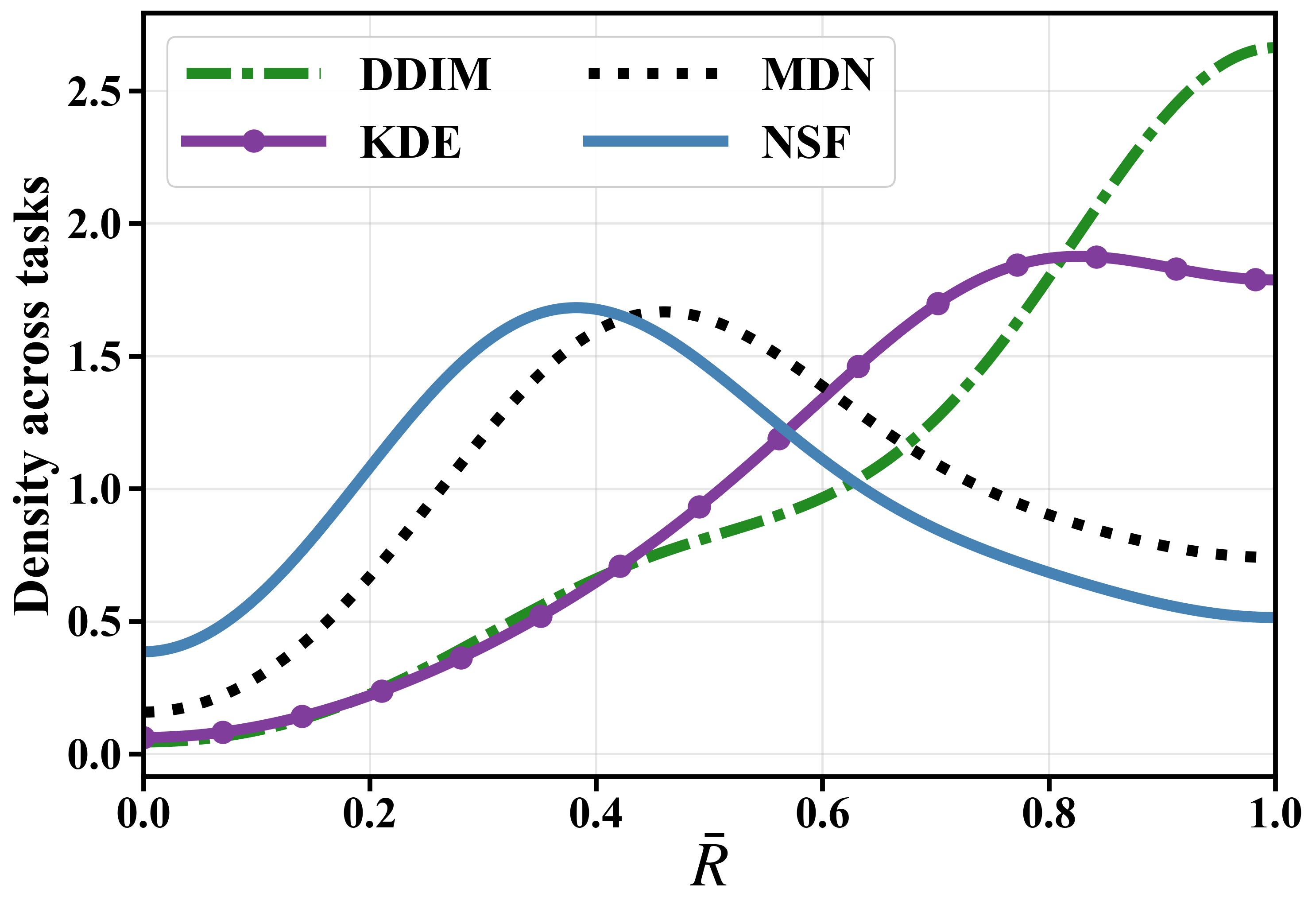}\label{fig:rdist:A:density}}\hfill
  \subfloat[Protocol-A: cumulative]{\includegraphics[width=0.49\columnwidth]{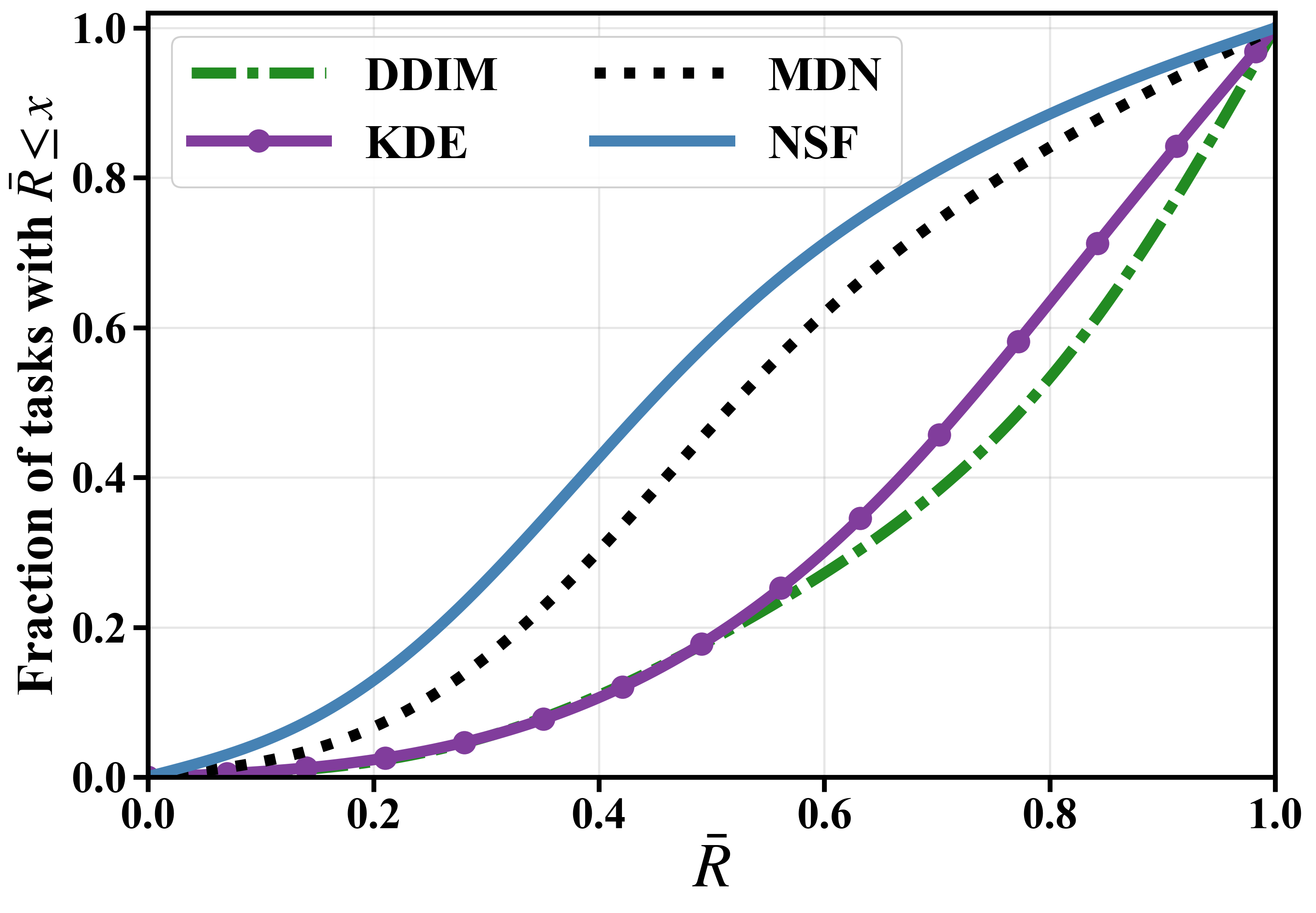}\label{fig:rdist:A:cdf}}\\
  \subfloat[Protocol-B: density]{\includegraphics[width=0.49\columnwidth]{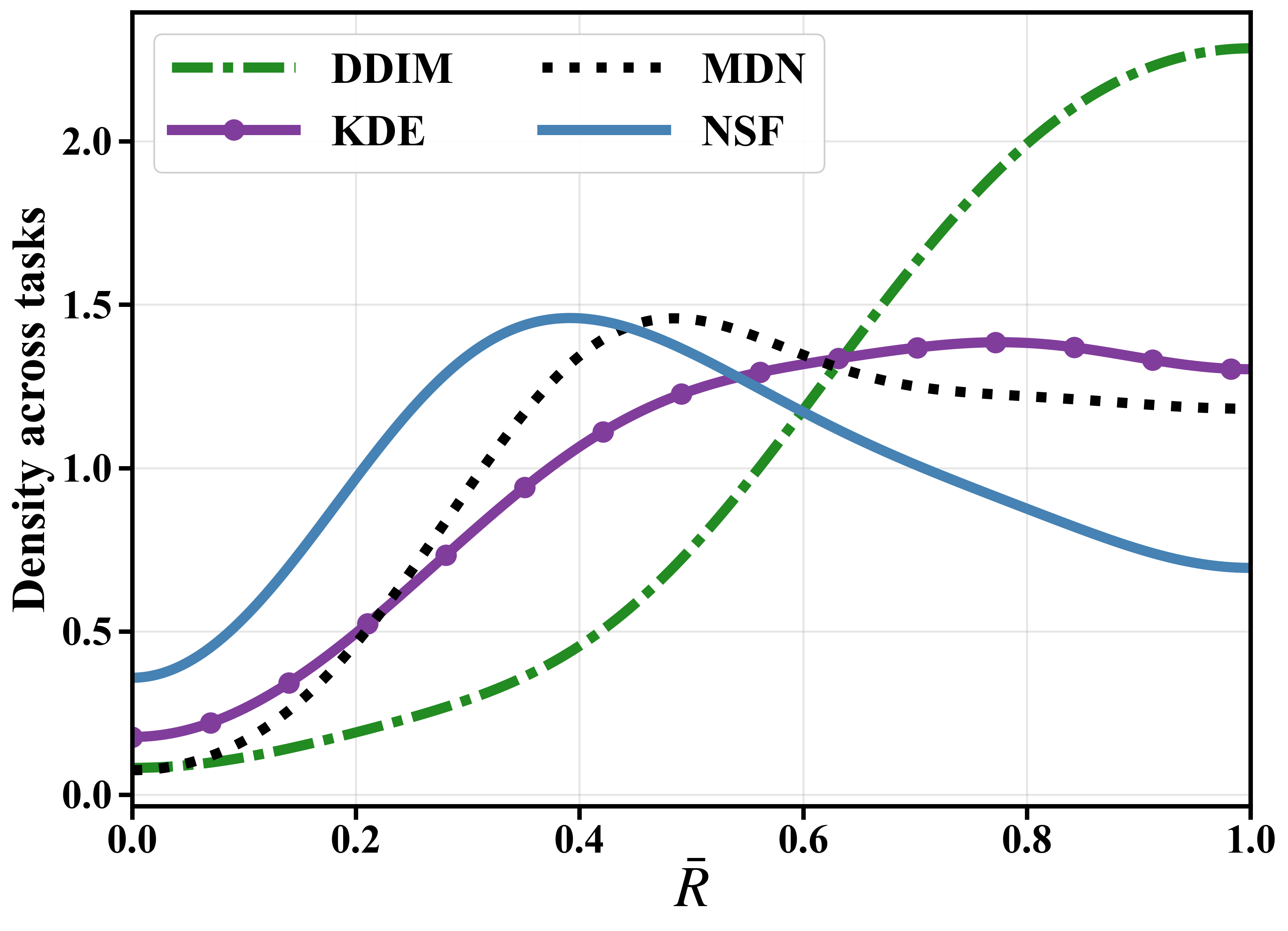}\label{fig:rdist:B:density}}\hfill
  \subfloat[Protocol-B: cumulative]{\includegraphics[width=0.49\columnwidth]{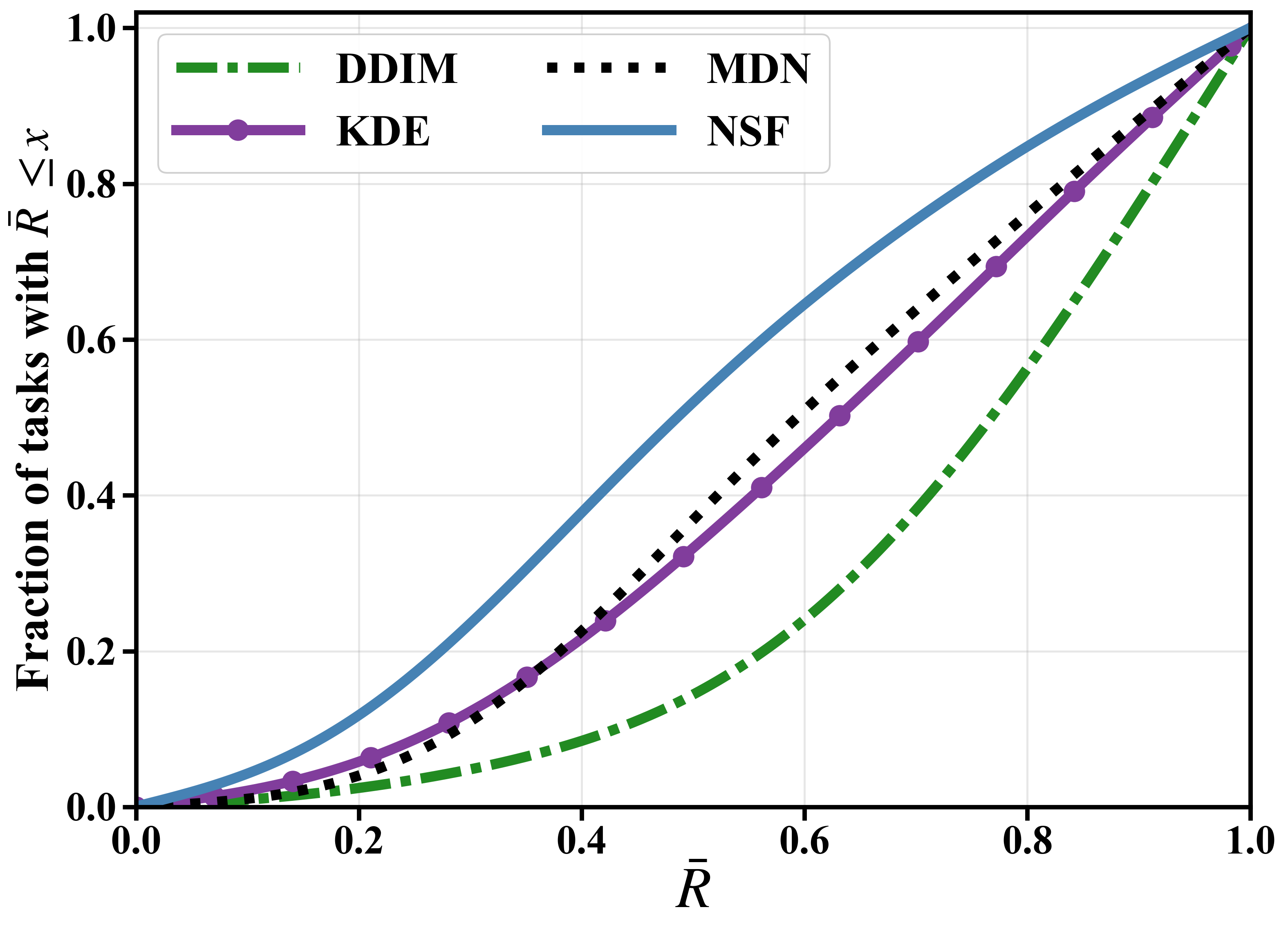}\label{fig:rdist:B:cdf}}\\
  \subfloat[Conditional model: density]{\includegraphics[width=0.49\columnwidth]{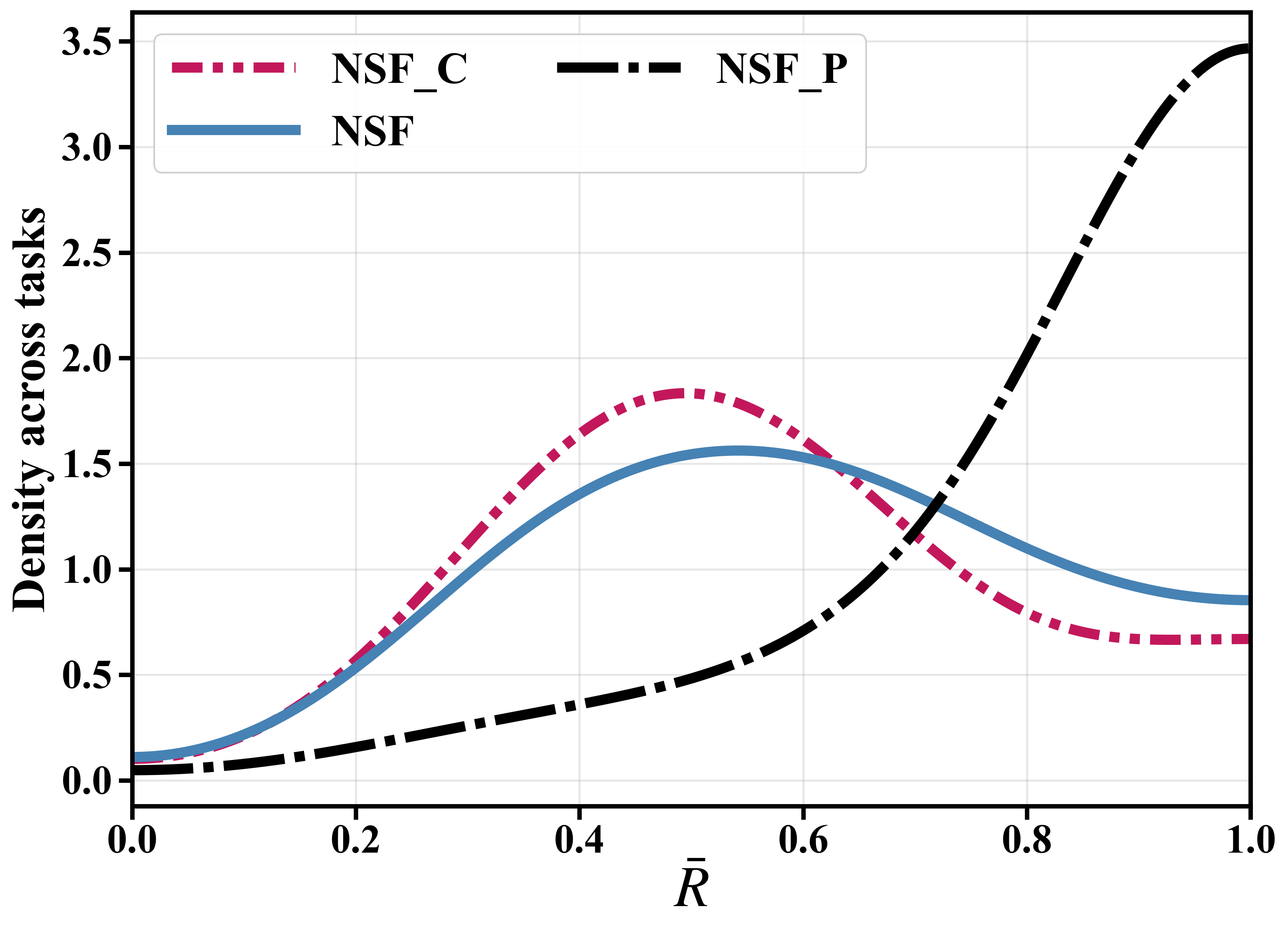}\label{fig:rdist:C:density}}\hfill
  \subfloat[Conditional model: cumulative]{\includegraphics[width=0.49\columnwidth]{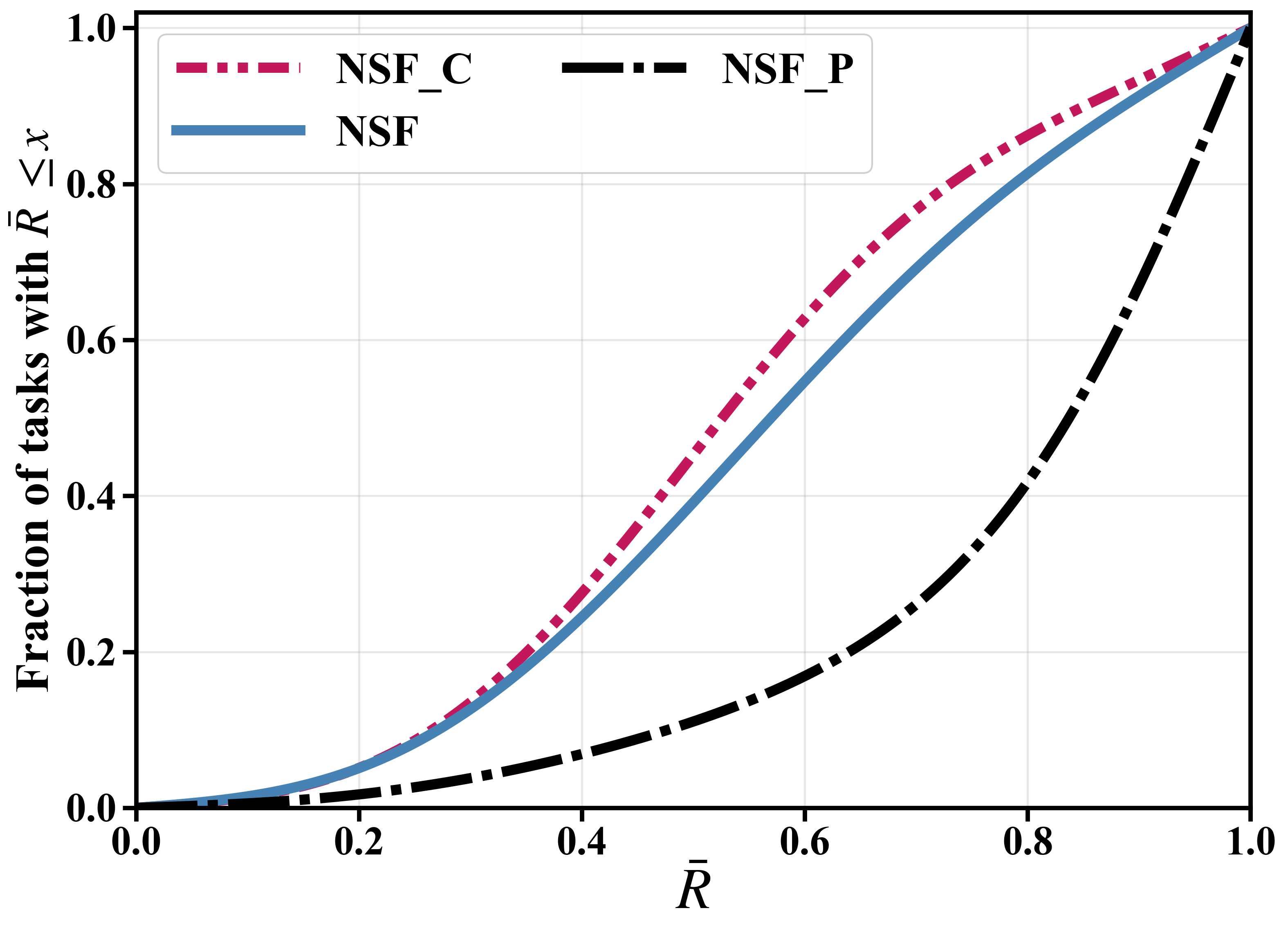}\label{fig:rdist:C:cdf}}
  \caption{Distribution of the per-task composite rank $\bar{R}$
  across the $n=28$ tasks. 
  (a, b)~Protocol-A: each model trained on all available data and
  evaluated on the $16.95$\,GHz target.
  (c, d)~Protocol-B: five-fold cross-validation on the $16.95$\,GHz target.
  (e, f)~Per-task NSF, NSF\textsubscript{C}, and
  NSF\textsubscript{P}
  under Protocol-A.}
  \label{fig:rdist-all}
  \vspace{-0.5cm}
\end{figure}

\begin{table}[!t]
\centering
\caption{Parametric distributions for the channel parameters at $16.95$\,GHz InH. 
$\mathcal{N}(\mu,\sigma)$ Normal with mean $\mu$ and standard
deviation $\sigma$; $\mathcal{LN}(\mu_{LN},\sigma_{LN})$
Log-Normal with scale $\exp(\mu_{LN})$ and shape $\sigma_{LN}$;
$\Gamma(\alpha,\beta)$ Gamma with shape $\alpha$ and scale
$\beta$; $\text{Lap}(\mu,b)$ Laplace with location $\mu$ and
scale $b$; $\text{C-Exp}(\beta_s,\mu_s)$ Composite Exponential $(1-\beta_s)\,\delta(1) + \beta_s\bigl(\text{Exp}(\mu_s)+1\bigr)$. }
\label{tab:parametric-fits}
\renewcommand{\arraystretch}{1.15}
\setlength{\tabcolsep}{3pt}
\setlength{\arrayrulewidth}{1pt}
\scriptsize
\begin{tabular}{|l|c|c|c|c|}
\hline
\textbf{Channel Parameter} & \textbf{Distribution} & \textbf{Parameters} &
\textbf{LOS} & \textbf{NLOS} \\
\hline
\# of AOA lobes $L^{\mathrm{AOA}}$       & $\text{Poisson}+1$ & $\lambda_{l,\mathrm{AOA}}$ & $0.49$ & $1.52$ \\
\hline
\# of AOD lobes $L^{\mathrm{AOD}}$       & $\text{Poisson}+1$ & $\lambda_{l,\mathrm{AOD}}$ & $1.11$ & $1.34$ \\
\hline
\# of time clusters $N$                  & $\text{Poisson}+1$ & $\lambda_c$              & $3.27$ & $3.61$ \\
\hline
\multirow{2}{*}{\# of subpaths $M$} & \multirow{2}{*}{$\text{C-Exp}$} & $\beta_s$ & $0.87$ & $0.90$ \\
\cline{3-5}
                                 &                                  & $\mu_s$   & $14.8$ & $55.0$ \\
\hline
\multirow{2}{*}{Inter-cluster delay $\Delta\tau$ [ns]} & \multirow{2}{*}{$\mathcal{LN}$} & $\mu_{LN,\tau}$    & $2.75$ & $2.08$ \\
\cline{3-5}
                                                          &                                  & $\sigma_{LN,\tau}$ & $1.17$ & $1.42$ \\
\hline
\multirow{2}{*}{Intra-cluster delay $\rho$ [ns]} & \multirow{2}{*}{$\Gamma$} & $\alpha_\rho$ & $1.74$ & $1.62$ \\
\cline{3-5}
                                                       &                            & $\beta_\rho$  & $25.1$ & $54.4$ \\
\hline
\multirow{2}{*}{Subpath AOA offset $\Delta\phi^{\mathrm{AOA}}$ [\textdegree]} & \multirow{2}{*}{$\text{Lap}$} & $\mu_{\phi,\mathrm{AOA}}$ & $0.00$ & $0.00$  \\
\cline{3-5}
                                                                                      &                                & $b_{\phi,\mathrm{AOA}}$   & $3.24$ & $13.57$ \\
\hline
\multirow{2}{*}{Subpath AOD offset $\Delta\phi^{\mathrm{AOD}}$ [\textdegree]} & \multirow{2}{*}{$\text{Lap}$} & $\mu_{\phi,\mathrm{AOD}}$ & $0.00$ & $0.00$ \\
\cline{3-5}
                                                                                      &                                & $b_{\phi,\mathrm{AOD}}$   & $1.35$ & $8.60$ \\
\hline
\multirow{2}{*}{Subpath ZOA offset $\Delta\theta^{\mathrm{ZOA}}$ [\textdegree]} & \multirow{2}{*}{$\text{Lap}$} & $\mu_{\theta,\mathrm{ZOA}}$ & $1.81$ & $0.97$ \\
\cline{3-5}
                                                                                       &                                & $b_{\theta,\mathrm{ZOA}}$   & $1.81$ & $2.49$ \\
\hline
\multirow{2}{*}{Subpath ZOD offset $\Delta\theta^{\mathrm{ZOD}}$ [\textdegree]} & \multirow{2}{*}{$\mathcal{N}$} & $\mu_{\theta,\mathrm{ZOD}}$    & $0.30$ & $0.57$ \\
\cline{3-5}
                                                                                       &                                 & $\sigma_{\theta,\mathrm{ZOD}}$ & $0.67$ & $0.77$ \\
\hline
\multirow{2}{*}{Mean ZOA $\theta^{\mathrm{ZOA}}$ [\textdegree]} & \multirow{2}{*}{$\text{Lap}$} & $\mu_{l,\mathrm{ZOA}}$ & $0.17$ & $0.57$ \\
\cline{3-5}
                                                                &                                & $b_{l,\mathrm{ZOA}}$   & $0.91$ & $1.22$ \\
\hline
\multirow{2}{*}{Mean ZOD $\theta^{\mathrm{ZOD}}$ [\textdegree]} & \multirow{2}{*}{$\mathcal{N}$} & $\mu_{l,\mathrm{ZOD}}$    & $-2.82$ & $-2.67$ \\
\cline{3-5}
                                                                &                                 & $\sigma_{l,\mathrm{ZOD}}$ & $1.48$  & $1.55$  \\
\hline
\multirow{2}{*}{Cluster power $P$ [dBm]} & \multirow{2}{*}{$\mathcal{N}$} & $\mu_P$    & $-54.6$ & $-73.45$ \\
\cline{3-5}
                                            &                                 & $\sigma_P$ & $19.27$  & $14.34$  \\
\hline
\multirow{2}{*}{Subpath power $\Pi$ [dBm]} & \multirow{2}{*}{$\mathcal{N}$} & $\mu_\Pi$    & $-61.07$ & $-81.42$ \\
\cline{3-5}
                                                   &                                 & $\sigma_\Pi$ & $14.74$  & $10.80$  \\
\hline
\end{tabular}
\end{table}

Fig.~\ref{fig:rdist-all}(a, b) plots the density and CDF of the
per-task composite rank $\bar{R}$ across the $28$ tasks. NSF concentrates
the most mass at low $\bar{R}$ (across-task mean $0.47$), with
MDN second ($0.54$); KDE and DDIM are shifted higher ($0.71$ and
$0.77$), consistent with Fig.~\ref{fig:forest_all}.

Fig.~\ref{fig:rdist-all}(c, d) repeats the comparison under Protocol-B
(five-fold cross-validation on the $16.95$\,GHz target), with tasks below $N=50$ falling back to their Protocol-A values. NSF
remains the leader (across-task mean $0.51$) and DDIM remains
last ($0.76$). The Protocol-A gap between MDN ($0.54$) and KDE
($0.71$) collapses under cross-validation to $0.61$ versus $0.62$.
Most of the change is on the KDE side (an improvement of $0.09$
against a $0.07$ decline for MDN), because Protocol-A evaluates
each model on the same target samples it was trained on, which
rewards the in-sample fit that flexible models like MDN can
achieve.
Protocol-B's held-out folds remove the in-sample advantage. KDE places one Gaussian kernel at each training sample, producing a smooth density that remains accurate on the held-out data. MDN learns its mixture parameters from the training samples, producing a fit that matches the training distribution closely but is less accurate on the held-out data. NSF's lead over the other three estimators is preserved across both protocols.

\begin{figure*}[!t]
  \centering
  \captionsetup[subfloat]{font=scriptsize,labelfont=scriptsize,labelsep=space,skip=0pt}
  \subfloat[$L^{\mathrm{AOA}}$ (LOS KS $=0.10$, NLOS KS $=0.06$)]{\includegraphics[width=0.43\textwidth]{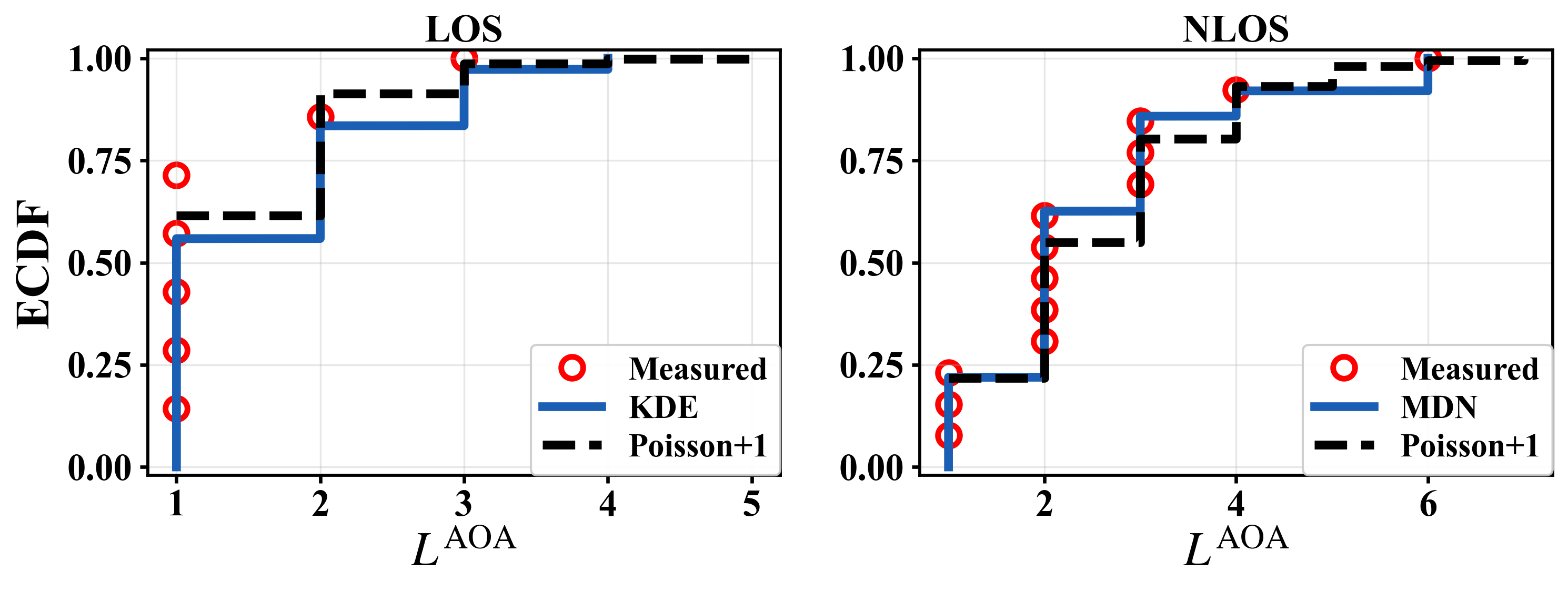}\label{fig:pf:aoa_lobe}}\hfill
  \subfloat[$L^{\mathrm{AOD}}$ (LOS KS $=0.10$, NLOS KS $=0.04$)]{\includegraphics[width=0.43\textwidth]{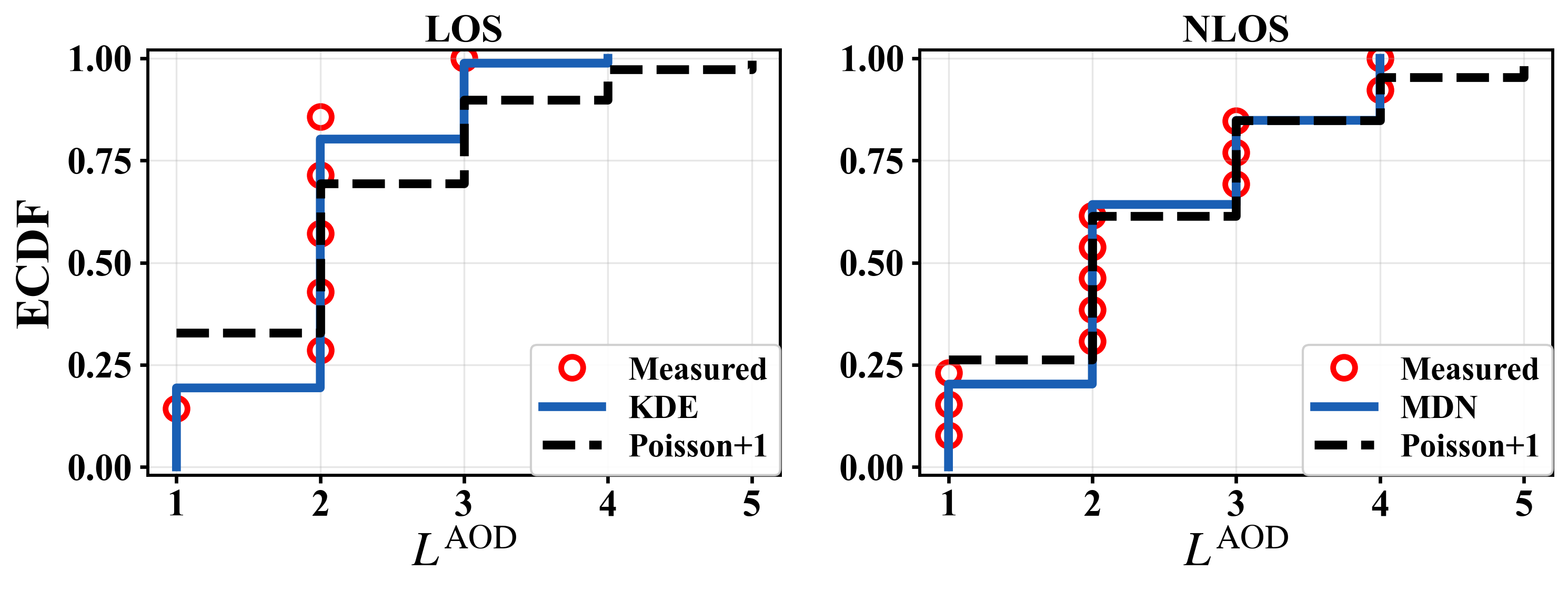}\label{fig:pf:aod_lobe}}\\
  \subfloat[$N$ (LOS KS $=0.17$, NLOS KS $=0.15$)]{\includegraphics[width=0.43\textwidth]{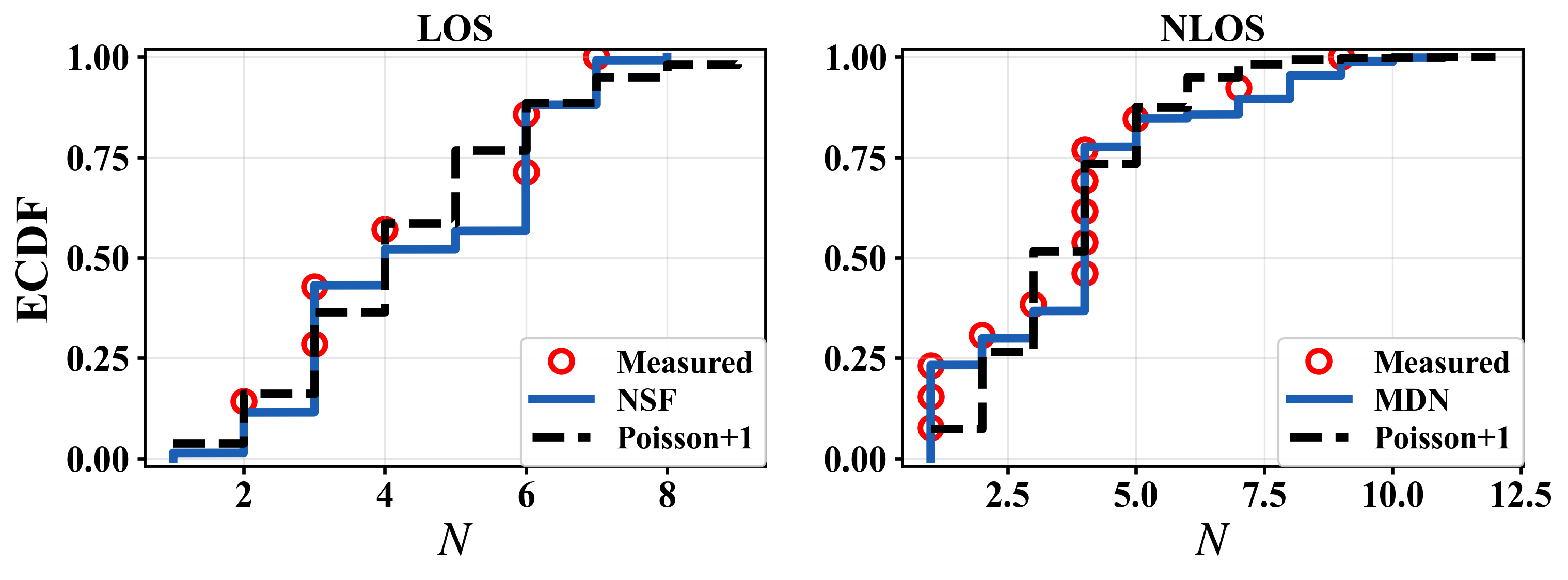}\label{fig:pf:tc}}\hfill
  \subfloat[$M$ (LOS KS $=0.18$, NLOS KS $=0.15$)]{\includegraphics[width=0.43\textwidth]{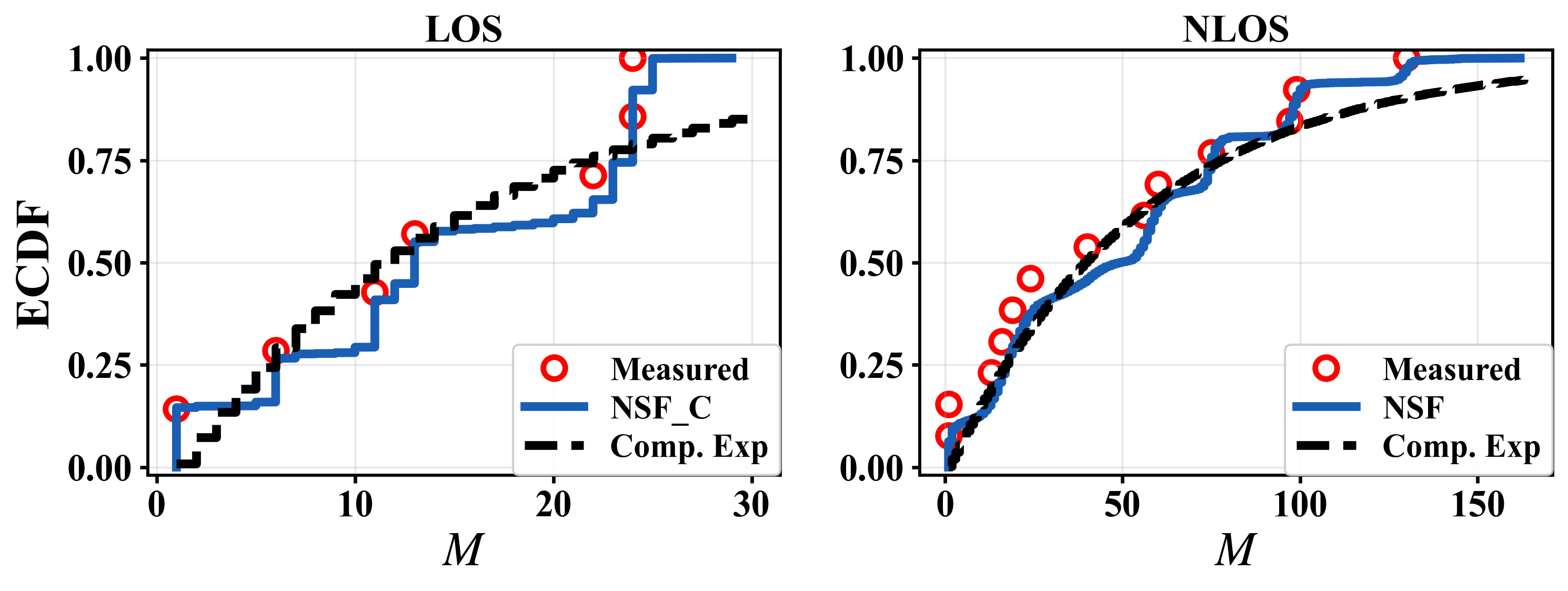}\label{fig:pf:sp}}\\
  \subfloat[$\Delta\tau$ (LOS KS $=0.18$, NLOS KS $=0.18$)]{\includegraphics[width=0.43\textwidth]{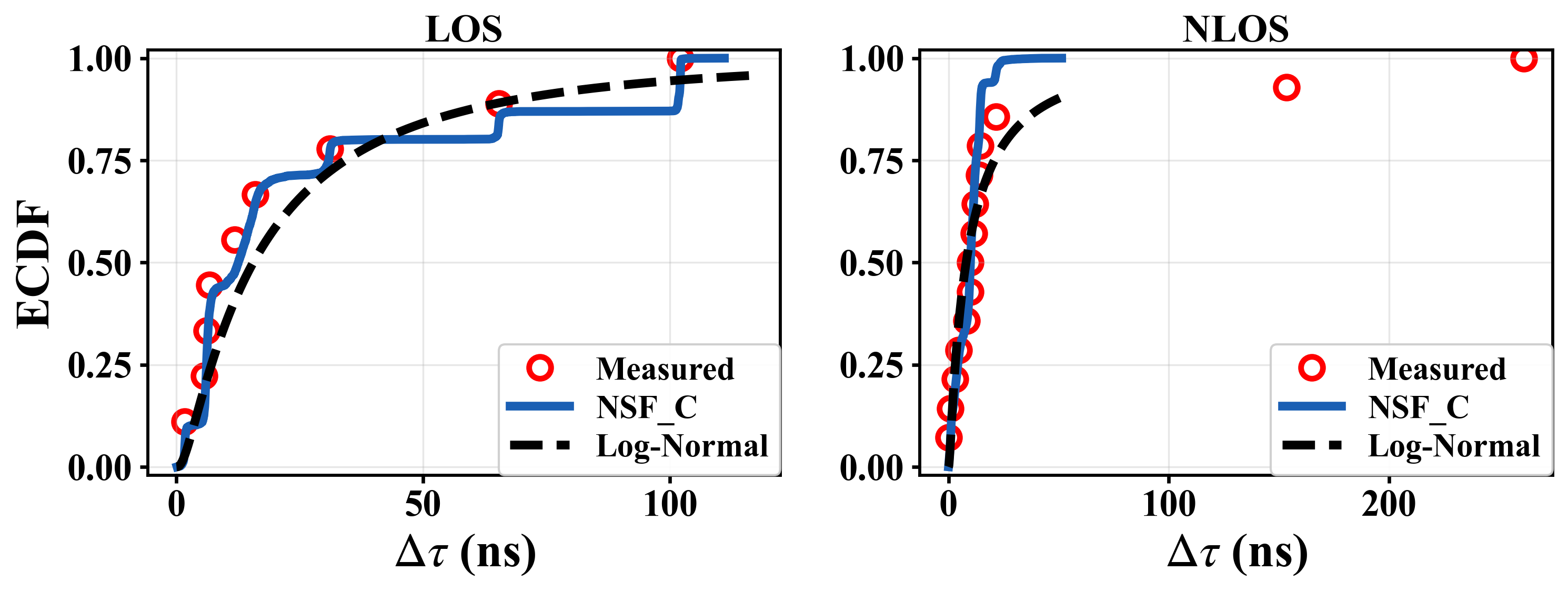}\label{fig:pf:icd}}\hfill
  \subfloat[$\rho$ (LOS KS $=0.05$, NLOS KS $=0.07$)]{\includegraphics[width=0.43\textwidth]{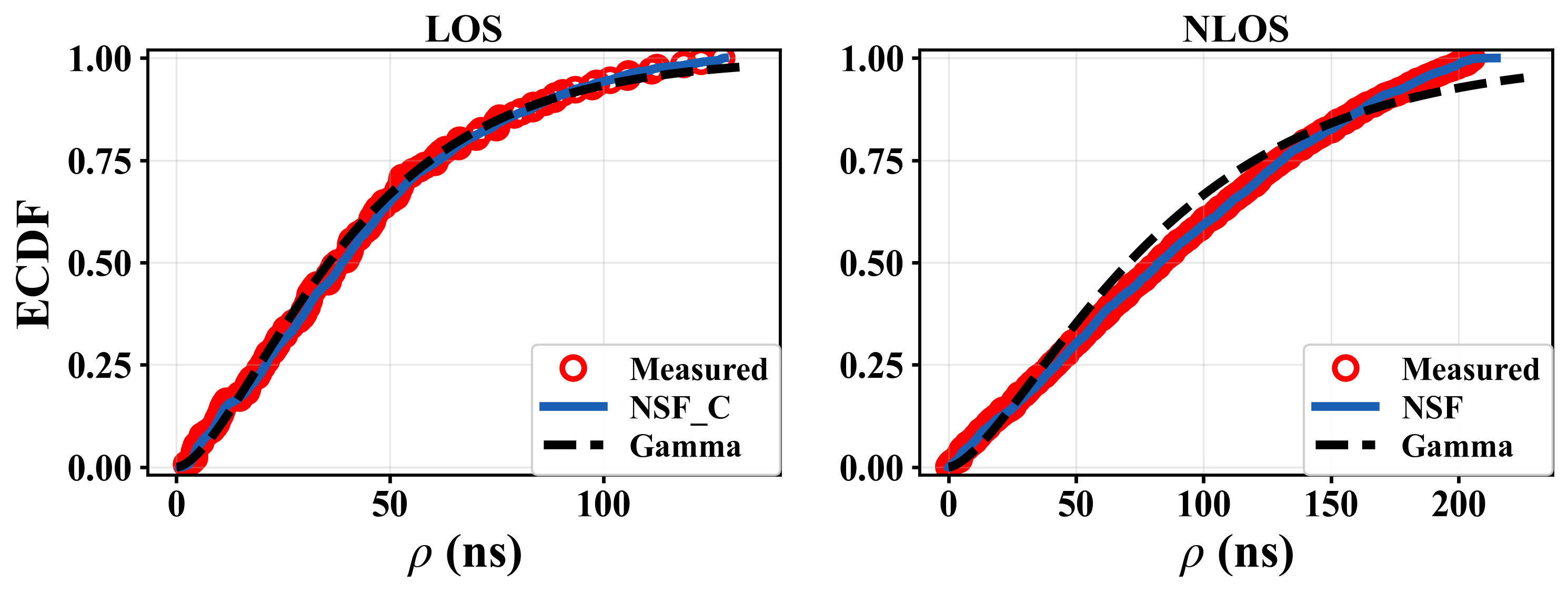}\label{fig:pf:ind}}\\
  \subfloat[$P$ (LOS KS $=0.11$, NLOS KS $=0.14$)]{\includegraphics[width=0.43\textwidth]{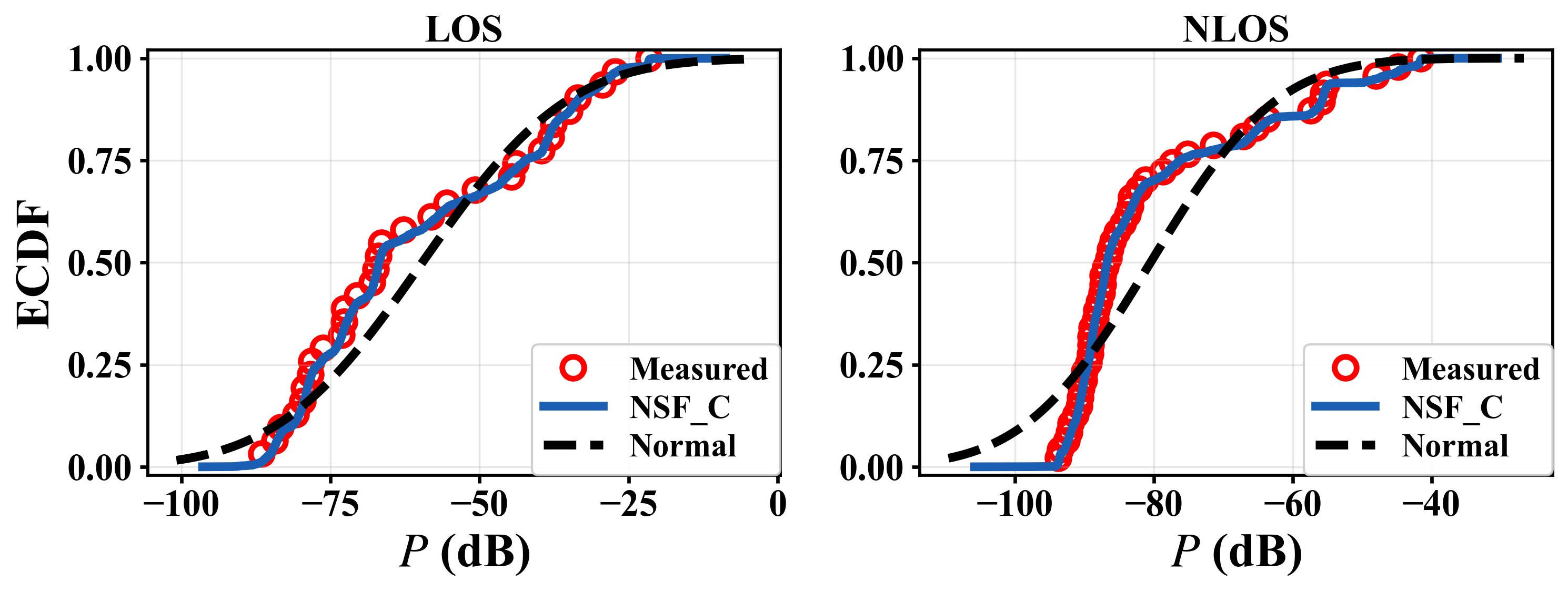}\label{fig:pf:cpwr}}\hfill
  \subfloat[$\Pi$ (LOS KS $=0.08$, NLOS KS $=0.07$)]{\includegraphics[width=0.43\textwidth]{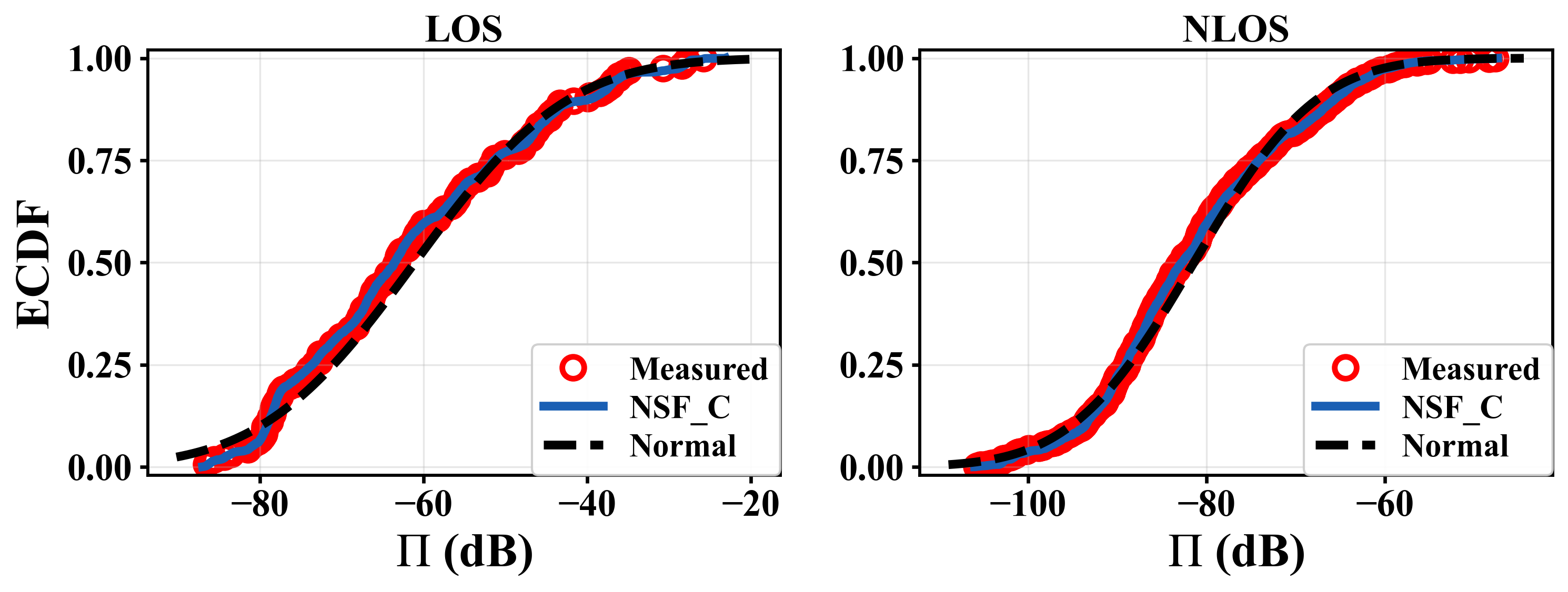}\label{fig:pf:spwr}}
  \caption{Per-parameter PF CDFs against the measured ECDFs for the count, temporal, and power parameters, namely $L^{\mathrm{AOA}}$ in (a), $L^{\mathrm{AOD}}$ in (b), $N$ in (c), $M$ in (d), $\Delta\tau$ in (e), $\rho$ in (f), $P$ in (g), and $\Pi$ in (h). Sub-caption labels list the LOS and NLOS KS distances between the fitted parametric CDF and the measured ECDF.}
  \label{fig:pf-counts}
  \vspace{-0.5cm}
\end{figure*}

\begin{figure}[!t]
  \centering
  \captionsetup[subfloat]{font=scriptsize,labelfont=scriptsize,labelsep=space,skip=0pt}
  \subfloat[$\Delta\phi^{\mathrm{AOA}}$ (LOS KS $=0.12$, NLOS KS $=0.09$)]{\includegraphics[width=0.43\textwidth]{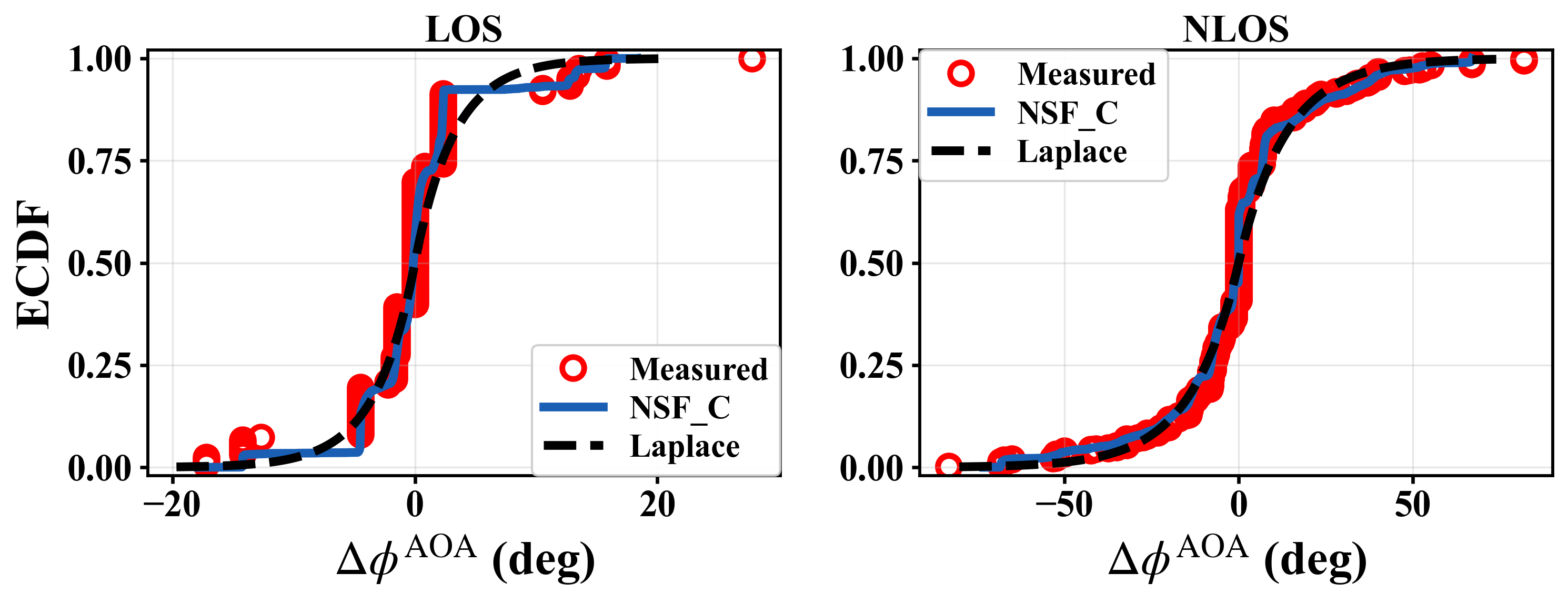}\label{fig:pf:daoa}}\hfill
  \subfloat[$\Delta\phi^{\mathrm{AOD}}$ (LOS KS $=0.17$, NLOS KS $=0.14$)]{\includegraphics[width=0.43\textwidth]{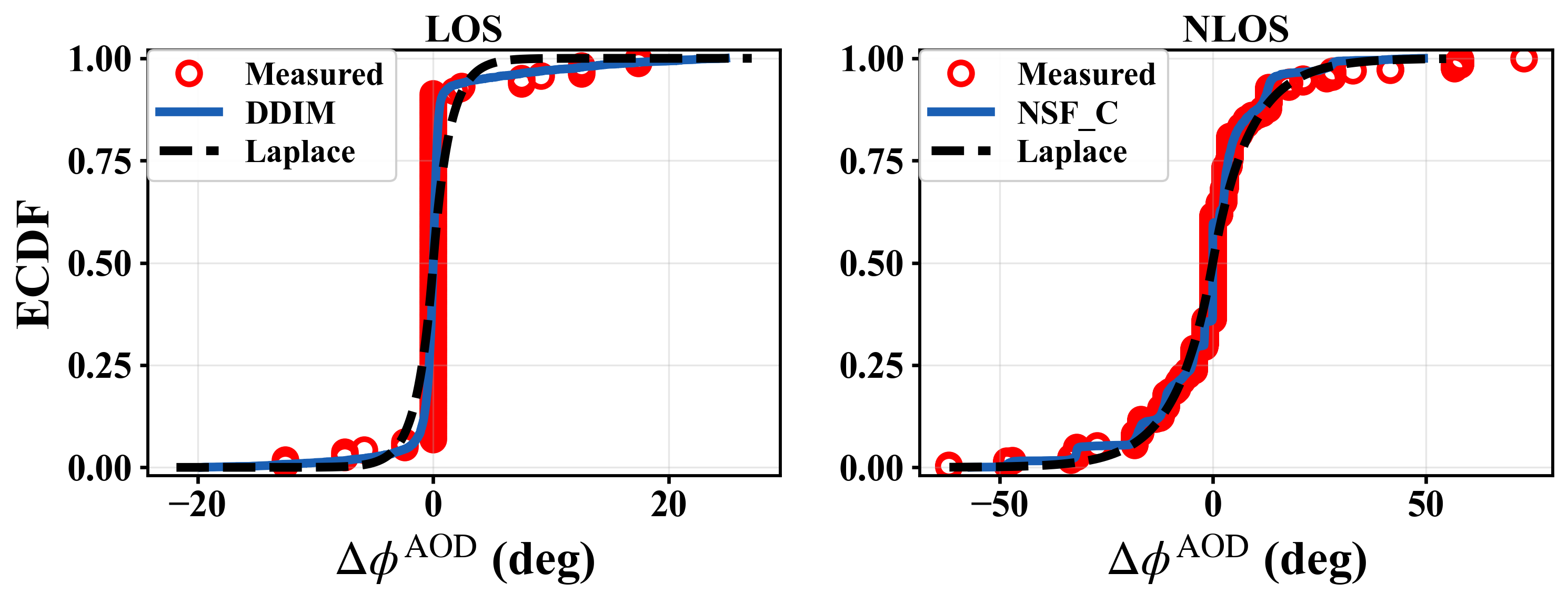}\label{fig:pf:daod}}\\
  \subfloat[$\Delta\theta^{\mathrm{ZOA}}$ (LOS KS $=0.17$, NLOS KS $=0.14$)]{\includegraphics[width=0.43\textwidth]{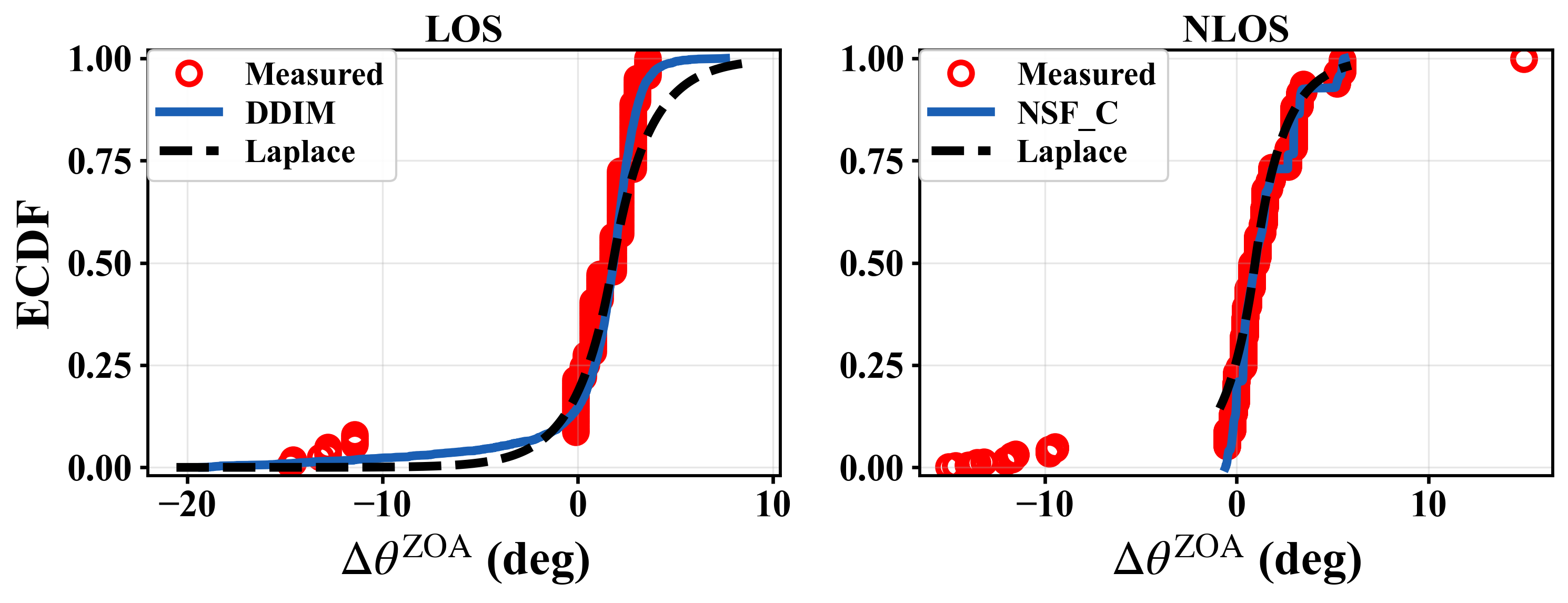}\label{fig:pf:dzoa}}\hfill
  \subfloat[$\Delta\theta^{\mathrm{ZOD}}$ (LOS KS $=0.19$, NLOS KS $=0.16$)]{\includegraphics[width=0.43\textwidth]{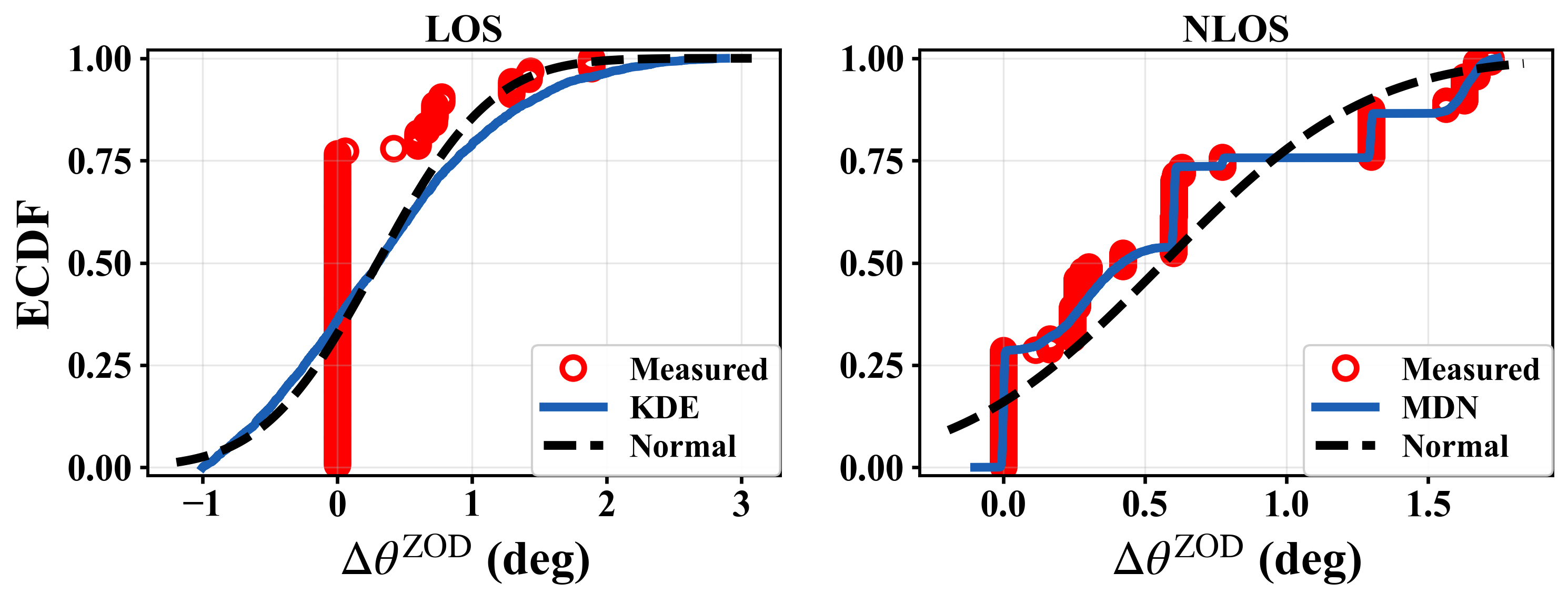}\label{fig:pf:dzod}}\\
  \subfloat[$\theta^{\mathrm{ZOA}}$ (LOS KS $=0.17$, NLOS KS $=0.13$)]{\includegraphics[width=0.43\textwidth]{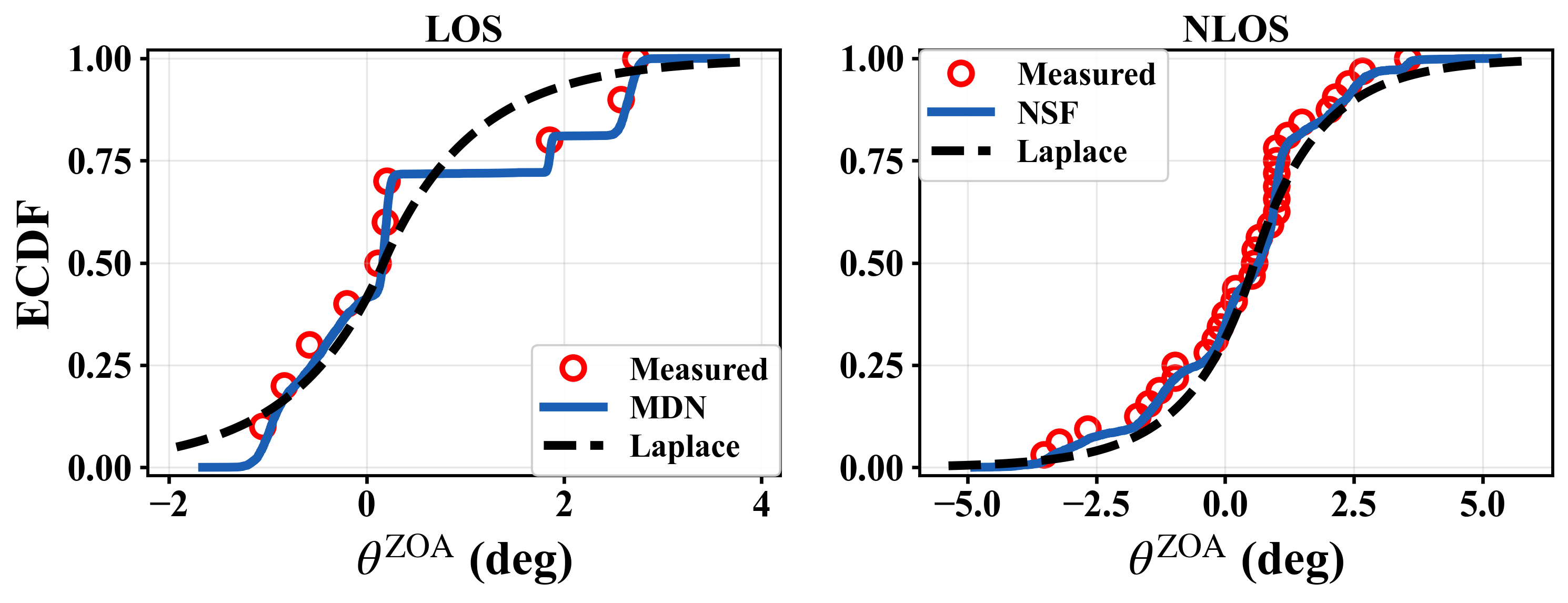}\label{fig:pf:mzoa}}\hfill
  \subfloat[$\theta^{\mathrm{ZOD}}$ (LOS KS $=0.13$, NLOS KS $=0.17$)]{\includegraphics[width=0.43\textwidth]{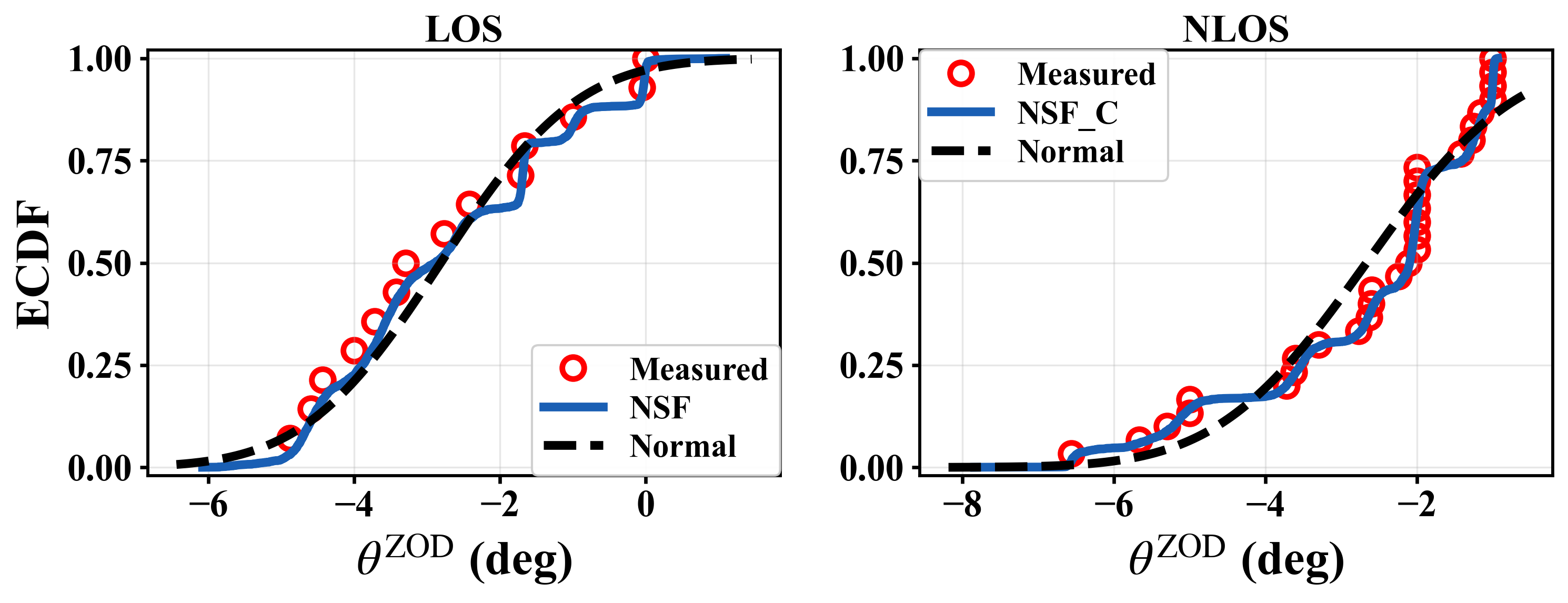}\label{fig:pf:mzod}}
  \caption{Per-parameter PF CDFs against the measured ECDFs for the angular parameters, namely $\Delta\phi^{\mathrm{AOA}}$ in (a), $\Delta\phi^{\mathrm{AOD}}$ in (b), $\Delta\theta^{\mathrm{ZOA}}$ in (c), $\Delta\theta^{\mathrm{ZOD}}$ in (d), $\theta^{\mathrm{ZOA}}$ in (e), and $\theta^{\mathrm{ZOD}}$ in (f).}
  \label{fig:pf-angles}
  \vspace{-0.5cm}
\end{figure}

\begin{figure*}[!t]
  \centering
  \subfloat[Omni RMS ASD]{\includegraphics[width=0.2\textwidth]{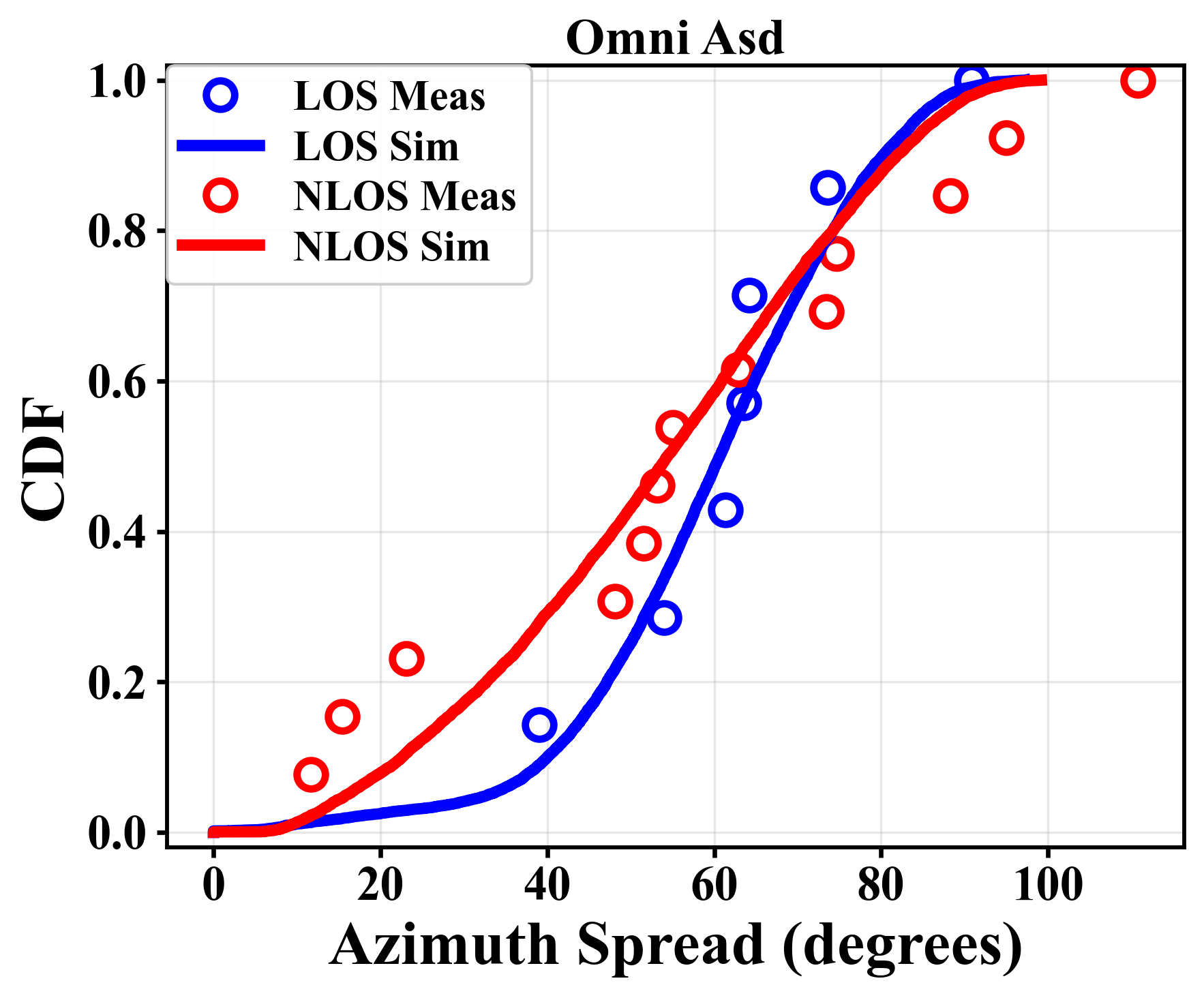}\label{fig:cm:omni-asd}}\hfill
  \subfloat[Omni RMS ASA]{\includegraphics[width=0.2\textwidth]{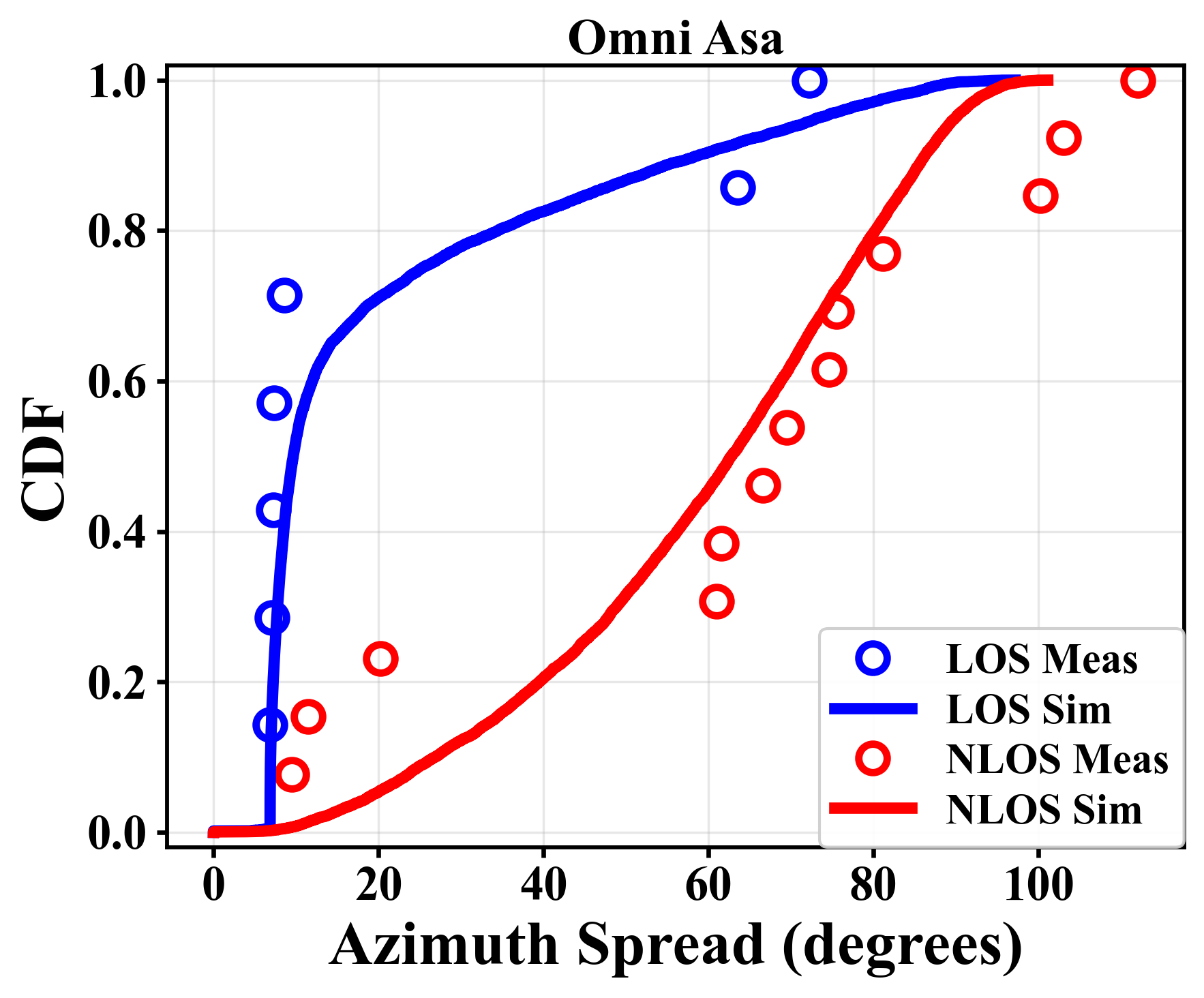}\label{fig:cm:omni-asa}}\hfill
  \subfloat[Omni RMS ZSD]{\includegraphics[width=0.2\textwidth]{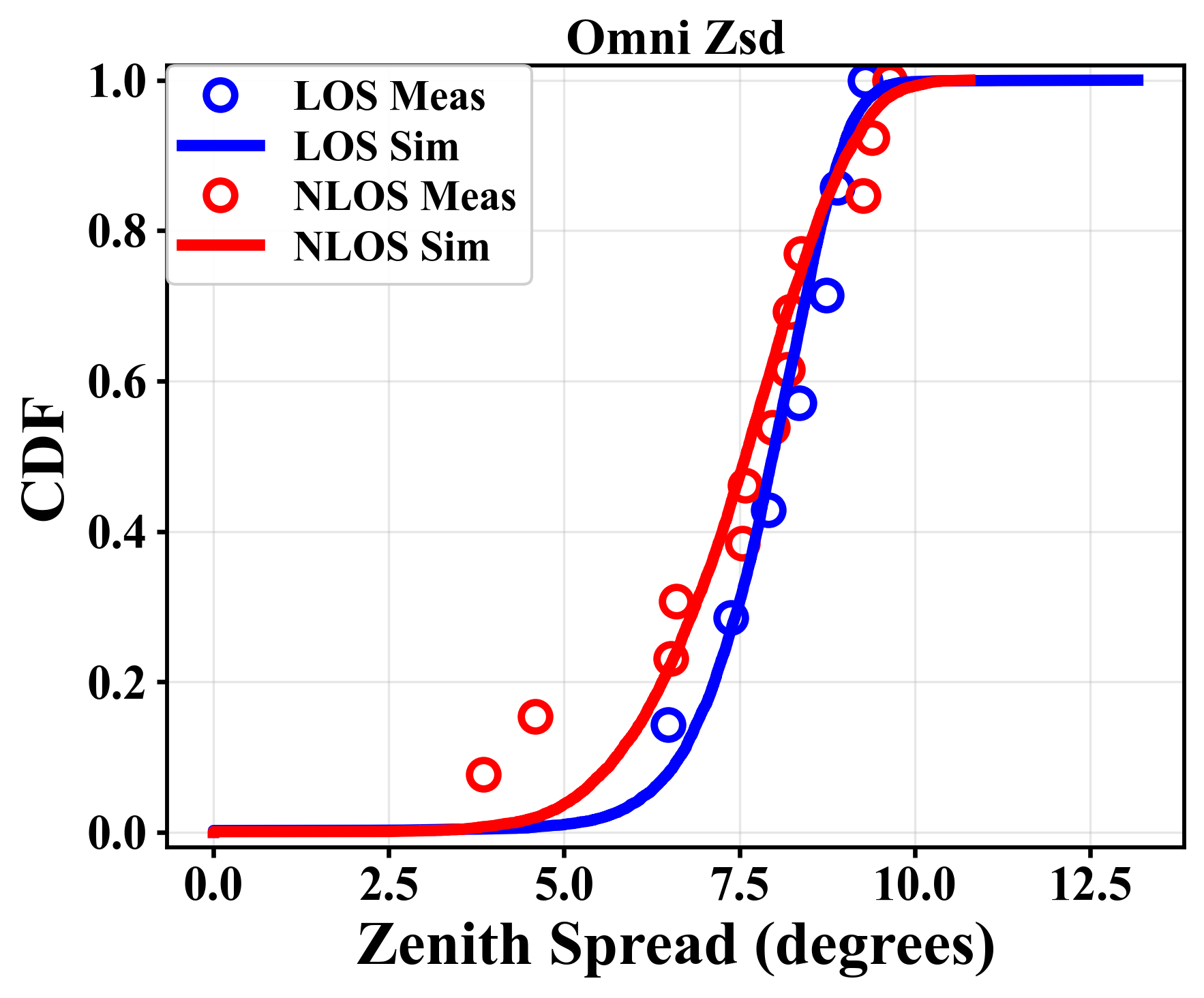}\label{fig:cm:omni-zsd}}\hfill
  \subfloat[Omni RMS ZSA]{\includegraphics[width=0.2\textwidth]{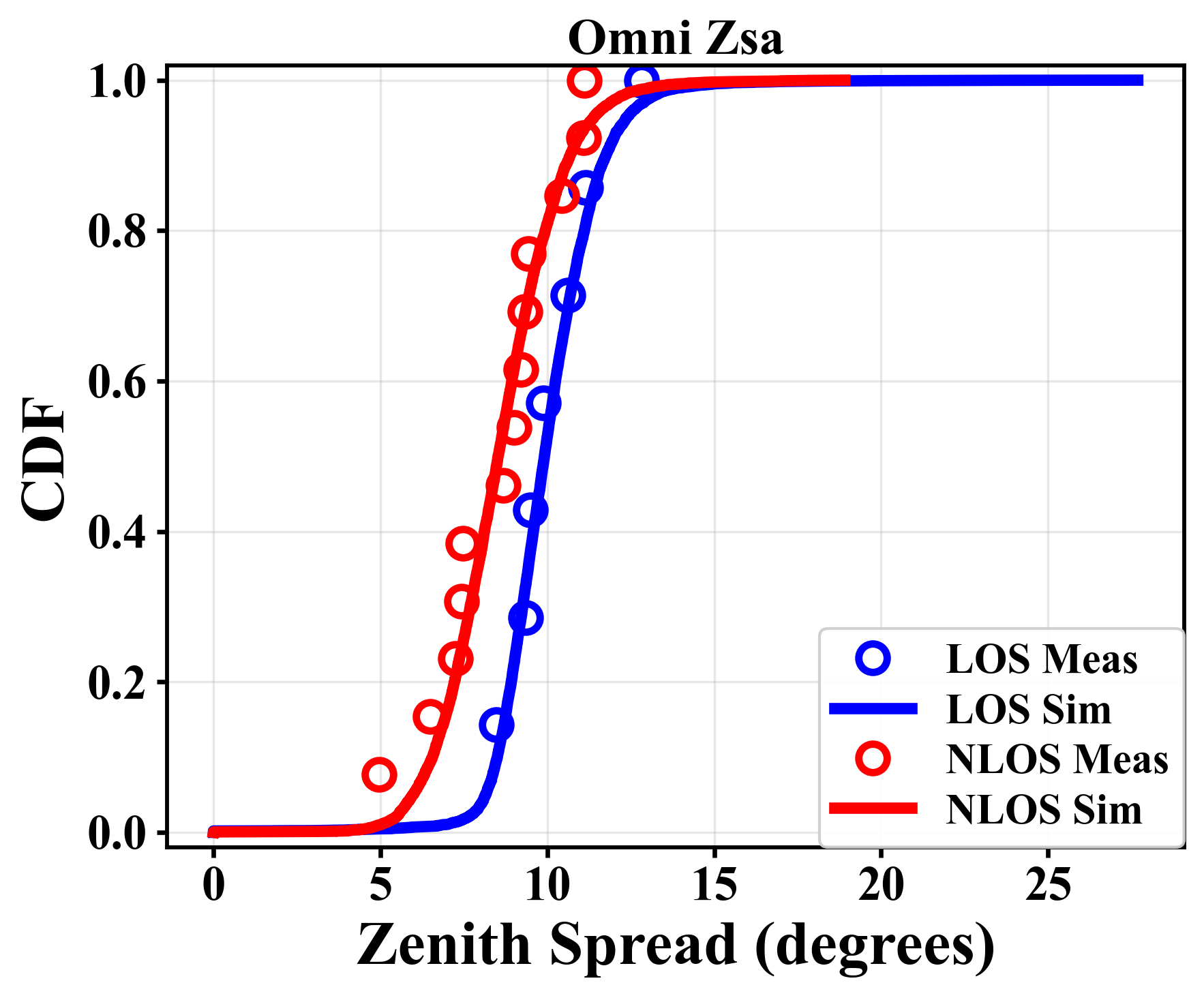}\label{fig:cm:omni-zsa}}\\[-2pt]
  \subfloat[Lobe RMS ASD]{\includegraphics[width=0.2\textwidth]{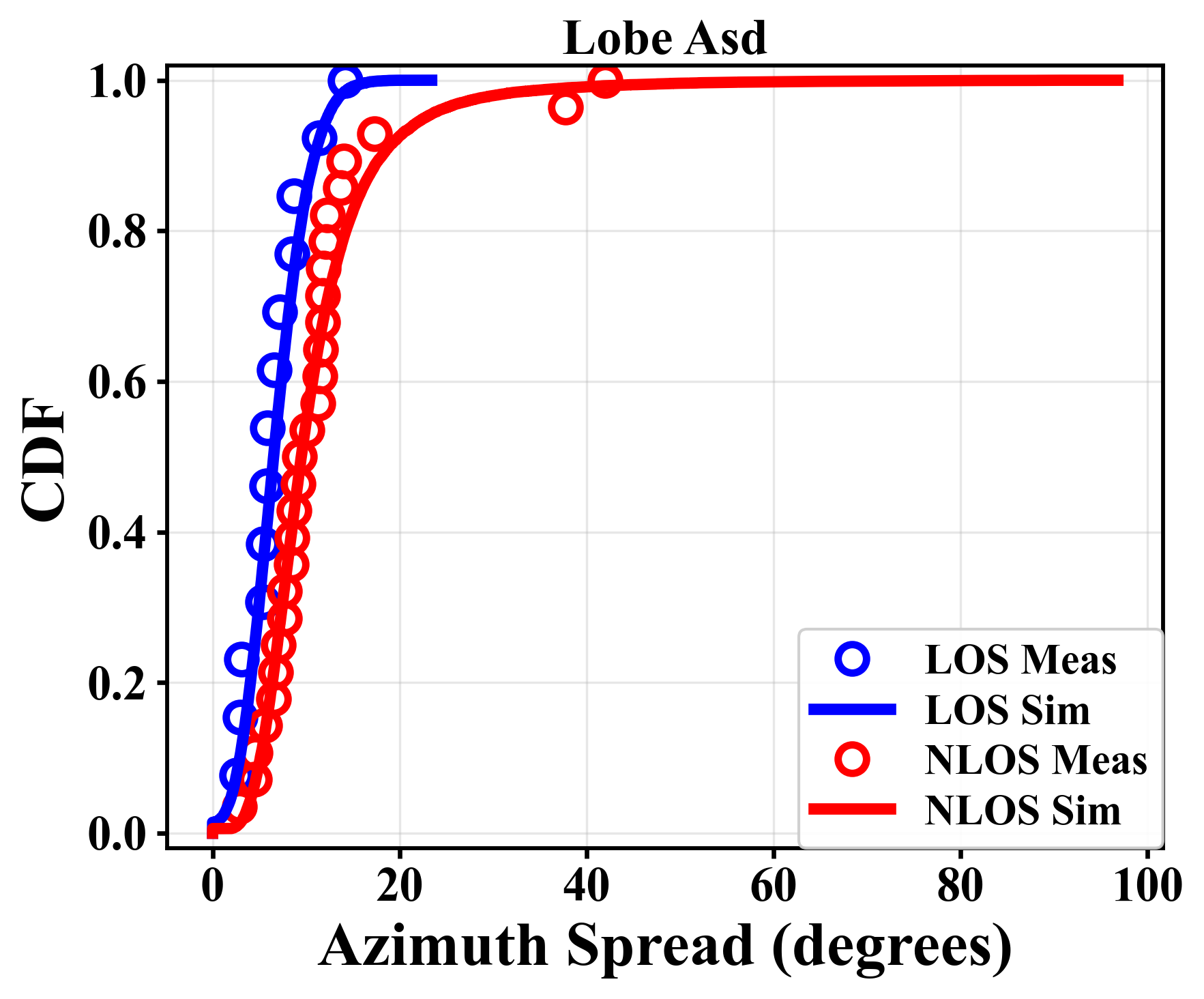}\label{fig:cm:lobe-asd}}\hfill
  \subfloat[Lobe RMS ASA]{\includegraphics[width=0.2\textwidth]{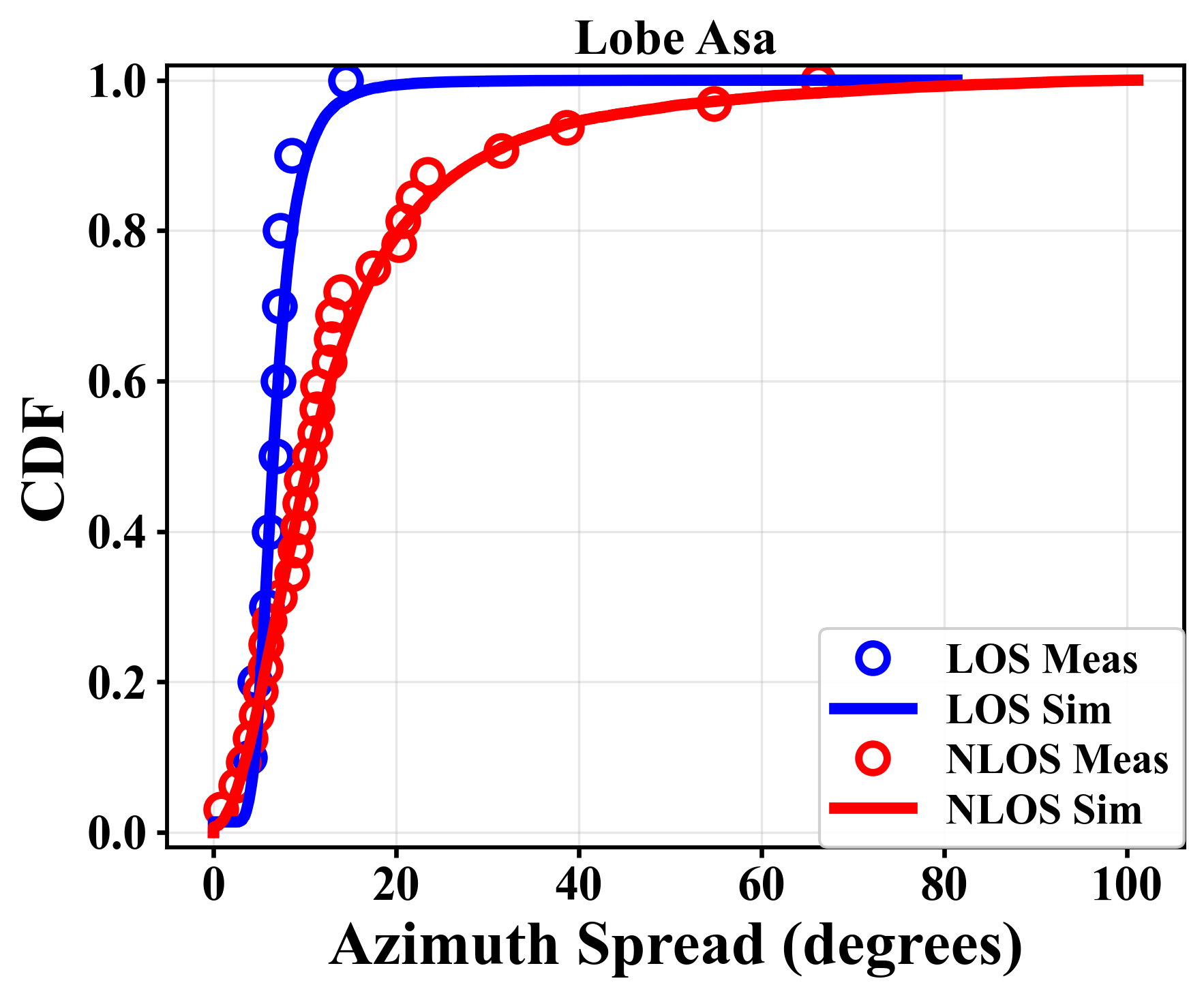}\label{fig:cm:lobe-asa}}\hfill
  \subfloat[Lobe RMS ZSD]{\includegraphics[width=0.2\textwidth]{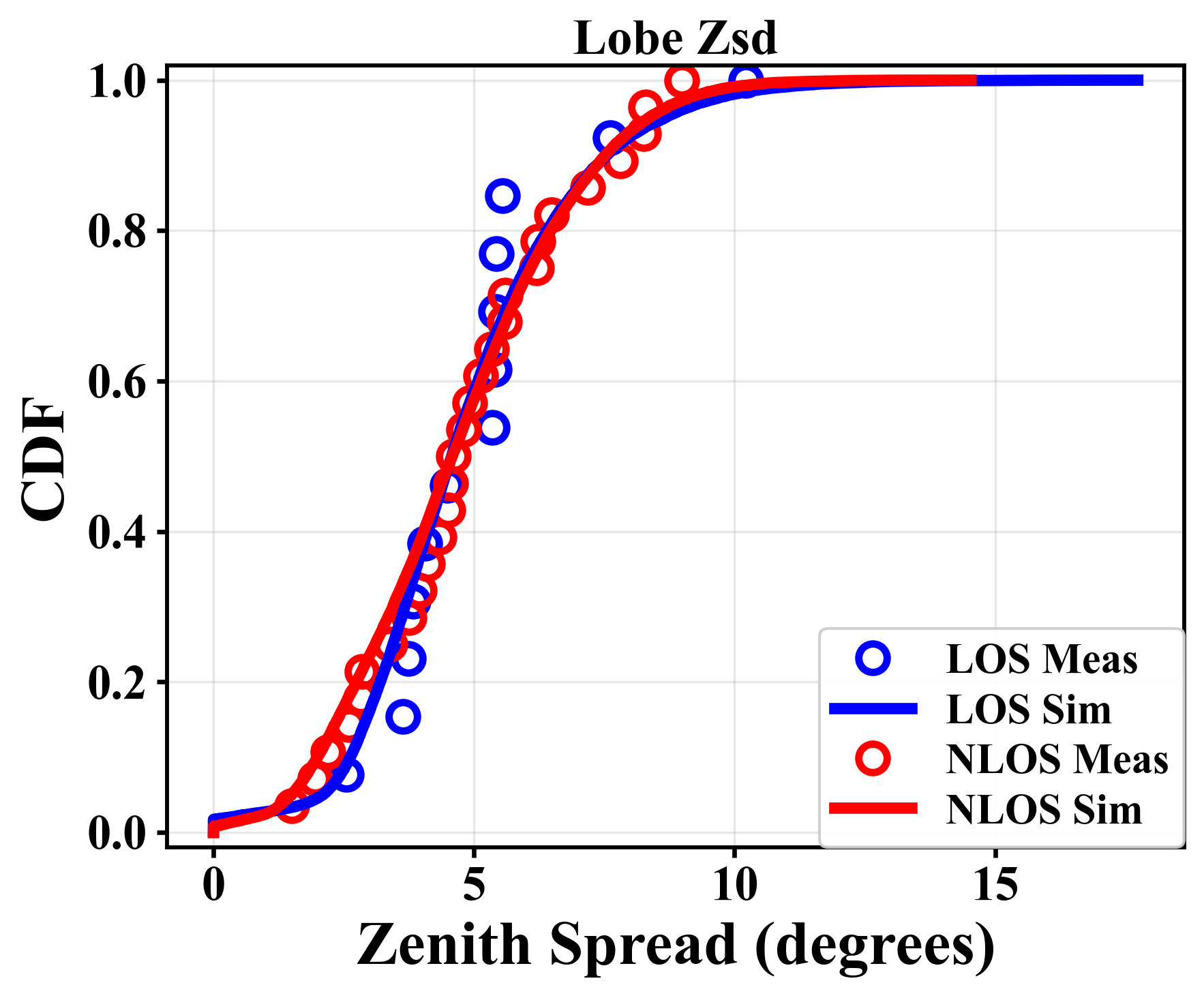}\label{fig:cm:lobe-zsd}}\hfill
  \subfloat[Lobe RMS ZSA]{\includegraphics[width=0.2\textwidth]{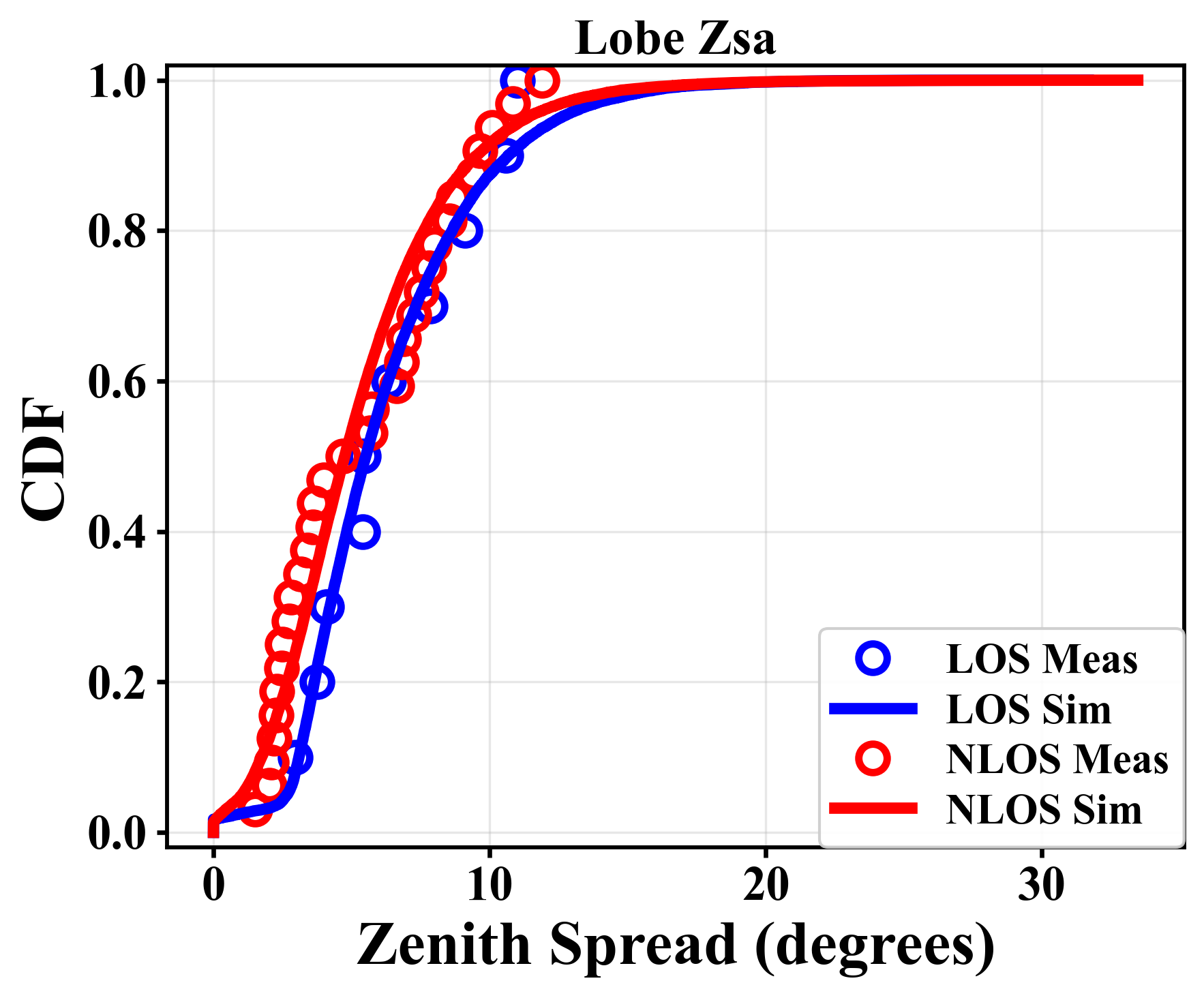}\label{fig:cm:lobe-zsa}}\\[-2pt]
  \hspace*{0.12\textwidth}%
  \subfloat[Omni RMS DS]{\includegraphics[width=0.2\textwidth]{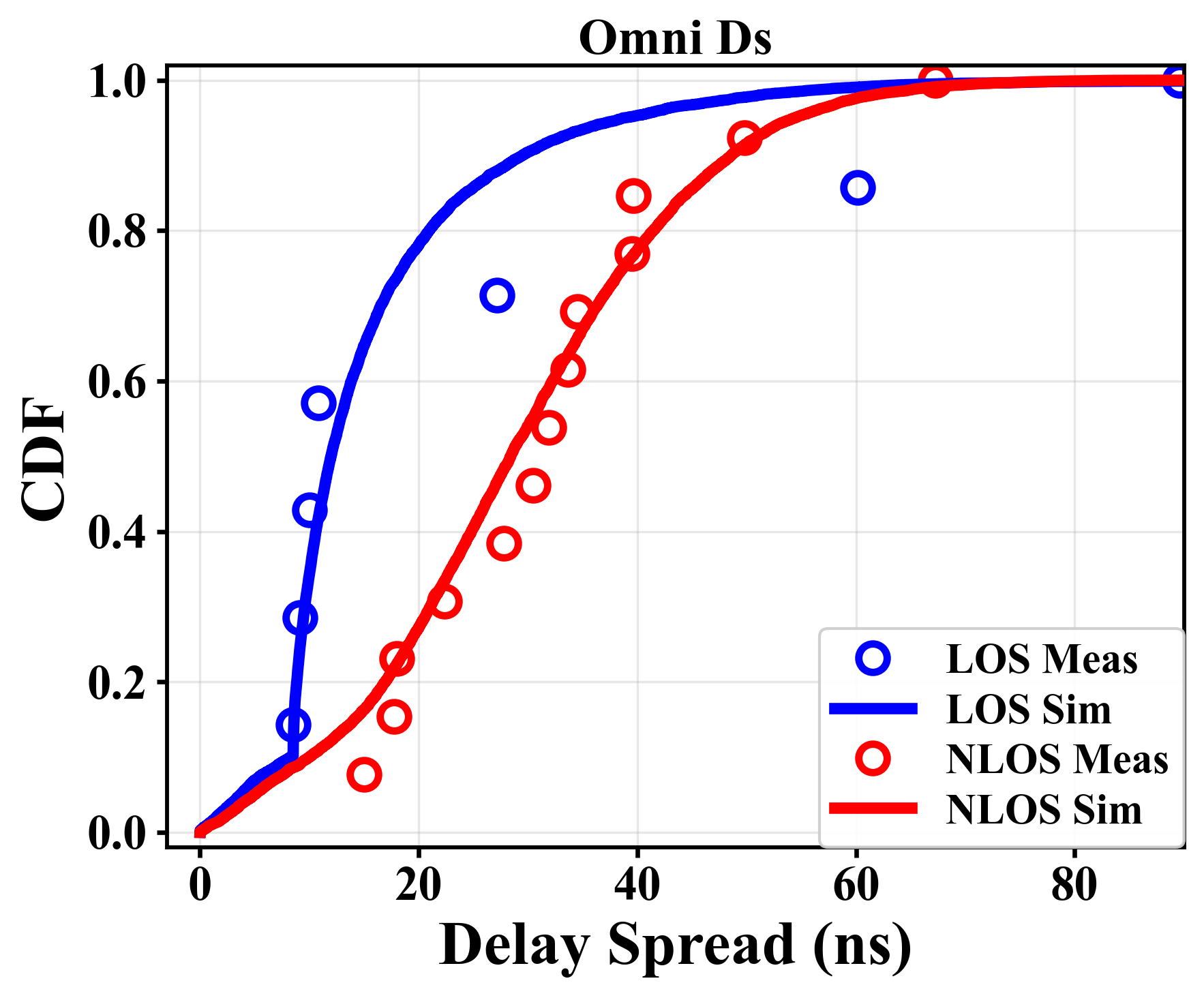}\label{fig:cm:omni-ds}}\hfill
  \subfloat[Dir RMS DS]{\includegraphics[width=0.2\textwidth]{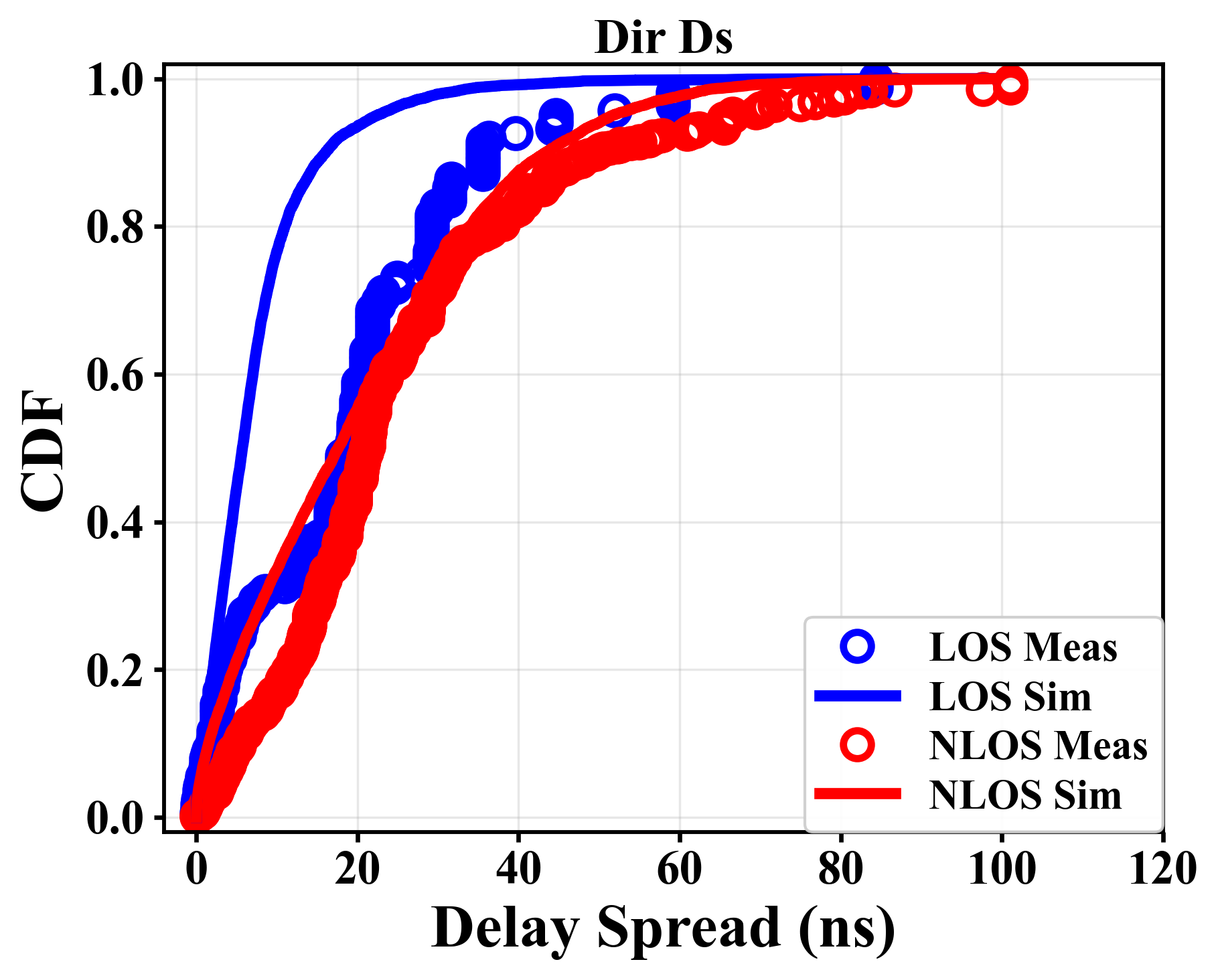}\label{fig:cm:dir-ds}}\hfill
  \subfloat[Cluster RMS DS]{\includegraphics[width=0.2\textwidth]{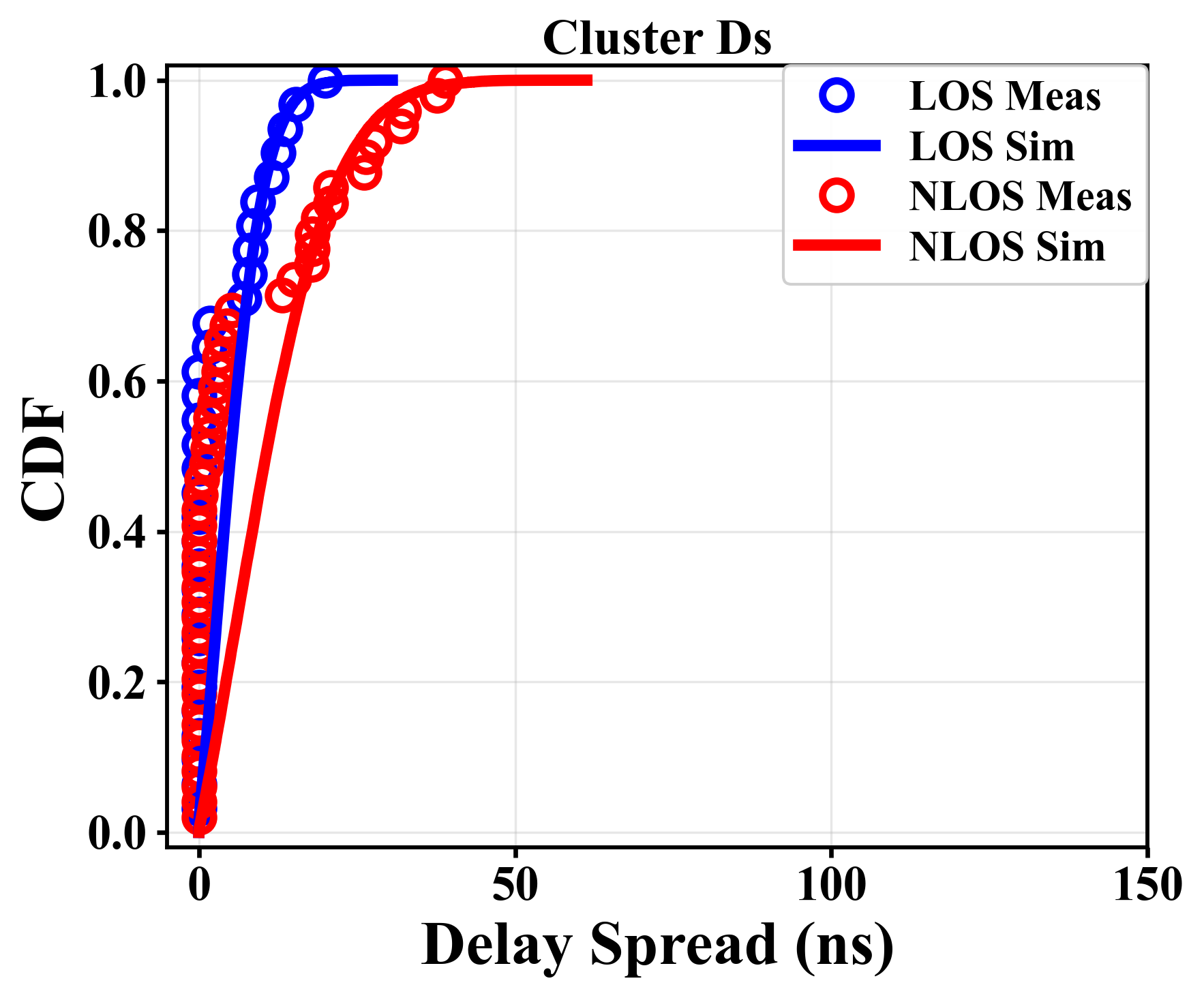}\label{fig:cm:cluster-ds}}%
  \hspace*{0.12\textwidth}
  \caption{Channel-modeling validation on the $16.95$\,GHz InH
  dataset. ECDFs from $10{,}000$ GenAI-driven channel
  realizations (solid lines) are overlaid on the measured ECDFs
  (circles); blue is LOS, red is NLOS.}
  \label{fig:cm-validation}
  \vspace{-0.1cm}
\end{figure*}

\subsection{NDE: Multi-Frequency Fitting}

To assess whether explicit task conditioning or multi-frequency
pooling improves NSF on the single-frequency $16.95$\,GHz InH
task, we compare three NSF variants on the same $n=28$ tasks. The
per-task baseline (NSF) is trained on the $16.95$\,GHz InH
measurements alone. The two multi-frequency variants additionally
use a $6.75$\,GHz InH campaign collected in the FR1(C) band at the same site as the $16.95$\,GHz measurements~\cite{Shakya2025InH}, providing auxiliary data from the same propagation environment at
a different carrier frequency.
NSF\textsubscript{C}, the FiLM-conditional variant, trains a single flow per task jointly on the $\{6.75,\,16.95\}$\,GHz data with FiLM modulation on the carrier frequency. NSF\textsubscript{P}, the pooled variant, trains a single flow per task on the union of the two bands with no frequency identifier provided to the network.

A Friedman omnibus test rejects equality of the three variants on
every metric ($p<2\!\times\!10^{-3}$). Cross-metric mean
normalized ranks are $0.583$ for NSF, $0.552$ for
NSF\textsubscript{C}, and $0.864$ for
NSF\textsubscript{P}, with Nemenyi critical difference
$\mathrm{CD}=0.209$ on the normalized scale for $k=3,\;n=28$.
NSF\textsubscript{C} attains the lowest mean rank on every metric, leading NSF individually on all five.
The composite gap of $0.031$ nevertheless stays within
$\mathrm{CD}$, and Fig.~\ref{fig:rdist-all}(e, f) shows
the NSF\textsubscript{C} density and CDF curves shifted to the
left of NSF on the rank axis.

NSF\textsubscript{C}'s advantage over NSF comes from the effective per-task training-set size. NSF observes only the $16.95$\,GHz samples per task, which is sparse for low-sample tasks such as LOS lobe counts and LOS inter-cluster delay. NSF\textsubscript{C} doubles the per-task sample count by training on both the $6.75$\,GHz and $16.95$\,GHz samples, while FiLM conditioning on the carrier frequency makes the spline output frequency-dependent and fits a different distribution at each band.

NSF\textsubscript{P} exhibits the opposite trend even though it
has access to the same auxiliary data. Without a per-sample
identifier, the pooled flow must fit a single distribution to the
union of the two campaigns. Channel parameters such as inter- and
intra-cluster delays, subpath angular spreads, and cluster-power
decay vary between the FR1(C) and FR3 bands, and the unconditional fit therefore converges to an aggregate distribution that
matches neither band accurately. The across-task mean rank of
$0.864$ exceeds that of NSF ($0.583$) by a margin larger than
$\mathrm{CD}$ on every metric, and the post-hoc test declares
NSF\textsubscript{P} significantly worse than both NSF and
NSF\textsubscript{C} on KS, CvM, AD, $W_1$, and Energy.
Concatenating the two campaigns and training one unconditional
NSF on the pool therefore degrades the $16.95$\,GHz fit despite
doubling the training data; the auxiliary $6.75$\,GHz samples
contribute to a better fit only when the model is informed of the
per-sample carrier frequency, as in NSF\textsubscript{C}.
\subsection{Parametric Fit and Exported Distributions}
\label{sec:results-parametric}

Table~\ref{tab:parametric-fits} lists the parametric family and fitted parameters selected by the augmented MLE step for each of the $14$ primary channel parameters at $16.95$\,GHz InH (LOS and NLOS). The candidate set for the MLE step comprises the four unconditional models (KDE, MDN, NSF, DDIM) and NSF\textsubscript{C}, and the per-task winner model generates the $5N$ augmentation samples.
A single parametric family is selected per parameter by minimum KS distance against the measurements. Figs.~\ref{fig:pf-counts} and~\ref{fig:pf-angles} show the corresponding parametric CDFs and the measured ECDFs. The fits match the measured shape closely, with KS $\leq 0.1$ on the subpath powers, the intra-cluster delay, and the lobe counts.
The remaining parameters stay within KS $\leq 0.19$, which is the range commonly accepted for statistical channel modeling.
The KS values confirm that the augmented MLE step delivers accurate closed-form fits for all $14$ channel parameters.

Cluster and subpath powers decay with their delays, making the joint distribution impossible to recover from independent marginal fits. ARIA therefore fits the powers conditionally on their delays using the conditional NSF mechanism of Section~\ref{sec:multi_freq}, with the delay replacing the carrier frequency as the conditioning input. One flow per scenario (LOS and NLOS) learns $p(P_n \mid \tau_n)$ for cluster powers and $p(\Pi_{m,n} \mid \rho_{m,n})$ for subpath powers. In our simulations, delays are drawn from their marginals first and the corresponding powers are then sampled from the trained conditional NSF, preserving the empirical delay--power coupling.

\begin{table*}[!h]
\centering
\caption{Measured and GenAI-simulated medians of the
second-order statistics at $16.95$\,GHz InH.}
\label{tab:second-order}
\setlength{\tabcolsep}{4pt}
\setlength{\arrayrulewidth}{1pt}
\footnotesize
{\renewcommand{\arraystretch}{1.34}%
\begin{tabular}[c]{|l|c|c|c|c|}
\hline
\multicolumn{5}{|c|}{\textbf{RMS DS [ns]}} \\
\hline
\multirow{2}{*}{\textbf{Param}} &
\multicolumn{2}{c|}{\textbf{LOS}} & \multicolumn{2}{c|}{\textbf{NLOS}} \\
\cline{2-5}
 & Meas & Sim & Meas & Sim \\
\hline
Omni    & $10.83$ & $12.02$ & $31.90$ & $29.82$ \\ \hline
Dir.    & $19.30$ & $13.18$ & $21.05$ & $18.82$ \\ \hline
Cluster & $0.00$  & $4.52$  & $1.41$  & $7.46$  \\ \hline
\end{tabular}}\hspace{4pt}%
{\renewcommand{\arraystretch}{1.15}%
\begin{tabular}[c]{|l|c|c|c|c|}
\hline
\multicolumn{5}{|c|}{\textbf{RMS AS [\textdegree]}} \\
\hline
\multirow{2}{*}{\textbf{Param}} &
\multicolumn{2}{c|}{\textbf{LOS}} & \multicolumn{2}{c|}{\textbf{NLOS}} \\
\cline{2-5}
 & Meas & Sim & Meas & Sim \\
\hline
Omni AOA & $7.33$  & $9.67$  & $69.49$ & $64.57$ \\ \hline
Omni AOD & $63.59$ & $59.19$ & $55.03$ & $54.76$ \\ \hline
Lobe AOA & $6.99$  & $6.55$  & $10.79$ & $10.55$ \\ \hline
Lobe AOD & $5.85$  & $6.46$  & $9.67$  & $9.43$  \\ \hline
\end{tabular}}\hspace{4pt}%
{\renewcommand{\arraystretch}{1.15}%
\begin{tabular}[c]{|l|c|c|c|c|}
\hline
\multicolumn{5}{|c|}{\textbf{RMS ZS [\textdegree]}} \\
\hline
\multirow{2}{*}{\textbf{Param}} &
\multicolumn{2}{c|}{\textbf{LOS}} & \multicolumn{2}{c|}{\textbf{NLOS}} \\
\cline{2-5}
 & Meas & Sim & Meas & Sim \\
\hline
Omni ZOA & $9.88$ & $9.82$ & $9.00$ & $8.51$ \\ \hline
Omni ZOD & $8.35$ & $7.89$ & $7.97$ & $7.55$ \\ \hline
Lobe ZOA & $5.89$ & $5.38$ & $5.18$ & $4.71$ \\ \hline
Lobe ZOD & $5.35$ & $4.48$ & $4.70$ & $4.54$ \\ \hline
\end{tabular}}
\vspace{-0.1cm}
\end{table*}

\section{Channel Modeling and Validation}
\label{sec:validation}

The parametric distributions in Table~\ref{tab:parametric-fits} feed the first-order parameters of a NYUSIM-style statistical channel model~\cite{Samimi2016, Shakya2024}, except for the cluster and subpath powers, which decay with delay and are therefore sampled from the delay-conditioned NSF of Section~\ref{sec:results-parametric}. The NYUSIM Monte Carlo generator draws channel realizations from the fitted distributions, and the resulting second-order statistics are compared against the measured channels. Because second-order statistics emerge from the joint sampling of multiple first-order parameters rather than being fitted directly, the comparison provides an end-to-end test of whether the GenAI-fitted distributions reproduce realistic channels.

The key second-order statistics used for validation are the omnidirectional and directional RMS delay spread and the omnidirectional and per-lobe RMS angular spread. The delay spreads depend on the inter-cluster delay, the intra-cluster delay, the cluster and subpath counts, and the cluster power decay. The per-lobe angular spread depends mainly on the subpath angular offsets; the omnidirectional version additionally depends on the number of spatial lobes and their mean-angle separation. Each spread therefore tests the joint consistency of several first-order parameters at once.

Fig.~\ref{fig:cm-validation} presents the ECDFs of
the measured and simulated second-order statistics on the $16.95$\,GHz InH campaign. Circles are the
measured ECDFs and solid lines are the corresponding ECDFs
computed from $10{,}000$ channel realizations for both LOS and NLOS scenarios. Table~\ref{tab:second-order} shows the medians of the measured
and GenAI-simulated ECDFs for the $11$ second-order statistics in Fig.~\ref{fig:cm-validation}. The omnidirectional RMS delay spread medians of the simulated channels match the measurements within about $2$\,ns, the per-lobe azimuth and zenith RMS spread medians within about $1^\circ$, and the omnidirectional zenith RMS spread medians within $0.5^\circ$. The close match between the measured and simulated ECDFs of the second-order statistics and their corresponding medians confirms that the GenAI-fitted distributions produce channels whose aggregate behavior matches the measurements.

\section{Conclusion}
\label{sec:conclusion}
We introduced \emph{ARIA}, an end-to-end generative AI pipeline that replaced the limited library of analytically chosen distribution families in statistical channel modeling with densities learned directly from the measurements. Starting from the raw measurement data, ARIA extracted the PDPs, recovered the primary channel parameters through TCSL analysis, fitted each parameter with four density estimators, and selected the per-task winner through a composite rank $\bar{R}$ over five goodness-of-fit metrics with Friedman--Nemenyi significance testing. An augmented MLE step, anchored to the measured samples, then converted each winning density into a closed-form parametric distribution of the same form that existing statistical channel simulators take as input. The fitted distributions fed a NYUSIM-style Monte Carlo channel generator, and the resulting second-order statistics were validated end-to-end against the measurements.

Instantiated on the NYU~WIRELESS $16.95$\,GHz InH campaign, NSF led the unconditional estimators with the lowest or tied-lowest mean rank on every metric, MDN followed within the Nemenyi critical difference, and KDE and DDIM formed the lower tier. Adding the $6.75$\,GHz campaign as a FiLM conditioning input lowered the mean rank on every metric, whereas pooling the two campaigns without the conditioning input degraded the $16.95$\,GHz fit. The augmented MLE step achieved KS~$\leq 0.1$ on the subpath powers, intra-cluster delay, and lobe counts, with the remaining parameters staying within KS~$\leq 0.19$. End-to-end, the GenAI-driven channel simulator reproduced the omnidirectional RMS delay spread medians within about $2$\,ns, the per-lobe azimuth and zenith spread medians within about $1^\circ$, and the omnidirectional zenith spread medians within $0.5^\circ$.

Future work can extend ARIA in two directions, namely FiLM-conditional synthesis of channel parameter distributions at unmeasured carrier frequencies, including the sparsely measured sub-THz bands, and conditioning on environment type and antenna configuration to merge scenario-specific models into a single generator.


\begin{thebibliography}{99}

\bibitem{3GPP_TS38101_1}
3GPP, ``NR; User equipment (UE) radio transmission and reception; Part~1: Range~1 Standalone,'' 3GPP TS~38.101-1, V18.5.0, 2024.

\bibitem{Bazzi2025}
A.~Bazzi, R.~Bomfin, M.~Mezzavilla, S.~Rangan, T.~S.~Rappaport, and M.~Chafii, ``Upper mid-band spectrum for 6G: Vision, opportunity and challenges,'' \textit{IEEE Commun. Mag.}, vol.~64, no.~1, pp.~206--212, Jan.~2026.

\bibitem{3GPP_TS38101_2}
3GPP, ``NR; User equipment (UE) radio transmission and reception; Part~2: Range~2 Standalone,'' 3GPP TS~38.101-2, V18.5.0, 2024.%

\bibitem{Rappaport2019}
T.~S.~Rappaport \textit{et al.}, ``Wireless communications and applications above 100~GHz: Opportunities and challenges for 6G and beyond,'' \textit{IEEE Access}, vol.~7, pp.~78729--78757, 2019.

\bibitem{Wang2025_6G}
H.~Gong, J.~Zhang, Y.~Zhang, and G.~Liu, ``New characteristics and modeling of 6G channels: Toward a unified channel model for standardization,'' \textit{IEEE Veh. Technol. Mag.}, vol.~20, no.~3, pp.~68--77, Sep.~2025.

\bibitem{IEEE80211ad}
IEEE, ``Enhancements for very high throughput in the 60~GHz band,'' IEEE Std~802.11ad-2012, Dec.~2012.

\bibitem{IEEE80211ay}
IEEE, ``Enhanced throughput for operation in license-exempt bands above 45~GHz,'' IEEE Std~802.11ay-2021, Jul.~2021.

\bibitem{Pi2011}
Z.~Pi and F.~Khan, ``An introduction to millimeter-wave mobile broadband systems,'' \textit{IEEE Commun. Mag.}, vol.~49, no.~6, pp.~101--107, Jun.~2011.

\bibitem{Rangan2014}
S.~Rangan, T.~S.~Rappaport, and E.~Erkip, ``Millimeter-wave cellular wireless networks: Potentials and challenges,'' \textit{Proc. IEEE}, vol.~102, no.~3, pp.~366--385, Mar.~2014.

\bibitem{Shakya2024}
D.~Shakya, S.~Ju, O.~Kanhere, H.~Poddar, Y.~Xing, and T.~S.~Rappaport, ``Radio propagation measurements and statistical channel models for outdoor urban microcells in open squares and streets at 142, 73, and 28~GHz,'' \textit{IEEE Trans. Antennas Propag.}, vol.~72, no.~4, pp.~3580--3595, Apr.~2024.

\bibitem{Shakya2025InH}
D.~Shakya \textit{et al.}, ``Comprehensive FR1(C) and FR3 lower and upper mid-band propagation and material penetration loss measurements and channel models in indoor environment for 5G and 6G,'' \textit{IEEE Open J. Commun. Soc.}, vol.~5, pp.~5192--5218, 2024.

\bibitem{3GPP38901}
3GPP, ``Study on channel model for frequencies from 0.5 to 100~GHz,'' 3GPP TR~38.901, V16.1.0, Dec.~2019.

\bibitem{Poddar2023}
H.~Poddar, S.~Ju, D.~Shakya, and T.~S.~Rappaport, ``A tutorial on NYUSIM: Sub-terahertz and millimeter-wave channel simulator for 5G, 6G, and beyond,'' \textit{IEEE Commun. Surveys Tuts.}, vol.~26, no.~2, pp.~824--857, 2nd Quart.,~2024.

\bibitem{Jariwala2026_NYUSIM}
I.~Jariwala \textit{et al.}, ``NYUSIM: A roadmap to AI-enabled statistical channel modeling and simulation,'' in \textit{Proc. IEEE Int. Conf. Commun. (ICC)}, Glasgow, UK, May~2026.

\bibitem{Samimi2016}
M.~K.~Samimi and T.~S.~Rappaport, ``3-D millimeter-wave statistical channel model for 5G wireless system design,'' \textit{IEEE Trans. Microw. Theory Techn.}, vol.~64, no.~7, pp.~2207--2225, Jul.~2016.

\bibitem{Sun2017}
S.~Sun, G.~R.~MacCartney, and T.~S.~Rappaport, ``A novel millimeter-wave channel simulator and applications for 5G wireless communications,'' in \textit{Proc. IEEE Int. Conf. Commun. (ICC)}, Paris, France, May~2017, pp.~1--7.

\bibitem{Jaeckel2014}
S.~Jaeckel, L.~Raschkowski, K.~B\"orner, and L.~Thiele, ``QuaDRiGa: A 3-D multi-cell channel model with time evolution for enabling virtual field trials,'' \textit{IEEE Trans. Antennas Propag.}, vol.~62, no.~6, pp.~3242--3256, Jun.~2014.

\bibitem{Liu2012_COST2100}
L.~Liu \textit{et al.}, ``The COST 2100 MIMO channel model,'' \textit{IEEE Wireless Commun.}, vol.~19, no.~6, pp.~92--99, Dec.~2012.

\bibitem{Bishop1994}
C.~M.~Bishop, ``Mixture density networks,'' Aston University, Birmingham, U.K., Tech. Rep. NCRG/94/004, 1994.

\bibitem{Rezende2015}
D.~J.~Rezende and S.~Mohamed, ``Variational inference with normalizing flows,'' in \textit{Proc. Int. Conf. Mach. Learn. (ICML)}, 2015, pp.~1530--1538.

\bibitem{Papamakarios2021}
G.~Papamakarios, E.~Nalisnick, D.~J.~Rezende, S.~Mohamed, and B.~Lakshminarayanan, ``Normalizing flows for probabilistic modeling and inference,'' \textit{J. Mach. Learn. Res.}, vol.~22, no.~57, pp.~1--64, 2021.

\bibitem{Durkan2019}
C.~Durkan, A.~Bekasov, I.~Murray, and G.~Papamakarios, ``Neural spline flows,'' in \textit{Proc. Adv. Neural Inf. Process. Syst. (NeurIPS)}, 2019, pp.~7509--7520.

\bibitem{SohlDickstein2015}
J.~Sohl-Dickstein, E.~Weiss, N.~Maheswaranathan, and S.~Ganguli, ``Deep unsupervised learning using nonequilibrium thermodynamics,'' in \textit{Proc. Int. Conf. Mach. Learn. (ICML)}, 2015, pp.~2256--2265.

\bibitem{Ho2020}
J.~Ho, A.~Jain, and P.~Abbeel, ``Denoising diffusion probabilistic models,'' in \textit{Proc. Adv. Neural Inf. Process. Syst. (NeurIPS)}, 2020, pp.~6840--6851.

\bibitem{Song2021}
J.~Song, C.~Meng, and S.~Ermon, ``Denoising diffusion implicit models,'' in \textit{Proc. Int. Conf. Learn. Representations (ICLR)}, 2021.

\bibitem{Lipman2023}
Y.~Lipman, R.~T.~Q.~Chen, H.~Ben-Hamu, M.~Nickel, and M.~Le, ``Flow matching for generative modeling,'' in \textit{Proc. Int. Conf. Learn. Representations (ICLR)}, 2023.

\bibitem{Fesl2024}
B.~Fesl, M.~Baur, F.~Strasser, M.~Joham, and W.~Utschick, ``Diffusion-based generative prior for low-complexity MIMO channel estimation,'' \textit{IEEE Wireless Commun. Lett.}, vol.~13, no.~12, pp.~3493--3497, Dec.~2024.

\bibitem{Orekondy2022}
T.~Orekondy, A.~Behboodi, and J.~B.~Soriaga, ``MIMO-GAN: Generative MIMO channel modeling,'' in \textit{Proc. IEEE Int. Conf. Commun. (ICC)}, Seoul, South Korea, May~2022, pp.~5322--5328.

\bibitem{Xia2024}
Z.~Hu, Y.~Li, and C.~Han, ``Transfer learning enabled transformer-based generative adversarial networks for modeling and generating terahertz channels,'' \textit{Commun. Eng.}, vol.~3, art.~no.~153, 2024.

\bibitem{Zhou2025_HighDim}
X.~Zhou, L.~Liang, J.~Zhang, P.~Jiang, Y.~Li, and S.~Jin, ``Generative diffusion models for high dimensional channel estimation,'' \textit{IEEE Trans. Wireless Commun.}, vol.~24, no.~7, pp.~5840--5854, Jul.~2025.

\bibitem{Huang2022_AIradioI}
C.~Huang \textit{et al.}, ``Artificial intelligence enabled radio propagation for communications---Part~I: Channel characterization and antenna-channel optimization,'' \textit{IEEE Trans. Antennas Propag.}, vol.~70, no.~6, pp.~3939--3954, Jun.~2022.

\bibitem{Huang2022_AIradioII}
C.~Huang \textit{et al.}, ``Artificial intelligence enabled radio propagation for communications---Part~II: Scenario identification and channel modeling,'' \textit{IEEE Trans. Antennas Propag.}, vol.~70, no.~6, pp.~3955--3969, Jun.~2022.

\bibitem{Zemen2025}
T.~Zemen \textit{et al.}, ``Site-specific radio channel representation for 5G and 6G,'' \textit{IEEE Commun. Mag.}, vol.~63, no.~6, pp.~106--113, Jun.~2025.

\bibitem{Samsung2025}
Samsung Research, ``Reimagining wireless channel modeling with generative AI,'' Dec.~2025. Available: \url{https://research.samsung.com/blog/Reimagining-Wireless-Channel-Modeling-with-Generative-AI}

\bibitem{MacCartney2017}
G.~R.~MacCartney and T.~S.~Rappaport, ``A flexible millimeter-wave channel sounder with absolute timing,'' \textit{IEEE J. Sel. Areas Commun.}, vol.~35, no.~6, pp.~1402--1418, Jun.~2017.

\bibitem{Ju2021}
S.~Ju, Y.~Xing, O.~Kanhere, and T.~S.~Rappaport, ``Millimeter wave and sub-terahertz spatial statistical channel model for an indoor office building,'' \textit{IEEE J. Sel. Areas Commun.}, vol.~39, no.~6, pp.~1561--1575, Jun.~2021.

\bibitem{Parzen1962}
E.~Parzen, ``On estimation of a probability density function and mode,'' \textit{Ann. Math. Statist.}, vol.~33, no.~3, pp.~1065--1076, Sep.~1962.

\bibitem{Silverman1986}
B.~W.~Silverman, \textit{Density Estimation for Statistics and Data Analysis}. London, U.K.: Chapman \& Hall, 1986.

\bibitem{Perez2018_FiLM}
E.~Perez, F.~Strub, H.~de~Vries, V.~Dumoulin, and A.~Courville, ``FiLM: Visual reasoning with a general conditioning layer,'' in \textit{Proc. AAAI Conf. Artif. Intell.}, vol.~32, no.~1, Feb.~2018, pp.~3942--3951.

\bibitem{Demsar2006}
J.~Dem\v{s}ar, ``Statistical comparisons of classifiers over multiple data sets,'' \textit{J. Mach. Learn. Res.}, vol.~7, pp.~1--30, Jan.~2006.

\bibitem{Kansal2023}
R.~Kansal \emph{et al.}, ``Evaluating generative models in high energy physics,'' \textit{Phys. Rev. D}, vol.~107, no.~7, p.~076017, Apr.~2023.

\bibitem{Massey1951}
F.~J.~Massey, ``The Kolmogorov--Smirnov test for goodness of fit,'' \textit{J. Amer. Statist. Assoc.}, vol.~46, no.~253, pp.~68--78, Mar.~1951.

\bibitem{Anderson1962}
T.~W.~Anderson, ``On the distribution of the two-sample Cram\'{e}r--von Mises criterion,'' \textit{Ann. Math. Statist.}, vol.~33, no.~3, pp.~1148--1159, Sep.~1962.

\bibitem{Anderson1954}
T.~W.~Anderson and D.~A.~Darling, ``A test of goodness of fit,'' \textit{J. Amer. Statist. Assoc.}, vol.~49, no.~268, pp.~765--769, Dec.~1954.

\bibitem{Villani2009}
C.~Villani, \textit{Optimal Transport: Old and New}. Berlin, Germany: Springer, 2009.

\bibitem{Szekely2013}
G.~J.~Sz\'{e}kely and M.~L.~Rizzo, ``Energy statistics: A class of statistics based on distances,'' \textit{J. Statist. Planning Inference}, vol.~143, no.~8, pp.~1249--1272, Aug.~2013.

\bibitem{Shakya2025UMi}
D.~Shakya \emph{et al.}, ``Urban outdoor propagation measurements and channel models at 6.75~GHz FR1(C) and 16.95~GHz FR3 upper mid-band spectrum for 5G and 6G,'' in \textit{Proc. IEEE Int. Conf. Commun. (ICC)}, Montreal, QC, Canada, Jun.~2025, pp.~3291--3296.

\bibitem{Shakya2025InF}
M.~Ying \emph{et al.}, ``Upper mid-band channel measurements and characterization at 6.75~GHz FR1(C) and 16.95~GHz FR3 in an indoor factory scenario,'' in \textit{Proc. IEEE Int. Conf. Commun. (ICC)}, Montreal, QC, Canada, Jun.~2025, pp.~3303--3308.

\bibitem{Steinbauer2001}
M.~Steinbauer, A.~F.~Molisch, and E.~Bonek, ``The double-directional radio channel,'' \textit{IEEE Antennas Propag. Mag.}, vol.~43, no.~4, pp.~51--63, Aug.~2001.

\bibitem{Scott1992}
D.~W.~Scott, \textit{Multivariate Density Estimation: Theory, Practice, and Visualization}. New York, NY, USA: Wiley, 1992.

\bibitem{Dumoulin2018}
V.~Dumoulin \emph{et al.}, ``Feature-wise transformations,'' \textit{Distill}, vol.~3, no.~7, p.~e11, Jul.~2018.

\bibitem{Friedman1937}
M.~Friedman, ``The use of ranks to avoid the assumption of normality implicit in the analysis of variance,'' \textit{J. Amer. Statist. Assoc.}, vol.~32, no.~200, pp.~675--701, Dec.~1937.

\bibitem{Nemenyi1963}
P.~B.~Nemenyi, ``Distribution-free multiple comparisons,'' Ph.D. dissertation, Princeton Univ., Princeton, NJ, USA, 1963.

\end{thebibliography}
\end{document}